\documentclass[fleqn,usenatbib]{mnras}

\usepackage{newtxtext,newtxmath}

\usepackage[T1]{fontenc}

\DeclareRobustCommand{\VAN}[3]{#2}
\let\VANthebibliography\thebibliography
\def\thebibliography{\DeclareRobustCommand{\VAN}[3]{##3}\VANthebibliography}

\usepackage{graphicx}	
\usepackage{amsmath}	
\usepackage{xspace}

\def\kms{km~s$^{-1}$}
\newcommand{\Gaia}{\emph{Gaia}\xspace}
\newcommand{\Hipparcos}{\textsc{Hipparcos}}
\newcommand{\Simbad}{\textsc{Simbad}}
\newcommand{\SBcat}{\emph{S$_{\mathrm{B}^9}$\xspace}}
\newcommand{\SBX}{\emph{S$_{\mathrm{B}^X}$\xspace}}
\newcommand{\jkt}[1]{{#1}}

\title[\SBcat\ catalogue]{The \SBcat\ catalogue\thanks{\url{https://www.astro.ulb.ac.be/sbx}}: status, comparison with non-single stars from \Gaia\ DR3 and evolution to \SBX}

\author[T. Merle et al.]{
T. Merle$^{1,2}$\thanks{E-mail: thibault.merle@ulb.be},
A. Jorissen$^1$, 
S. Alexandre$^1$,
J. Desuter$^1$, 
C. Loup$^3$, 
A. Tokovinin$^4$, 
G. Traven$^5$,
\newauthor 
M. Van der Swaelmen$^6$, 
S. Van Eck$^1$, 
G. Van de Steene$^2$,
J. Southworth$^7$, 
and G. Sadowski$^1$
\\
$^1$ BLU-ULB, Institut d'Astronomie et d'Astrophysique, Universit\'e Libre de Bruxelles, 1050 Brussels, Belgium\\
$^2$ Royal Observatory of Belgium, Avenue Circulaire 3, 1180 Brussels, Belgium\\
$^3$ Observatoire astronomique de Strasbourg, CDS, Université de Strasbourg, CNRS, UMR 7550, Strasbourg, France \\
$^4$ Cerro Tololo Inter-American Observatory, NSF’s NOIRLab Casilla 603, La Serena, Chile\\
$^5$ Faculty of Mathematics \& Physics, University of Ljubljana, Jadranska 19, 1000 Ljubljana, Slovenia\\
$^6$ INAF - Osservatorio Astrofisico di Arcetri, 50125, Firenze, Italy\\
$^7$ Astrophysics Group, Keele University, Keele, Staffordshire ST5 5BG, UK
}

\date{Accepted XXX. Received YYY; in original form ZZZ}

\pubyear{\the\year{}}

\begin{document}
\label{firstpage}
\pagerange{\pageref{firstpage}--\pageref{lastpage}}
\maketitle

\begin{abstract}
\emph{The Ninth Catalogue of Spectroscopic Binary Orbits} (\SBcat) is a comprehensive compilation of spectroscopic binaries (SBs) with orbital parameters sourced from the literature, comprising approximately 4000 systems with about 2800 single-lined and 1200 double-lined binaries. This work presents the latest status of the \SBcat\ catalogue after over two decades of development since its online inception in 2004. In particular, we expose the statistical properties of SBs in terms of orbital period distributions and eccentricity-period diagrams per spectral type and evolutionary stage. We perform a careful cross-match with the \Gaia\ Data Release 3 (DR3) to update astrometric parameters and compare with the \Gaia\ DR3 Non-Single Star (NSS) catalogue. Our cross-matching approach uses positional separations, magnitudes, and proper-motion back-propagation to identify counterparts. The final \SBcat\ version updated by D.~Pourbaix (2021-03-02) includes 4003~SB systems, some in higher-order multiples: 152 in triples, 71 in quadruples and 14 in higher-order systems. Among these 4003~SB, 3976 have matching \Gaia\ DR3 identifiers, while 21 are too bright and six too faint for \Gaia\ detection. red Ten \SBcat\ systems with periods larger than 1180~d (including a spectroscopic triple) have been spatially resolved by \Gaia~DR3. We identify a common sample of 827 \SBcat\ binaries cross-matched with \Gaia\ NSS, among which 655 are considered as reliable, based on relative period and absolute eccentricity differences not exceeding 10\%. The limited overlap (21\% of \SBcat) is primarily  due to selection cuts in NSS SB1 analysis, brightness limits, temporal baselines, and partial orbital solutions in the \Gaia\ NSS catalogue. This study highlights the strengths and limitations of both catalogues and establishes a clean benchmark sample for future binary star research. Our work marks the transition of \SBcat\ into \SBX, \emph{The eXtended Catalogue of Spectroscopic Binary Orbits}$^\star$, featuring a modern relational database, improved web interface, and Virtual Observatory access standards, aiming to enhance accessibility, data quality, and analysis capabilities for the binary and multiple star community.
\end{abstract}




\begin{figure*}
    \centering
    \includegraphics[width=0.47\linewidth]{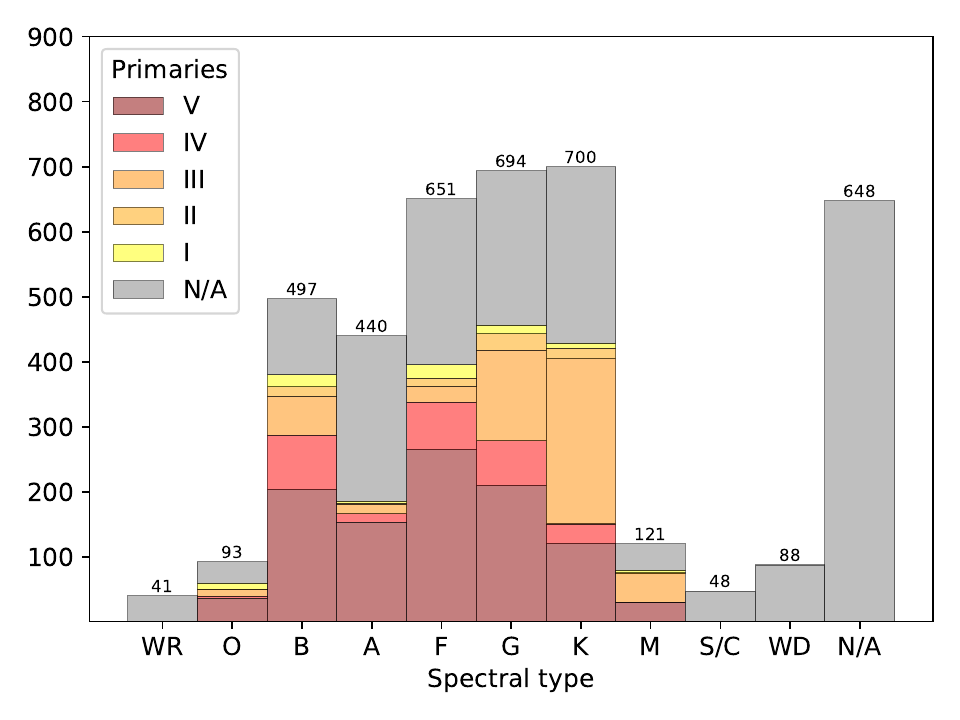}
    \includegraphics[width=0.50\linewidth]{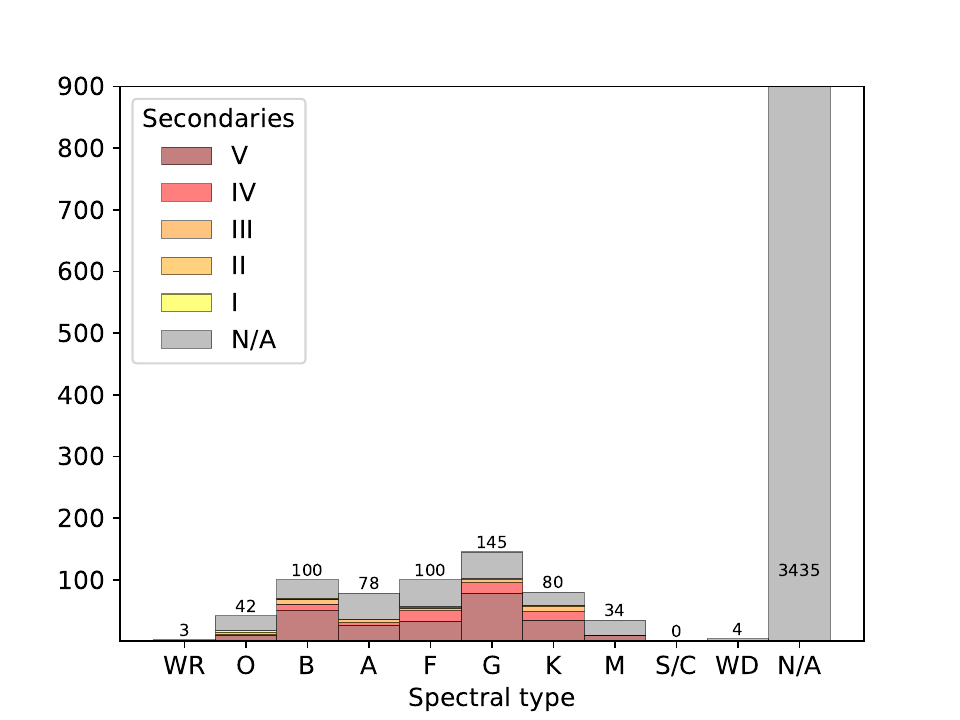}
    \caption{Distribution of the \SBcat\ binaries per spectral type and luminosity class for primaries (left) and secondaries (right). There are 16\% of primaries without spectral type (grey in the left panel) and only 9\% of secondaries with a spectral type (non-grey in the right panel). }
    \label{fig:histo_st_lc}
\end{figure*}

\section{Introduction}
Spectroscopic binaries (SB) are nowadays routinely detected in every large spectroscopic survey, by measuring radial-velocity (RV) variations either in temporal series of one visible component (SB1) or between two (or more) visible components (SB$n$, with $n\ge2$). Although detection is relatively easy, orbital characterisation requires a good sampling along the orbital cycle, which is only reachable for a marginal fraction of large spectroscopic-survey targets. For instance, there were only \jkt{two} SB1 (among 800 detected) for which we could determine an orbital solution from the observations obtained by the \Gaia-ESO Survey \citep{2020A&A...635A.155M}. In the infrared APOGEE survey DR16 and DR17, about 8\,000 SB candidates were identified, out of which a bit more than 300 have enough observations to constrain the orbital parameters \citep{2021AJ....162..184K}.
With the advent of the detection of dormant black holes in SB1 (with an early-type primary by \citealt{2022A&A...664A.159M, 2022NatAs...6.1085S}, and now even around late-type stars by \citealt{2023MNRAS.518.1057E,2024A&A...686L...2G}), it is of the uttermost importance to have a database of SB with reliable orbital and astrophysical parameters.

The \emph{Ninth Catalogue of Spectroscopic Binary Orbits} (\SBcat, \citealt{2004A&A...424..727P}) is \jkt{an} historical compilation of spectroscopic binaries with orbital parameters. It is the online continuation of the printed catalogue SB8 \citep{1989PDAO...17....1B} which contains 1\,469 SB systems and critical notes on most of the systems\footnote{Available on CDS Vizier at \url{https://cdsarc.cds.unistra.fr/viz-bin/cat/V/64}}. For completeness we recall the earlier versions of this catalogue, historically started at the Lick Observatory, with the five first versions of the SB catalogue \citep{1905LicOB...3..136C, 1910LicOB...6...17C, 1924LicOB..11..141M, 1936LicOB..18....1M, 1948LicOB..20....1M}; then pursued at the Dominion Astrophysical Observatory with the sixth \citep{1967PDAO...13..119B}, seventh \citep{1978PDAO...15..121B} and eighth \citep{1989PDAO...17....1B} versions of the catalogue. Since 2000, and under the advocacy of A. Tokovinin, the catalogue was sponsored by the IAU, through the former commission C30 (`Radial Velocities') inherited by commission G1 (`Binary and Multiple Star Systems') in 2015, led by D. Pourbaix\footnote{who deceased on Nov. 14, 2021} and hosted at the \textit{Université Libre de Bruxelles}, Belgium. The last update of the \SBcat\ catalogue by D.~Pourbaix was done on 2021-03-02, and contains more than 5\,000 orbits. The queryable database is available at \url{https://sb9.astro.ulb.ac.be/} with a mirror at CDS Vizier \url{https://cdsarc.cds.unistra.fr/viz-bin/cat/B/sb9}. We stress that this version was the last maintained by D.~Pourbaix and does not include the recently revised orbits using HERMES/Mercator \citep{2011A&A...526A..69R} spectroscopic data for about 50 systems \citep{2024A&A...684A..74M}, and about 56 spectroscopic orbits, many belonging to stellar triples from \citet{2020AJ....160..251T} and \citet{2020AJ....160...69T,2022AJ....163..161T}.

The \SBcat\ catalogue serves as a reference and comparison sample for numerous studies on stellar binaries. It is also highly valuable due to its copious notes. We try to compile here a non-exhaustive list of most important studies which are based on the \SBcat\ catalogue. It has been of fundamental importance for reviews on stellar multiplicity  \citep[\emph{e.g.}][]{2013ARA&A..51..269D, 2017ApJS..230...15M} and other disciplines like astroinformatics \citep{2022CoBAO..69..251A}. It was also used in large spectroscopic surveys  (\emph{e.g.}, Gaia-ESO: \citealt{2020A&A...635A.155M}, APOGEE: \citealt{2021AJ....162..184K}, LAMOST: \citealt{2025ApJS..276...11L}). The \SBcat\ catalogue can be used as well for selecting samples for astrometric studies \citep[\emph{e.g.}][]{2005A&A...442..365J, 2007A&A...464..377F, 2020MNRAS.496.1922B} and for feeding other databases \citep[\emph{e.g.} MSC: ][]{tokovinin2018}. It can serve for searching \jkt{for} companions to close SBs \citep{2006A&A...450..681T}, investigating solar-like oscillators \citep{2024A&A...682A...7B}, or solar twins \citep{2009AJ....137.3442S}, \jkt{and} studying stellar rotation in B-type stars \citep{2010ApJ...722..605H}, etc. It can serve as \jkt{a} reference for testing new techniques like Bayesian inference or Markov Chain Monte Carlo methods \citep[\emph{e.g.}][]{2022AJ....163..220V, 2022AJ....163..118A, 2023A&A...672A..82L}. It is also important for understanding the impact of the companions on the evolution of stars \citep[\emph{e.g.}][]{2008EAS....29....1M, 2008MNRAS.389..869E}, for searching for brown-dwarf companions \citep{2011A&A...525A..95S}, or compact companions \citep{2023MNRAS.521.5927J}, and even in exoplanetary science  \citep{2015ARA&A..53..409W}.

The \Gaia mission \citep{2016A&A...595A...1G} is an ambitious ESA spacecraft measuring the positions, parallaxes, proper motions and magnitudes of about 2 billion stars. In addition the on-board Radial Velocity Spectrometer \citep[RVS,][]{2023A&A...674A...5K} provides RV for more than 33 millions stars. The third Data Release \citep[DR3,][]{2023A&A...674A...1G} includes the Non-Single Star catalogue \citep[NSS,][]{2023A&A...674A..34G} which  outclasses, in a single release, all the binary catalogues known so far. In summary, the NSS catalogue includes 813\,000 sources, and among them about 277\,000 SB. The individual epoch spectra will be released in DR4.   

The goal of the present work is to present the status of the last version of the \SBcat\ catalogue (Sect.~\ref{sec:status}), to present its careful cross-match with \Gaia~DR3 (Sect.~\ref{sec:xmatch}), and to critically compare it with the \Gaia DR3 NSS catalogue (Sect.~\ref{sec:gaia}). Finally we discuss the long-term future improvements planned for the \SBcat\ database (Sect.~\ref{sec:improvements}).

\begin{figure}
    \includegraphics[width=\linewidth]{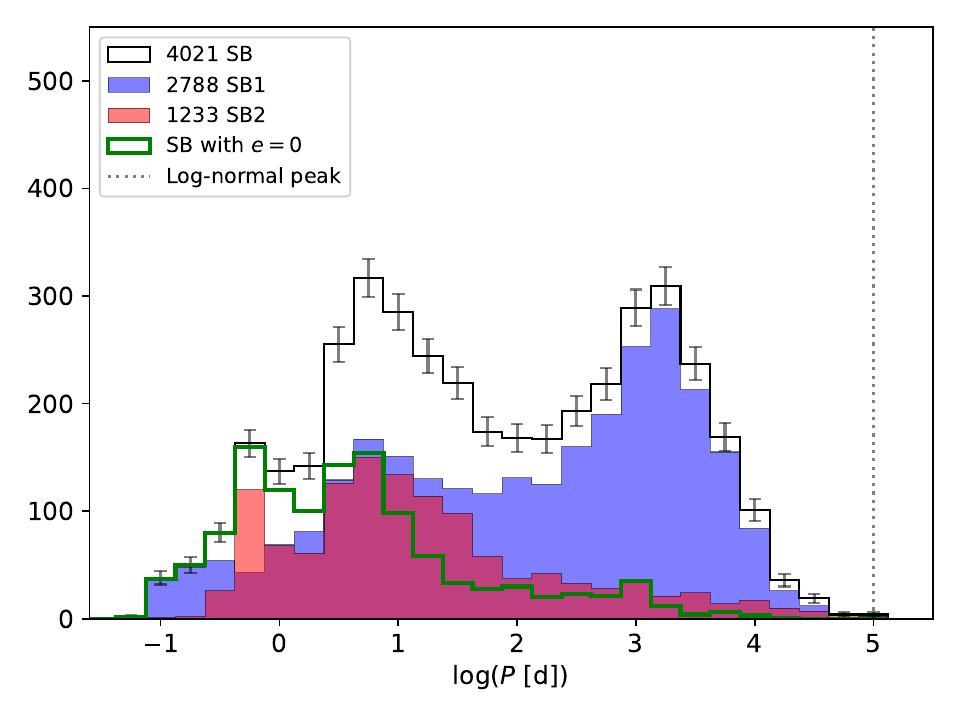}   
    \caption{Distribution of the orbital periods of the SB1, SB2 and combination of them. The distribution shows three peaks at 0.6, 7 and 1800~d which are discussed in the text. {\color{red} Error bars are taken as Poissonian.} The dotted line represents the mode of the log-normal distribution of binaries \citep{2017ApJS..230...15M} peaking around 270 a. SBs lie on the short-period side of this distribution.}
    \label{fig:histo_periods}
\end{figure}

\begin{figure*}
    \centering
    \includegraphics[width=0.49\linewidth]{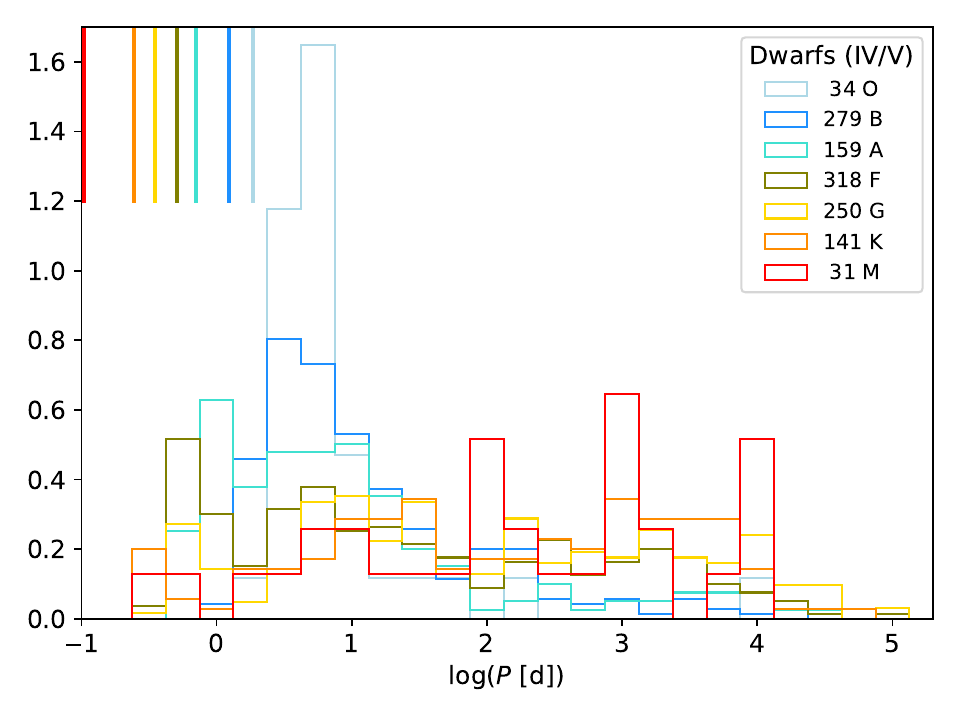}
    \includegraphics[width=0.49\linewidth]{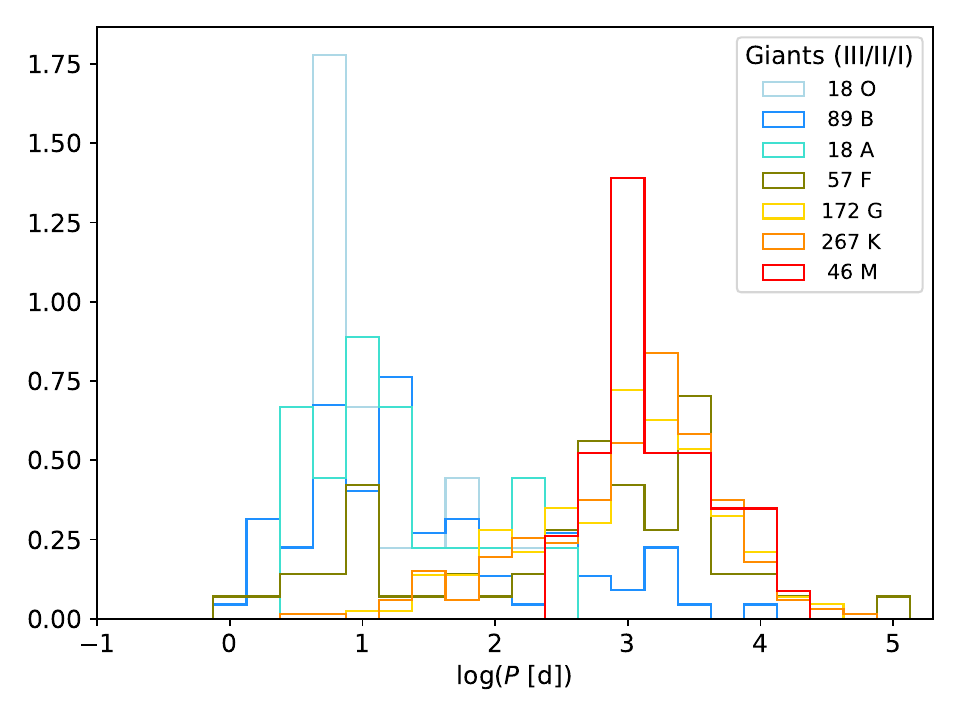}
    \caption{Distribution of the orbital periods per spectral type for dwarfs (left) and giants (right) of SB primary components. The $y$-axis corresponds to the period distribution normalised so that the integral of the frequency times the bin size equals unity.
    Colour codes the spectral types. In the left panel, vertical lines correspond to the minimum period for a given spectral-type from contact equal-mass binaries.}
    \label{fig:histo_periods_per_stlc}
\end{figure*}

\section{Current status}
\label{sec:status}

\subsection{SB1 and SB2}
In the \SBcat\ catalogue, each entry is identified by a \emph{system number} to which several orbits may be attached. A unique \emph{orbit number} is associated to the corresponding published reference. In the current version (2021-03-02), there are exactly 5042 orbits belonging to 4021 binaries (\emph{i.e.} \SBcat\ systems). We immediately note that some of these \SBcat\ systems can be part of higher-order stellar multiples, preventing us, at \jkt{present, from clearly stating} how many physical stellar multiples there are in the \SBcat\ catalogue. To identify SB1 and SB2, we look at orbits that have non-null semi-amplitudes $K_1$ \emph{or} $K_2$\footnote{There are 11 orbits for which $K_2$ alone is defined.} (SB1), or $K_1$ \emph{and} $K_2$ (SB2), respectively. It should be noted that there are 129 \SBcat\ systems which have both SB1 and SB2 orbital solutions. We consider them as SB2 in the subsequent analysis, since their composite nature has been revealed. In the \SBcat\ catalogue, 2/3 of the system numbers are SB1 (2\,788) and 1/3 are  SB2 (1\,233). RV data are included in the catalogue for 3\,111 \SBcat\ systems representing 77\% of all systems. Systems with missing RV data come from earlier versions of the catalogue, and a quick look at some of these systems reveals that these data are generally available in the original publications, but not easy to extract. Finally, a grade from 0 (tentative) to 5 (definitive)\footnote{A unique system has a grade of 5.1, it is \SBcat\ system 815, \emph{i.e.} $\alpha$ Cen.} was originally set for each orbit but was gradually abandoned in the last version of the catalogue. In fact, the grading scheme was not based in any way on objective criteria but was based on the subjective experience of the providers of the earliest versions. D. Pourbaix had started to think about a new grading scheme, but this scheme was never implemented. About half of the orbits have a grading mark.

\subsection{Statistical properties}
\subsubsection{Spectral types and luminosity classes}
The \SBcat\ catalogue lacks a quantitative astrophysical characterisation of the stars. The components of each SB are only characterised by a spectral type and a luminosity class, when available. 
The spectral type classification generally comes from the M-K scheme reported by the authors who published the orbital parameters, except for 145 SBs with A-F primaries \citep[from][]{2009ApJS..180..117A}. 
The distribution per spectral type and luminosity class of the \SBcat\ primaries and secondaries are displayed on Fig.~\ref{fig:histo_st_lc}. Neither the distribution of stars among spectral types nor the distribution of luminosity classes do reflect the stellar mass function. In the \SBcat\ catalogue, SBs with an early-type primary are over-represented while late-type dwarf primaries are under-represented as compared to giants. For GKM spectral types, main-sequence (V) stars are also largely under-represented. We stress that the fraction of primaries without a determined spectral type (16\%) or a luminosity class (51\%) is fairly large, increasing to 85\% and 91\% for secondaries, partially explained by the fact that there are approximately 2/3 of SB1 and 1/3 of SB2 in the \SBcat\ catalogue. The fraction of SB2 with a known spectral type or luminosity class is 46\% and 30\% respectively. This situation calls for improvements (see Sect.~\ref{sec:improvements}). 

\subsubsection{The orbital-period distribution}

The aggregated orbital-period distribution (black line) for all SBs in the \SBcat\ catalogue is presented in Fig.~\ref{fig:histo_periods}. In addition, we show the distributions for SB1s (blue shading), SB2s (red shading), and circular SBs (green line). Almost all SBs with periods shorter than 1~d have been circularised. The aggregated orbital-period distribution presents three peaks at $\sim$\,0.6, $\sim$\,7 and $1800$~d. Interpreting the characteristics of these distributions is challenging, as their properties emerge from multifaceted origins, encompassing both physical mechanisms and selection biases. In the following, we try to disentangle these influences.

We show on the left panel of Fig.~\ref{fig:histo_periods_per_stlc} SBs with dwarf primaries and on the right SBs with giant primaries colour-coded per spectral type. We also added the theoretical expectations for the minimum period for a given spectral type assuming equal-mass contact binaries on the main sequence, \emph{i.e.}, with the primary filling its Roche lobe. We use the formula from \citet{1983ApJ...268..368E} to express the Roche radius $R_{\max}$. From Kepler's third law, it is then possible to evaluate the minimum period $P_{\min}$ of a binary system able to host a Roche-lobe-filling star ($R_{\max} = R$) of a given spectral type characterized by radius $R$ and mass $M$:
\begin{equation}
    P_{\min} = 0.35\; R_{\max}^{3/2}\; M^{-1/2} 
    \label{Eq:PourbaixEnvelope}
\end{equation}
with $P_{\min}$ in days, and $M$ and $R_{\max}$ in solar units. We use an \href{https://www.pas.rochester.edu/~emamajek/EEM_dwarf_UBVIJHK_colors_Teff.txt}{updated} version of the calibrations of \citet{2013ApJS..208....9P} for estimating $M$ and $R_{\max}$ per spectral type. For clarity, we show individual panels per spectral type in Appendix~A (Fig.~\ref{fig:histo_periods_details}).
We now try to decode the shape of the orbital-period distribution with a mixture of physical and observational-bias effects.

The left-hand ridge of the first peak (around 0.6~d) is clearly associated with the physical limit of contact-binary orbits as demonstrated by the left panels of Figs.~\ref{fig:histo_periods_per_stlc} and \ref{fig:histo_periods_details}. The theoretical minimum-period expectations for contact binaries go from 0.1~d for M dwarfs to 1~d for O dwarfs which are well matched by the observed distributions, except for M dwarfs. The latter are much less numerous than the other spectral types in \SBcat\ despite constituting the vast majority of stars. The SB with the shortest orbit in the \SBcat\ catalogue is an SB1 cataclysmic variable: the dwarf nova GW~Lib with an orbital period of $1.280\pm0.001$~h \citep{2002PASP..114.1108T}.
On the contrary, the right-hand shallower slope of the first peak is probably an observational bias caused by (i) astronomers favouring studies of contact binaries (hence the number of contact binaries with a derived orbit is larger than their longer-period neighbours), and (ii) orbits of early-type binaries, with their few broad lines, are increasingly difficult to characterize at large periods when the RV amplitude  becomes small.

The second peak (around 7~d) is, like the first peak, probably a combination of observational and physical effects. Even though this second peak corresponds to the maximum of 
the orbital-period distribution of SB2 systems (i.e., with nearly twin components), a look at Fig.~\ref{fig:histo_periods_details} showing the distribution of SB2 systems among the different dwarf spectral types reveals that this correspondence seems to be purely coincidental with no background physical meaning.

The origin of the second peak is more difficult to trace than the first peak. 
On its left ridge, the peak probably results from the early-type-star (O and B) period distribution peaking at about 7~d
(close to their contact limit below which the period distribution cannot go) with the later types presenting a dip at periods slightly 
shorter than 7~d, before climbing up again towards their contact threshold (Fig.~\ref{fig:histo_periods_details}). 
On its right ridge, the
peak at 7~d seems to be the combination of 
the number of early-type orbits diminishing towards longer periods (as already argued in relation with the first peak, since for O-type stars, the first and second peaks are superimposed)
and the period distribution of late-type stars remaining flat. The above arguments alone would hint at the biased 
nature of the 7~d peak. However, several physical scenarios have been developed to explain the accumulation of binaries in 
the period range from 1 to 10~d \citep[][and references therein]{2024BSRSL..93..170M}. Among them, secular evolution through Kozai-Lidov cycles and tidal friction involving a third distant companion (\emph{i.e.} evolution of compact hierarchical triples) is appealing \citep[\emph{e.g.}][see also Fig.~25 in \citealt{2025A&A...693A.124G}]{2007ApJ...669.1298F,2018MNRAS.479.4749B, 2022IAUS..364...52L}. This scenario is also supported by the stellar triples recently found by comparing \Gaia\ NSS astrometric acceleration and SB solutions \citep{2023A&A...674A..34G}, and the ones found by cross-matching with the catalogue of wide binaries \citep{2021MNRAS.506.2269E}, which all show a peak between 5 and 10~d (see Figs. 52, 53, and 54 in \citealt{2023A&A...674A..34G}). Binaries with late-type (G, K, M) giant primaries (right panels of Figs.~\ref{fig:histo_periods_per_stlc} and \ref{fig:histo_periods_details}) do not contribute to the second peak, since the physical limit set by the  Roche-lobe radius makes $P_{\min}$ increase along the red and asymptotic giant branches.

The third peak at $\sim 1800$~d (5~a) is dominated by SB1 involving a late-type primary. It is clearly an observational bias because, for periods beyond this peak, the RV variations drop below the detection limit of current spectrographs, while orbital periods of the widest bound pairs may reach $10^8-10^9$~d (\emph{e.g.}, Proxima~Cen orbits $\alpha$~Cen in approximately half a million years; see, \emph{e.g.}, \citealt{2017A&A...598L...7K}). For SB1s in \Gaia\ DR3 NSS, this peak is just below 1000~d and only reflects the inability of Gaia DR3  to detect binaries with an orbital period much longer than the RV-monitoring time span (3~years for \Gaia\ DR3). It is indeed clear from studies gathering all kinds of binaries (thus including visual binaries -- VB) that the main reservoir of binaries (\emph{i.e.} at the peak of the log-normal distribution) is located around 230~a \citep{2017ApJS..230...15M}.
The SB with the longest period in the \SBcat\ catalogue is a twin VB-SB2 of K0V components with an estimated period of $320\pm20$~a \citep{2000A&AS..145..215P}, which is close to the peak position of the log-normal distribution of late-type stars \citep{2017ApJS..230...15M}. 
Therefore, the lack of long-period SBs is primarily due to the absence of long-term RV data series. Such data are available only for a modest number of relatively bright stars, and their uncertainties often exceed a few \kms. 

\begin{figure*}
    \centering
    \includegraphics[width=0.49\linewidth]{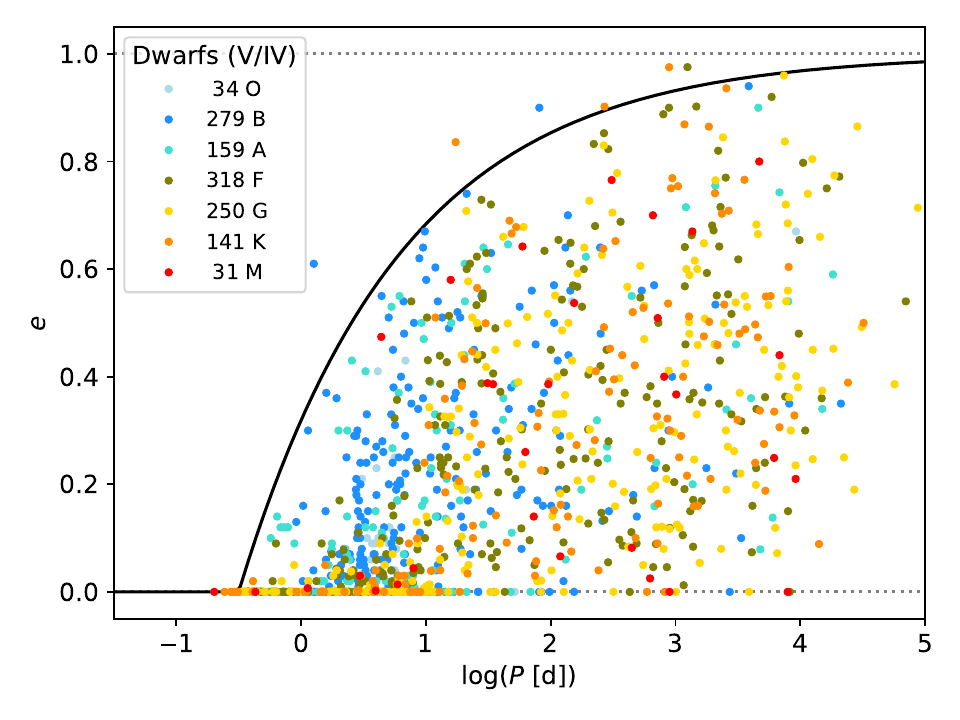}
    \includegraphics[width=0.49\linewidth]{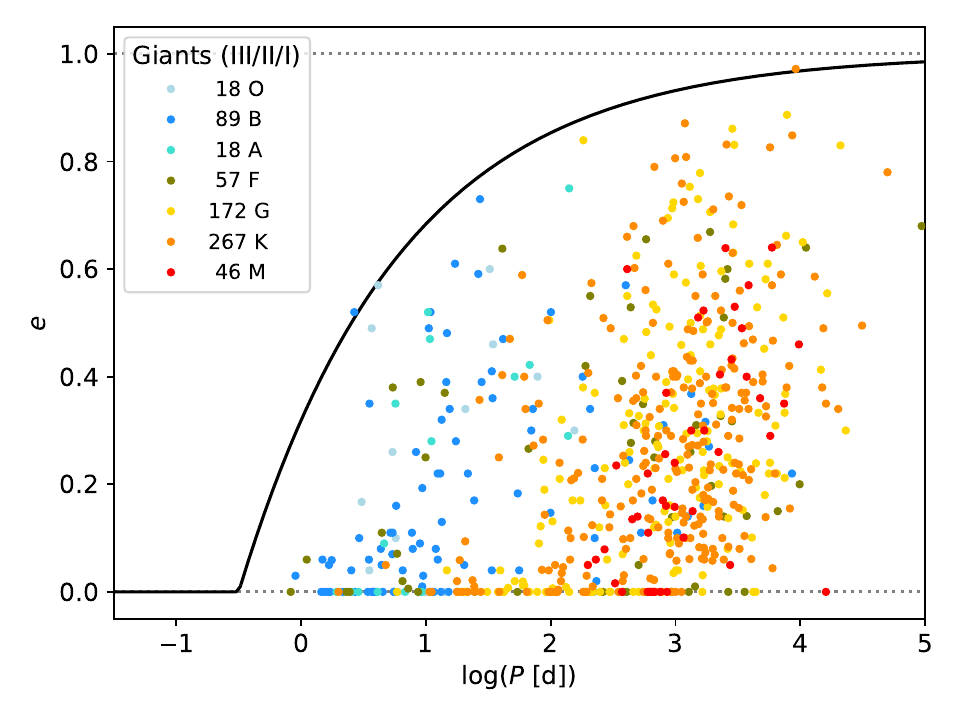}
    \caption{The $e - \log{P}$ diagram of the \SBcat\ binaries colour-coded by spectral type for dwarfs (left) and for giants (right panel). The solid line is the \emph{empirical} fit to the \SBcat\ data \citep{2004A&A...424..727P}.
    }
    \label{fig:e_logp}
\end{figure*}

\subsubsection{The eccentricity-period diagram}

The eccentricity-period diagram for the \SBcat\ catalogue is displayed in Fig.~\ref{fig:e_logp} for dwarf primaries (left) and giant primaries (right) colour-coded by spectral type.
Since models predict that binaries can form at all eccentricities (except may be at the smallest ones: $e \le 0.05$), the presence of an upper envelope to the data points may be interpreted as the signature of the tidal processes leading to the many circular orbits accumulating at the shortest periods, and depleting in priority the upper left corner of the eccentricity-period diagram. 
Circularisation mechanisms are still highly debated, and the corresponding theoretical and observed behaviour is best illustrated in \citet{2022ApJ...927L..36B} and \citet{2023MNRAS.524.3978M}. 
In Fig.~\ref{fig:e_logp}, we show the \emph{empirical} envelope $e_{\max} = 1-(P_0/P)^{1/3}$ fitted by \citet{2004A&A...424..727P} on \SBcat\ data with the circular period $P_0=0.35$~d roughly corresponding to contact systems of solar-type stars (Table~\ref{tab:Pc_st}).
Comparing the left and right panels of Fig.~\ref{fig:e_logp}, we notice a shift of the envelopes towards longer periods with later spectral types, much better marked for giants as compared to dwarfs. This shift has not the same origin for dwarf and giant primaries though, as we now discuss. For clarity, we show individual panels (for dwarf and giant primaries per spectral type) in Appendix A, Fig.~\ref{fig:elogP_details}. There we used instead the \emph{periastron} envelope\footnote{
For the sake of completeness, we also mention the existence of the \emph{circularisation} envelope $e_{\max} = [1-(P_{\mathrm c}/P)^{2/3}]^{1/2}$ derived from equilibrium-tide theory \citep{1981A&A....99..126H} that we do not use in the present work.}
$e_{\max}=1-(P_{\mathrm c}/P)^{2/3}$ derived from the Roche lobe theory \citep{2008EAS....29....1M,2009A&A...498..489J}. We may generalise Eq.~\ref{Eq:PourbaixEnvelope} by introducing the filling factor $f=R/R_{\max}$:
\begin{equation}
\label{eq:Pc_f}
    P_{\mathrm c} = 0.35 (R/f)^{3/2} M^{-1/2},
\end{equation}
thus defining a critical period for circularisation in systems which reach a filling factor $f$ at periastron,
with $P_{\mathrm c}$ in days and $M$ and $R_{\max}$ in solar units. Typical values per spectral type are provided in Table~\ref{tab:Pc_st} together with circular periods for the three filling factor values of 1, 0.5 and 0.25, displayed on the left panels of  Fig.~\ref{fig:elogP_details}.

\begin{table}
    \centering
    \caption{Circular periods $P_\mathrm{c}$ as a function of the filling factor $f$ for a typical giant primary of 2~M$_\odot$ and for typical main sequence primaries. Components of equal masses have been assumed.}
    \begin{tabular}{cccccc}
\hline
Spectral Type & $M$  & $R$  & \multicolumn{3}{c}{$P_\mathrm{c}$ [d]} \\
 & [M$_\odot$] & [R$_\odot$] & $f=1.0$ & $f=0.5$ & $f=0.25$ \\
\hline
\noalign{Giants}
\hline 
  & 2.0 & 500& 2800 & 7800 & 22000 \\
  & 2.0 & 100 & 250 & 700 & 2000 \\
  & 2.0 & 10  & 7.8 & 22 & 63 \\ 
\hline
\noalign{Dwarfs}
\hline
O & 35  & 10  & 1.9 & 5.3 & 15.0\\
B & 9.9 & 5.0 & 1.2 & 3.5 & 10.0\\
A & 2.0 & 2.0 & 0.7 & 2.0 &  5.6\\
F & 1.3 & 1.4 & 0.5 & 1.4 &  4.1\\
G & 1.0 & 1.0 & 0.4 & 1.0 &  2.8\\
K & 0.7 & 0.7 & 0.2 & 0.7 &  2.0\\
M & 0.3 & 0.3 & 0.1 & 0.3 &  0.8\\
\hline
    \end{tabular}
    \label{tab:Pc_st}
\end{table}

For systems with late-type dwarf primaries (left panels of Figs.~\ref{fig:e_logp} and \ref{fig:elogP_details}), the absence of eccentric systems in the $e-\log P$ diagram at short periods does not result from an observational bias, but rather is an evolutionary effect, as we now explain. In Fig.~\ref{fig:elogP_details}, we first draw the periastron envelope 
corresponding to the circular period for a filling factor of 1 (Table~\ref{tab:Pc_st} and solid line in Fig.~\ref{fig:elogP_details}), which is similar to the period threshold for contact systems. There are thus very few systems falling to the left of this envelope. Then we draw with dashed and dotted lines the envelopes corresponding to filling factors of 0.5 and 0.25, respectively. It clearly appears that the region enclosed by the $f = 0.25$ and $f = 1$ envelopes are less populated as one considers later spectral types. 
This results from the fact that, because late-type stars live longer, they have
more time to circularise, and hence, circularisation may involve systems with smaller filling factors where tides (and thus circularisation) are less efficient and circularisation thus takes longer (see, \emph{e.g.}, \citealt{2005ApJ...620..970M}).

For systems with giant primaries (right panels of Figs.~\ref{fig:e_logp} and \ref{fig:elogP_details}), the increase from a few days to hundreds of days in the circularisation period is related to the increasingly large radii associated with giants of later spectral types since the radius increases as the stars evolve along the giant branch, ending up as very large asymptotic-giant-branch stars of late M types – besides our Figs.~\ref{fig:e_logp} and \ref{fig:elogP_details}, this has been illustrated as well in Figs.~24 and 25 of \citet{2023A&A...674A..34G}. Early-giant primaries are of similar size as their dwarf counterparts, which  is why they have a similar circularisation period (a few days) as dwarfs. In contrast, late-giant primaries can reach hundreds of solar radii and imply a circularisation period of two orders of magnitude larger than that of dwarf primaries.

\begin{figure*}
    \centering
    \includegraphics[trim=1cm 0 2.5cm 0, clip, width=0.33\linewidth]{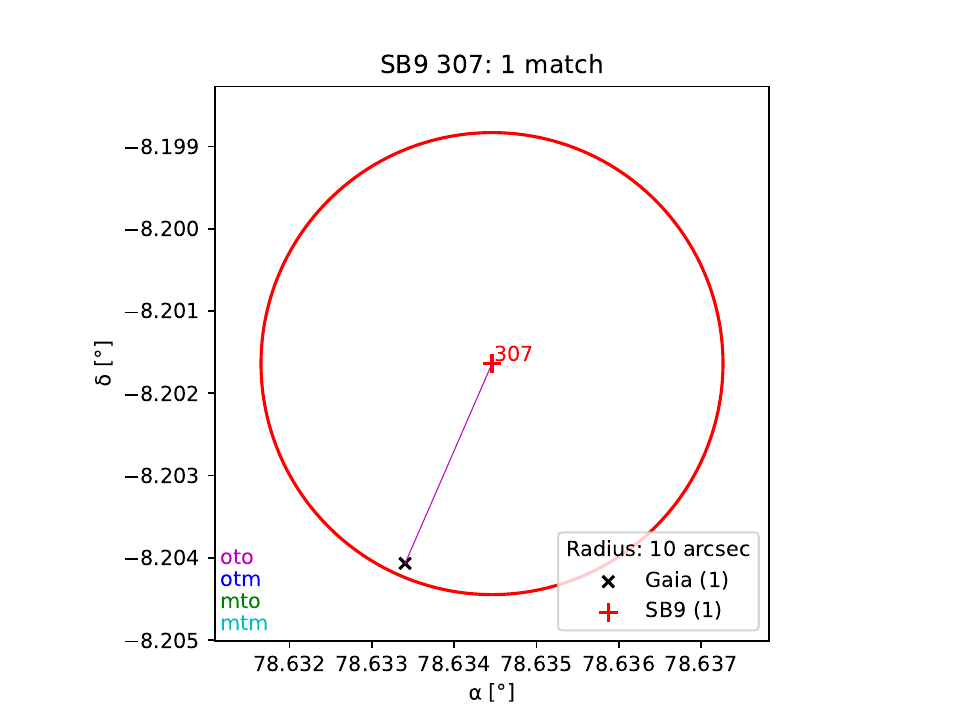}
    \includegraphics[trim=1cm 0 2.5cm 0, clip, width=0.33\linewidth]{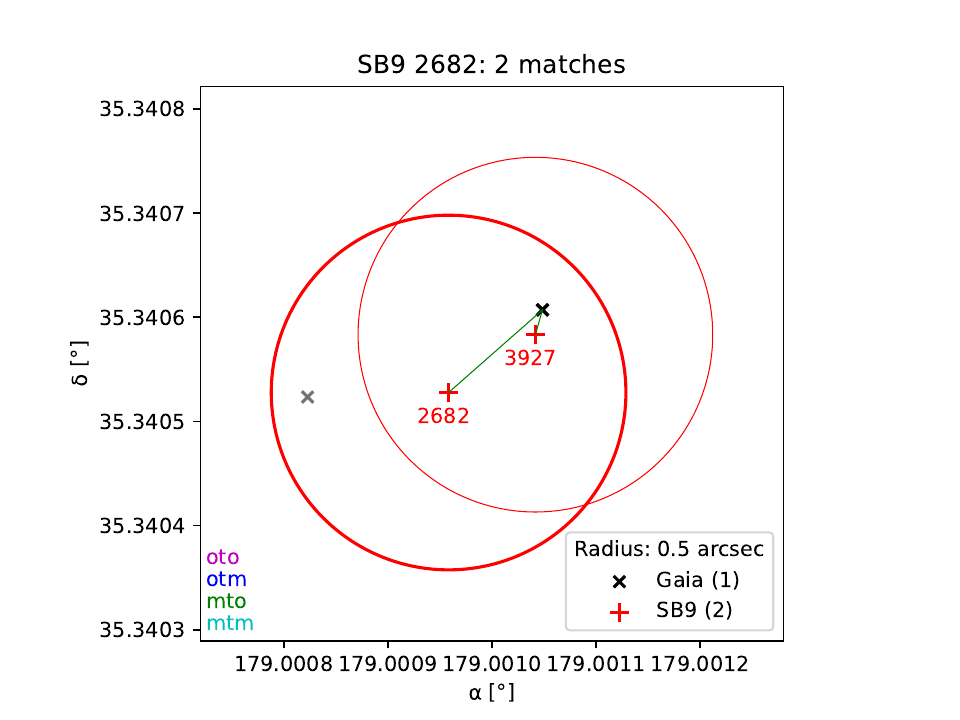}
    \includegraphics[trim=1cm 0 2.5cm 0, clip, width=0.33\linewidth]{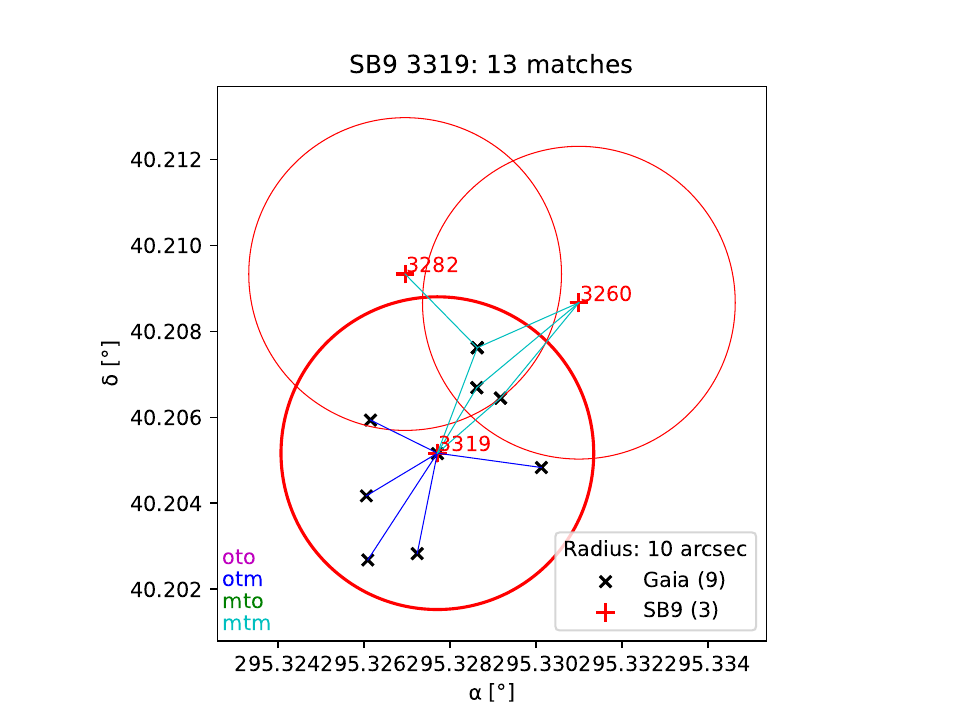}
    \caption{\textbf{Left}: \emph{one-to-one} cross-match for the \SBcat\ system 
    307 ($\beta$ Ori B, $G=6.6$). $\beta$~Ori~A is too bright \jkt{to be} observed by \Gaia. The \SBcat\ system 307 was wrongly assigned the coordinates of $\beta$~Ori~A  (in \SBcat) explaining such a large separation in the match.
    This example illustrates the need to have a large cone search radius value of 10 arcsec. \textbf{Middle}: zoom on \emph{many-to-one} matches for the \SBcat\ systems 2682 and 3927 separated by 0.14~arcsec, not resolvable by \Gaia. For readability, the radius in the plot has been decreased to 0.5 arcsec. The grey cross corresponds to the \Gaia\ source at epoch J2016.0. \textbf{Right}: \emph{one-to-many} and \emph{many-to-many} cross-matches for the \SBcat\ system 3319 (TIC 139154020, $G=15.7$), member of the open cluster NGC~6819. This cross-match involves \jkt{two} other \SBcat\  systems and \jkt{three} \Gaia~DR3 sources leading to \jkt{six} one-to-many and \jkt{three} many-to-many matches.}
    \label{fig:xmatches}
\end{figure*}

\section{Cross-match with \Gaia DR3}
\label{sec:xmatch}
We performed a cross-match of the \SBcat\ catalogue with \Gaia DR3 \citep{2016A&A...595A...1G, 2023A&A...674A...1G}. Cross-matching catalogues is a complex and challenging problem, because there are no general rules to determine whether two observations made with different telescopes, at different wavelengths, with different resolutions, and at different epochs, are physically related to the same target. In other words, looking only at the closest separations is not always a guarantee to get the proper match. Different approaches exist: for instance using positions, position uncertainties and proper motions is an option, see, \emph{e.g.} the cross-matched catalogues of \Gaia DR1 and DR2 with external catalogues like 2MASS, AllWISE, \Hipparcos, etc. \citep{marrese2017, marrese2019}. They promote homogeneity and consistency and avoid the use of \emph{a~priori} knowledge based on photometry for instance. Nevertheless, including proper motions does not address the astrometric-binary problem (the disturbance introduced by the orbital motion on the proper motion) as it would be necessary to include the binary orbits when available. Other approaches use combinations of requirements based on photometry and radial velocity \citep[e.g.][]{tsantaki2022}. A cross-match process is always a compromise between the number of requirements and the fraction of mismatching objects. Here we aim at minimising this fraction.

Within \SBcat, \Gaia DR2 and DR3 identifiers were already implemented by D. Pourbaix, but unfortunately, we did not have access to the methodology of his assignments. That is why we decided to make our own cross-match, and compare them with Pourbaix's and \Simbad\ classification. `\Simbad\ focuses on the objects of interest'\footnote{\url{http://simbad.cds.unistra.fr/guide/catalogues.htx}} and at the time of the starting of this project, not all \SBcat\ systems were added to it. The cross-identification between the \Simbad\ database and \Gaia DR3 has been performed in May 2022, involving 6.5 millions of objects, excluding objects with coordinates less precise than the sub-arcsecond level (quality D and E).

Before cross-matching, we uploaded the \SBcat\ catalogue to the \Gaia\ archive, selected systems using ADQL queries, and performed an epoch back-propagation to account for proper motion, resulting in 3981 matched systems out of 4021, while identifying 21 overly bright (Table~\ref{tab:too_bright}) and six overly faint systems (Table~\ref{tab:too_faint}) not present in \Gaia~DR3.The details of the adopted strategy is given in Appendix~\ref{sec:strategy}. We describe in the following section the workfow to determine the best matches.

\begin{table*}
\centering
\footnotesize
\caption{The 21 brightest \SBcat\ systems without any match in \Gaia\ DR3, ordered by decreasing brightness.}
\begin{tabular}{llrrrrl}
\hline
$V$ [mag] & \SBcat & Type & Bayer name & Common name & HIP & Remark \\
\hline
$-1.46$ & 416 & SB1 & $\alpha$ CMa & Sirius & 32349 & Brightest star, with a WD companion \\
$-0.10$ & 815\footnotemark[1] & SB2 & $\alpha$ Cen & Rigel Kentaurus & 71683 & Resolved components (16.5 arcsec) with distant companion Proxima Cen \\
 & & & & \& Toliman & 71681 & \\
0.08 & 306 & SB1 & $\alpha$ Aur & Capella & 24608 & Brightest subsystem of a stellar 2+2 quadruplet\footnotemark[2] \\
0.37 & 467\footnotemark[1] & SB1 & $\alpha$ CMi & Procyon & 37279 & WD companion \\
0.97 & 766 & SB2 & $\alpha$ Vir & Spica & 65474 & $\beta$ Cep variable; innermost subsystem of 3+1 quadruplet\footnotemark[2] \\
1.25 & 2446 & SB1 & $\beta$ Cru & Mimosa & 62434 & Upper Sco–Cen association \\
1.33\footnotemark[3] & 725 & SB1 & $\alpha$ Cru & Acrux & 60718 & Brightest subsystem of a stellar 3+3 sextuplet\footnotemark[2] \\
1.79 & 648 & SB1 & $\alpha$ UMa & Dubhe & 54061 & Resolved (0.7 arcsec) subsystem of 3+2 quintuplet\footnotemark[2] \\
1.83 & 500 & SB2 & $\gamma^2$ Vel & Regor & 39953 & Wolf–Rayet \\
1.90 & 366 & SB2 & $\beta$ Aur & Menkalinan & 28360 &  \\
1.90 & 462 & SB1 & $\alpha$ Gem A & Castor A & 36850 A & Forms a quadruple with Castor B at 3 arcsec \\
1.92 & 410 & SB1 & $\gamma$ Gem & Alhena & 31681 &  \\
1.92 & 1234 & SB1 & $\alpha$ Pav & Peacock & 100751 & Tucana Horologium association \\
2.02 & 76 & SB1 & $\alpha$ UMi & Polaris & 11767 & Cepheid variable, innermost subsystem of a triplet\footnotemark[2] \\
2.06 & 4 & SB2 & $\alpha$ And/$\delta$ Peg\footnotemark[4] & Alpheratz/Sirrah & 677 & RS CVn-type variable, resolved (24 mas) binary with companion at 6.7 arcsec\footnotemark[2] \\
2.22 & 158 & SB2 & $\beta$ Per & Algol & 14576AB & Outer pair (0.1 arcsec; 680 d) \\
2.25 & 157 & SB2 & $\beta$ Per & Algol & 14576A & Inner pair (2.3 mas) eclipsing binary of 2.8 d \\
2.32 & 1837 & SB1 & $\delta$ Sco & Dschubba & 78401 & Upper Sco association \\
2.55 & 793 & SB2 & $\zeta$ Cen &  & 68002 &  \\
2.65 & 98 & SB2 & $\beta$ Ari & Sheratan & 8903 &  \\
3.00 & 461 & SB1 & $\alpha$ Gem B & Castor B & 36850 B & Forms a quadruple with Castor A at 3 arcsec \\
\hline
\end{tabular}
\begin{flushleft}
Notes. Apparent $V$ magnitudes come from \Simbad\ unless specified. The type distinguishes whether the system has been identified as SB1 or SB2 in \SBcat. \\
$^1$ Rejected as best match with unrelated \Gaia\ source.\\
$^2$ From the Multiple Star Catalogue \citep{tokovinin2018}.\\
$^3$ Magnitude from \citet{1995yCat.5050....0H}.\\
$^4$ This system belongs to two constellations. 
\end{flushleft}
\label{tab:too_bright}
\end{table*}

\begin{table}
\centering
\scriptsize
\caption{The six \SBcat\ systems without any match in \Gaia\ DR3 because they are too faint in $G$. All are SB1.}
\begin{tabular}{lllll}
\hline
$V$ [mag] & \SBcat & Name & Remark \\
\hline
17.5\footnotemark[1] & 2469 & SDSS J022502.06+001541.0\footnotemark[3] & Period of 3.6 d \\
22.5 & 1137 & PSR J1915+1606 & Hulse–Taylor pulsar, period 0.32 d \\
22.8 & 2467 & SDSS J005406.06+003432.0 & Period of 15.9 d \\
>23\footnotemark[5] & 480 & UY Vol & Low-mass X-ray binary, period 0.15 d \\
23.3\footnotemark[2] & 511 & PSR J0823+0159 & Pulsar, period 1710 d \\
26.9\footnotemark[4] & 1421 & PSR J2305+4707 & NS+WD binary, $P=12.3$ d, $e=0.66$ \\
\hline
\end{tabular}
\begin{flushleft}
Notes. Apparent magnitudes are taken from \Simbad\ unless otherwise specified.\\  
$^1$ Magnitude in $r$ band from \SBcat.\\  
$^2$ Magnitude in $r$ band from \Simbad.\\  
$^3$ Only two identifiers in \Simbad, from SDSS and \SBcat.\\  
$^4$ From \citet{van_kerkwijk1999}.\\  
$^5$ Quiescent $V$ magnitude from \citet{1992ApJ...385..314T}; $V=16.9$ reported by \Simbad.\\  
\end{flushleft}
\label{tab:too_faint}
\end{table}

\subsection{Different types of resulting matches}
Ideally, we should obtain one \Gaia\ target match per \SBcat\ system (one-to-one). But within a cone search of 10 arcsec, centred on each \SBcat\ system, several matches with the dense \Gaia\ catalogue may occur (one-to-many). Conversely, less often, several \SBcat\ systems can match a unique \Gaia\ source (many-to-one). In some additional cases, several \SBcat\ systems (closer than 20 arcsec from each other) can match similar \Gaia targets (many-to-many). We may summarise the matches in these four categories as follows:
\begin{itemize}
    \item 2068 one-to-one (\emph{oto}) matches for 2068 \SBcat\ systems;
    \item 6327 one-to-many (\emph{otm}) matches for 1651 systems; the maximum number of matches for one \SBcat\ system is 50, this system being in the bulge (\SBcat\ 2452).
    \item 141 many-to-one (\emph{mto}) matches for 141 systems; the maximum number of \SBcat\ identifiers for one \Gaia\ source is 4.  
    \item 504 many-to-many (\emph{mtm}) matches for 144 systems.
\end{itemize}
Note that 23 \SBcat\ systems fall both in the \emph{otm} and \emph{mtm} matching categories. Illustrating examples are provided in Fig.~\ref{fig:xmatches} showing how complicated the cross-matches can be: on the left panel, the ideal case of \emph{oto} match (in magenta) between \SBcat\ and \Gaia; on the middle panel, the \emph{mto} case (in green) where two \SBcat\ systems match the same \Gaia\ source, in that case corresponding to a spectroscopic triple unresolved by \Gaia; and on the right a complicated case combining \emph{otm} and \emph{mtm} where several \Gaia\ sources match the \SBcat\ system (\emph{otm}, in blue) while other \Gaia\ sources also match several other \SBcat\ systems (\emph{mtm}, in cyan).

\subsection{Filtering the cross-matches}
\subsubsection{Euclidean distance}  
For each \SBcat\ system, we use an Euclidean distance on angular separation $\rho$ and difference of $G$-band magnitudes $\Delta G$ of the matches to automatically select the one with minimum $d$, where:
\begin{equation}
 d = \sqrt{\left( \frac{\rho}{\langle \rho\rangle} \right)^2 + \left( \frac{\Delta G}{\langle \Delta G \rangle} \right)^2}
\end{equation}
with $\langle \rho \rangle = 0.06$~arcsec and $\langle \Delta G \rangle = 0.04 $ are the median angular separation and magnitude differences of all the matches at J2000.0 in a cone search radius restricted to one arcsec. Nevertheless the magnitudes provided in the \SBcat\ catalogue are not the $G$ magnitudes of \Gaia. Transformation to the \Gaia\ $G$ magnitude thus needs to be done, as we describe in the next section.

\subsubsection{Transformation from \SBcat\ to \Gaia\ magnitudes}
\begin{figure}
    \centering
    \includegraphics[width=\linewidth]{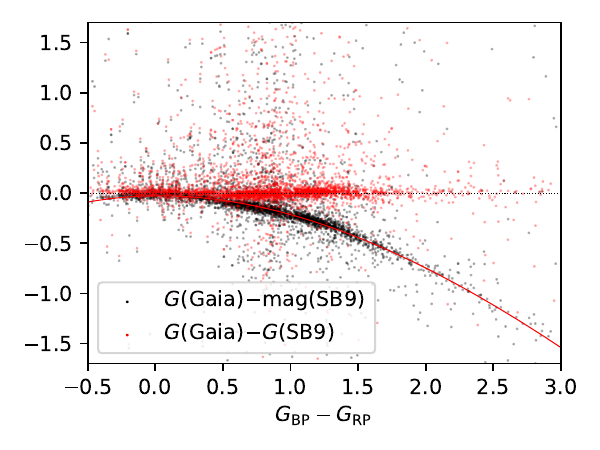}
    \caption{Magnitude differences before (black) and after (red) transformation of \SBcat\ into \Gaia\ magnitudes using magnitude-colour relations provided by the \Gaia\ documentation.}
    \label{fig:dG_bp_rp}
\end{figure}

In the \SBcat\ catalogue, an integrated magnitude is given with its associated filter. It is mainly given in $V$ band (89\%), but 6\%  of systems have no magnitudes, 2\% have $B$ magnitudes, 1\% have `$P$' magnitudes, meaning `Photographic' historically coming from SB8 \citep{1989PDAO...17....1B}, and the remaining 2\% are in filters $J$, $G$, $K$, $r$, $R$, $y$, or $I$. We transformed the $B$, $V$, $R$, $I$, $J$, $K$ or $r$ magnitudes into $G$ ones using relations provided in Tables 5.8 and 5.9 of the \Gaia~DR3 documentation\footnote{Section~\href{https://gea.esac.esa.int/archive/documentation/GDR3/Data_processing/chap_cu5pho/cu5pho_sec_photSystem/cu5pho_ssec_photRelations.html}{5.5.1 Photometric relationships with other photometric systems} of Gaia DR3 documentation.}. For `$P$' and $y$ bands we used the same relations as for transforming $B$ and $J$ bands, respectively, which are the closest in terms of wavelength coverage. Figure~\ref{fig:dG_bp_rp} presents the results of this transformation: the significant magnitude differences clearly disappear at large colour index, and the mean $G$ difference after transformation is $0.02 \pm 0.12$ mag.

\subsubsection{Sanity checks}
The process is described in details in Appendix~\ref{sec:sanity_checks}. In the following subsections, we present the interesting outcomes of this process.

\paragraph{\SBcat\ orbits in higher-order systems.}
\label{sec:hom}
From the manual checking of systems potentially not resolved by \Gaia, because having multiple \SBcat\ system numbers for a given \Gaia source, we identified:
\begin{itemize}
 \item 76 triple systems (152 system numbers, Table~\ref{tab:triple});
 \item 34 quadruple systems: 32 true quadruples (65 system numbers) and possibly \jkt{two} `optical' ones (\jkt{four} system numbers), all reported in Table~\ref{tab:quadruple};
 \item \jkt{six} multiple systems including the Trapezium in the Orion Nebula (14 system numbers, Table~\ref{tab:hom}).
 \end{itemize}
 
Regularly, such hierarchies could be part of higher-order systems, \emph{e.g.} by having visible common proper-motion companions that can themselves be SB. We did not extensively search for them but they can easily be checked with the Multiple Star Catalogue \citep[\href{http://www.ctio.noirlab.edu/~atokovin/stars/stars.php}{MSC},][]{tokovinin2018}.
We identified six stellar systems having a multiplicity larger than four and with at least two subsystems with orbits reported in \SBcat (see also Table~\ref{tab:hom}):
\begin{itemize}
\item in the Orion Trapezium cluster, which is a crowded, optically thick region, we have $\theta^1$~Ori A, a triple system, whose inner binary, \SBcat\ 340, is an eclipsing SB1; and, located at $\sim 9$~arcsec, $\theta^1$~Ori B, (\SBcat\ 341), an eclipsing SB2
(\href{https://simbad.cds.unistra.fr/simbad/sim-coo?Coord=05+35+16.5-05+23+14&CooFrame=ICRS&CooEqui=2000.0&CooEpoch=J2000&Radius.unit=arcmin&submit=Query+around&Radius=0.18}{\Simbad},
\href{https://www.ctio.noirlab.edu/~atokovin/stars/stars.php?cat=HIP&number=26220%20}{MSC only for $\theta^1$~Ori A},
\href{https://sb9.astro.ulb.ac.be//ProcessMainform.cgi?Catalog=ADS&Id=4186&Coord=&Epoch=2000&radius=10&unit=arc+min}{\SBcat}).
\item WDS~J15382+3615: a sextuplet (3+3) where the primary triple, HIP 76563, has the inner pair  being \SBcat\ 1510, and the secondary triple, HIP 76566, has the inner pair being \SBcat\ 1511; the outer orbit has an estimated period of about 15 ka 
(\href{http://simbad.u-strasbg.fr/simbad/sim-coo?Coord=15+38+12.9143683608%2B36+14+48.687871596&CooFrame=ICRS&CooEqui=2000.0&CooEpoch=J2000&Radius.unit=arcmin&submit=Query+around&Radius=1}{\Simbad},
\href{http://www.ctio.noirlab.edu/~atokovin/stars/stars.php?ids=15382%2B3615}{MSC},
\href{https://sb9.astro.ulb.ac.be//ProcessMainform.cgi?Catalog=HD&Id=139691&Coord=&Epoch=2000&radius=10&unit=arc+min}{\SBcat}).
\item WDS~J04357+1010: a sextuplet (4+2), where the primary is a 2+2 quadruple (88~Tau) with one of the inner orbit being \SBcat~257, an astro-spectro-eclipsing binary, and at 70 arcsec, 88~Tau~B is an SB1 (\SBcat~1527) 
(\href{http://simbad.u-strasbg.fr/simbad/sim-coo?Coord=04+35+39.2662111032%2B10+09+38.711644992&CooFrame=ICRS&CooEqui=2000.0&CooEpoch=J2000&Radius.unit=arcmin&submit=Query+around&Radius=2}
{\Simbad},
\href{https://www.ctio.noirlab.edu/~atokovin/stars/stars.php?cat=HD&number=29140}{MSC},
\href{https://sb9.astro.ulb.ac.be//ProcessMainform.cgi?Catalog=HD&Id=29140&Coord=&Epoch=2000&radius=10&unit=arc+min}{\SBcat}).
\item WDS~J18058+2127: a quintuplet (3+2), where the primary is a triple, HIP 88637, with inner pair (\SBcat\ 1023, an eclipsing SB1) and outer pair  (\SBcat\ 1024); and the secondary is an SB2, HIP~88639  (\SBcat\ 1986) 
(\href{http://simbad.cds.unistra.fr/simbad/sim-coo?Coord=18+05+49.710403%2B21+26+45.21588&CooFrame=ICRS&CooEqui=2000.0&CooEpoch=J2000&Radius.unit=arcmin&submit=Query+around&Radius=0.6}{\Simbad},
\href{https://www.ctio.noirlab.edu/~atokovin/stars/stars.php?cat=HD&number=165590}{MSC},
\href{https://sb9.astro.ulb.ac.be//ProcessMainform.cgi?Catalog=HD&Id=165590&Coord=&Epoch=2000&radius=10&unit=arc+min}{\SBcat}).
\item WDS~J05154+3241: a quintuplet (2+3), where the primary, HIP 24504, is an SB1 (\SBcat\ 305) of $\delta$~Sct type, and the secondary, HIP 24502, located at 14.25 arsec has an SB1 in the inner orbit (\SBcat\ 1484) 
(\href{http://simbad.u-strasbg.fr/simbad/sim-coo?Coord=05+15+24.3941965368%2B32+41+15.359679780&CooFrame=ICRS&CooEqui=2000.0&CooEpoch=J2000&Radius.unit=arcmin&submit=Query+around&Radius=1}{\Simbad},
\href{http://www.ctio.noirlab.edu/~atokovin/stars/stars.php?cat=HD&number=33959}{MSC},
\href{https://sb9.astro.ulb.ac.be//ProcessMainform.cgi?Catalog=GAIADR3&Id=180702438220217344&Coord=&Epoch=2000&radius=10&unit=arc+min}{\SBcat}).
\item WDS~J07346+3153: a septuplet (4+3), where the primary, Castor A, is \jkt{an} SB1 (\SBcat\ 462) and \jkt{forms} a 2+2 quadruple with Castor B, another SB1 (\SBcat\ 461); YY Gem, located at 70.5 arcsec, is a triple whose inner binary is an SB2 (\SBcat\ 463)
(
\href{http://simbad.cds.unistra.fr/simbad/sim-id?Ident=CCDM+J07346%2B3153AB&NbIdent=query_hlinks&Coord=07+34+35.87319%2B31+53+17.8160&children=2&submit=children&hlinksdisplay=h_all}{\Simbad},
\href{https://www.ctio.noirlab.edu/~atokovin/stars/stars.php?cat=HIP&number=36850}{MSC},
\href{https://sb9.astro.ulb.ac.be//ProcessMainform.cgi?Catalog=HIP&Id=36850&Coord=&Epoch=2000&radius=10&unit=arc+min}{\SBcat}
). 
\end{itemize}
This investigation of high-order systems revealed that, among the 31 quadruple systems identified in \SBcat, and reported in Table~\ref{tab:quadruple}, 3 are complete spectroscopic quadruples, with the 3 orbits derived while the other have only 2 orbits referenced, as summarized in Table~\ref{tab:sb4}: 
\begin{itemize}
    \item $\mu$~Ori \citep{2000A&A...354..881D, 2002AJ....123.1723F}: differential astrometry by \citet{2008AJ....135..766M} allowed them to determine mutual inclinations between the orbits and to completely characterize this quadruple system.
    \item HD~117078 \citep{2011Obs...131..351G}: the pair of pairs was visually resolved with speckle interferometry \citep{2022AJ....164...58T}.
    \item VW~LMi \citep{2008MNRAS.390..798P}, one of the tightest quadruple system known: the outer orbit period is smaller than a year. It includes a contact EB-SB2 and a detached SB2. The secular evolution of such a compact quadruple system shows an apsidal motion of the inner contact EB-SB2 \citep{2020MNRAS.494..178P}. 
\end{itemize} 

\begin{table*}
    \centering
    \footnotesize
    \caption{Details of the three complete spectroscopic quadruples in \SBcat. They all have a 2+2 architecture and are not resolved by \Gaia\ DR3.}
    \begin{tabular}{llcrllclll}
    \hline
    Name  & & Type & \SBcat & $P$ [d] & $e$ & \Gaia & $G$ & RUWE & Separation [arcsec] \\
    \hline
    \multicolumn{1}{l|}{}          & A  & SB1 & 372 & $4.4475858 \pm 0.0000012$ & $0.0044 \pm 0.0014$ & & & & \\
    \multicolumn{1}{l|}{$\mu$ Ori} & B  & SB2 & 373 & $4.7835361 \pm 0.0000028$ & $0$ (Fixed)          & & & & \\
    \multicolumn{1}{l|}{}          & AB & SB2 & 374 & $6809.9 \pm 3.4$          & $0.7426 \pm 0.0020$  & \multicolumn{1}{|l}{3329650894896257024} & 4.674 & & 0.27 \\
    \hline
    \multicolumn{1}{l|}{}          & A  & SB1 & 3875 & $5.837891 \pm 0.000008$  & $0.0144 \pm 0.0027$  & & & & \\
    \multicolumn{1}{l|}{HD 117078} & B  & SB1 & 3876 & $204.256 \pm 0.010$      & $0.727 \pm 0.007$    & & & & \\
    \multicolumn{1}{l|}{}          & AB & SB2 & 3877 & $15000$ (Fixed)          & $0.374$ (Fixed)      & \multicolumn{1}{|l}{1441569596393755648} & 9.468 & 27.1 & 0.12 \\
    \hline
    \multicolumn{1}{l|}{}          & A  & SB2 & 3213 & $0.47755106 \pm 0.00000003$ & $0$ (Fixed)        & & & & \\
    \multicolumn{1}{l|}{VW LMi}    & B  & SB2 & 4003 & $7.93063 \pm 0.00003$      & $0.035 \pm 0.003$  & & & & \\
    \multicolumn{1}{l|}{}          & AB & SB2 & 4004 & $355.02 \pm 0.17$          & $0.097 \pm 0.011$  & \multicolumn{1}{|l}{733549185449902208} & 8.035 & 3.6 & 0.15 \\
    \hline
    \end{tabular}
    \begin{flushleft}
    Notes. Separations are given between A and B systems and come from
    \href{http://www.ctio.noirlab.edu/~atokovin/stars/stars.php}{MSC}
    \citep{tokovinin2018}.
    \end{flushleft}
    \label{tab:sb4}
\end{table*}

\begin{figure}
    \centering
    \includegraphics[width=\linewidth]{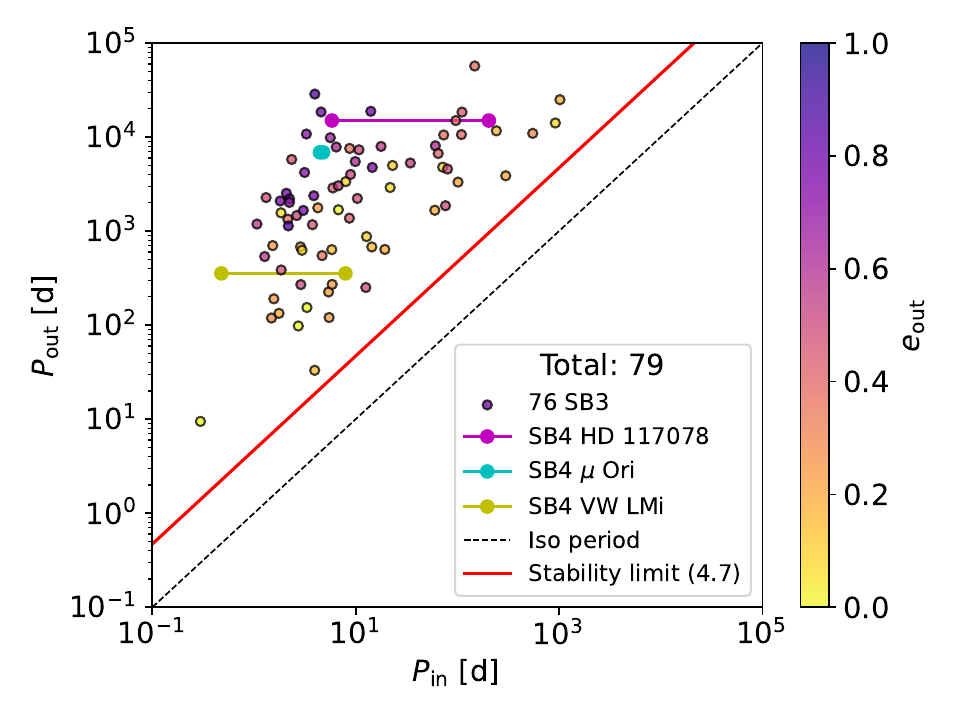}
    \caption{Outer periods vs inner period(s) of spectroscopic triple and quadruple systems in \SBcat, colour-coded by the eccentricity of the outer orbits. The iso-period is shown by the black dashed line. The stability limit is represented by a period ratio $P_\mathrm{out}/P_\mathrm{in}=4.7$.}
    \label{fig:sb3}
\end{figure}

All of these quadruple systems have a 2+2 architecture, \emph{i.e.} two binaries orbiting each other. The two first in Table~\ref{tab:sb4} are VB-SB, while the VW~LMi system will be difficult to resolve astrometrically, maybe by \Gaia\ at the end of the mission  \citep{2020MNRAS.494..178P}. However, none of them has been identified as a binary in \Gaia\ DR3, with angular separations between each pair being smaller than 0.45~arcsec.

The 76 spectroscopic triple systems identified in \SBcat\ are reported in Table~\ref{tab:triple}.
Figure~\ref{fig:sb3} presents their outer periods as a function of the inner ones, and colour-coded by the eccentricity of the outer orbit.  The dynamical stability criterion $P_\mathrm{out}/P_\mathrm{in} > 4.7 $ is represented by the red line and given by \citet[MSC,][]{tokovinin2018} based on \citet{mardling2001} for hierarchical triples with coplanar orbits. This figure reveals the hierarchical nature of high-order systems.
Most of them are already \jkt{in}, or have been added (\jkt{three} of them) \jkt{to}, the MSC \citep{tokovinin2018}. We also added on this figure the three spectroscopic quadruples where inner and outer orbits are in the \SBcat. \jkt{Contrary} to the two other spectroscopic quadruples, $\mu$~Ori has two inner orbits with very similar periods, thus having a 1:1 resonant configuration. These  possible mean motion resonances (1:1:, 3:2, etc.) could be important to constrain evolutionary channels to form them \citep{2019A&A...630A.128Z}.

Finally, we illustrate how complicated the situation can be when cross-matching and looking for high-order multiples. When multiple errors occur, it is not straightforward to disentangle them. Investigating two \SBcat\ systems (1559 \& 1607) with the same name in \SBcat\ (BD~+21~255), a potential triple or quadruple, these two systems appear to have periods of 4137~d and 1273~d, leading to a ratio of 3.2, too small to be hierarchical according to \citet{mardling2001}. Component A (HIP 8876) corresponds to \SBcat~1559 \citep{1998A&AS..131...25U}. Nevertheless, as many high-order systems included in \SBcat, component B has the coordinates of component A, mixing the \SBcat\ identification names. The issue here (reported in the notes of \SBcat), is that \citet{1992A&A...260..115J}, while willing to observe BD~+21~255 for their study on S-type stars, actually observed a K5 star located South-West at 54 arcsec (namely TYC~1212-473-1). The authors published two different orbits assigned to BD~+21~255 a and b but the suffix got lost during inclusion in \SBcat\ and because of this, later on \Simbad\ assigned \SBcat\ systems 1559 \& 1607 to the same target  \SBcat\ BD~+21$^\circ$~255, while \SBcat 1607 should point toward TYC~1212-473-1. So here we have two SB1, do they constitute a physical quadruple system? When looking at their \Gaia~DR3 parallaxes ($\varpi_\mathrm{A}=1.086\pm0.033$ and $\varpi_\mathrm{B}=1.531\pm0.024$~mas) which are significantly different, it seems that they form an optical quadruple, \emph{i.e.} the two SB1 are not physically bound, or at least they form a very loosely bound pair of binaries. Similarly, the visual pair formed by HD~179332 (\SBcat\ 3839) and BD+60 1892 (\SBcat\ 3840) have significantly different \Gaia\ DR3 parallaxes of $5.986\pm0.062$ and $3.2511\pm0.0307$~mas, respectively. These two apparent quadruples are reported as `optical?' in Table~\ref{tab:quadruple}.   
While a similar conclusion is initially reached for HD~110318 (\SBcat\ 735) and VV~Crv (\SBcat\ 734), with parallaxes of $13.895\pm0.163$ and $12.269\pm0.053$~mas, respectively, it transpires that HD~110318 and VV~Crv, separated by 5~arcseconds, together form the physical binary STF~1669, which has been known for nearly 200 years \citep{1827cnsd.book.....S}. The proper motion is substantial, at $-0.13$~arcsec~a$^{-1}$, so the pair would rapidly move apart if it were merely an optical alignment. The discrepant parallax of VV~Crv is biased and its uncertainty probably underestimated in \Gaia DR3. There is also a third star, C (BD$-12$~3675, located at 59~arcsec), resulting in a total system that is quintuple.

\begin{figure*}
    \includegraphics[width=0.9\linewidth]{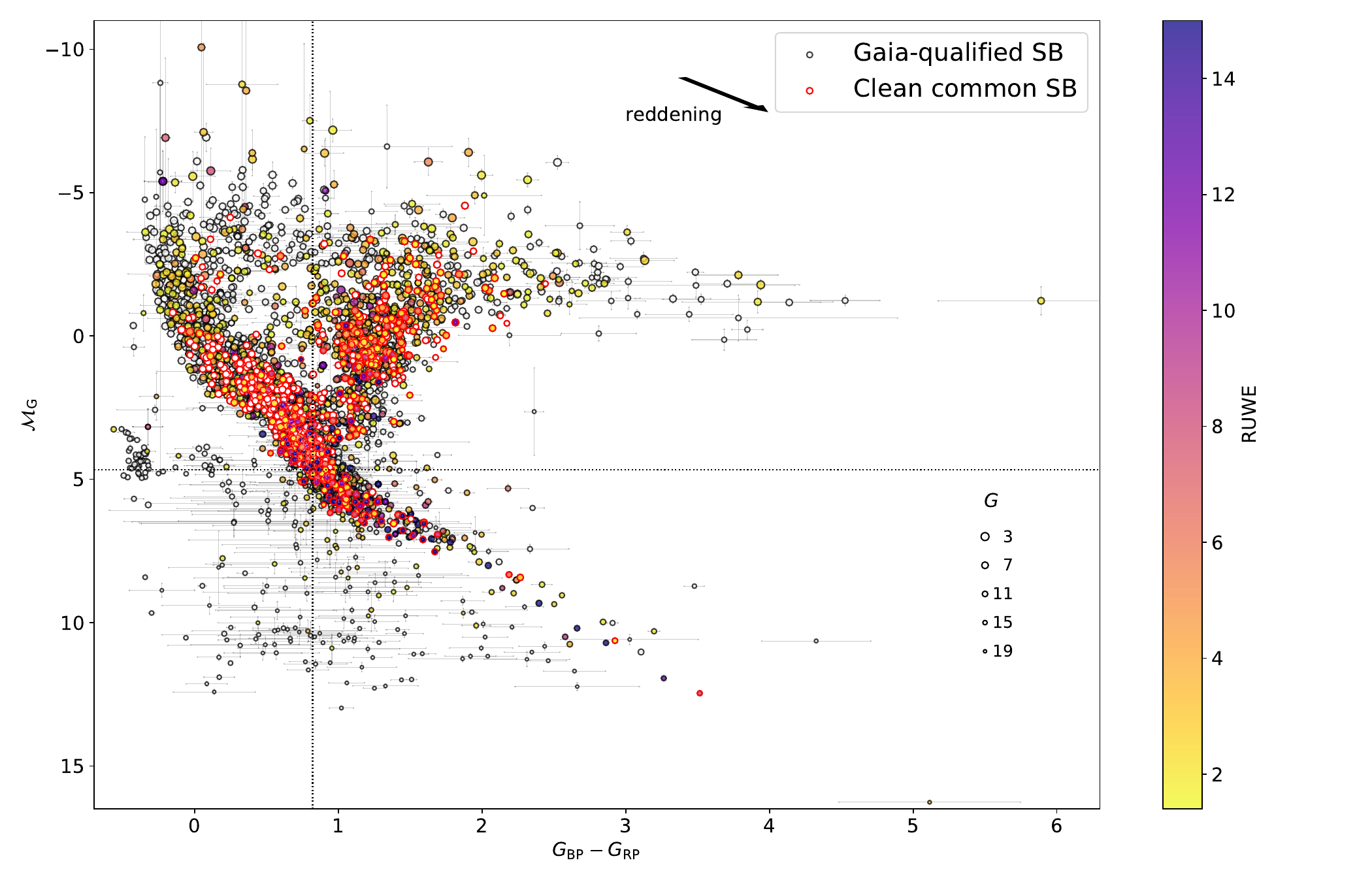}
    \caption{Colour-absolute magnitude diagram for 3\,781 SB  (94\%) of the \SBcat\ catalogue with \Gaia\ parallaxes, magnitudes and colors (\Gaia-qualified SB), colour-coded by the \Gaia\ DR3 RUWE value. SB with RUWE lower than 1.4 are displayed in white. The size of the symbols are related to the apparent magnitude in the $G$ band. The position of the Sun would be at the intersection of the dotted lines. The clean common sample of 655 SB between \SBcat\ and \Gaia~NSS  (defined in Sect.~\ref{sect:outliers}) is shown with red circles.}
    \label{fig:cmd}
\end{figure*}

\begin{table}
\centering
\caption{Common sample of binaries in \SBcat~$\cap$~\Gaia\ and \SBcat~$\cap$~NSS.}
\begin{tabular}{lrrrrr}
\hline
Catalogue & Binary & in triple & in quad & in higher-order & Total \\
\hline
\SBcat    & 3\,747  & 150 & 69  & 12\footnotemark[2] & 3\,976 \\
\Gaia\ DR3 & 3\,756\footnotemark[4] & 76\footnotemark[5] & 50 & 11\footnotemark[6] & 3\,892 \\
\hline
\SBcat    & 800 & 22 & 4 & 1 & 827 \\
NSS       & 800 & 11 & 2 & 1 & 814 \\
\\
\multicolumn{6}{l}{Clean common sample} \\
\\
\SBcat    & 645 & 8 & 2 & 0 & 655 \\
NSS       & 645 & 8 & 2 & 0 & 655 \\
\hline
\SBcat\footnotemark[1] & 3\,770 & 152 & 71 & 14\footnotemark[3] & 4\,003 \\
\hline
\end{tabular}
\begin{flushleft}
Notes. \\
$^1$The numbers for \SBcat\ alone are also given for reference. \\
$^2$Two systems also appear in triples (\SBcat\ 1023 and 1024), that is why the total is not 3978. \\
$^3$Four systems also appear in triples (\SBcat\ 1023 and 1024) and quadruples (\SBcat\ 461 and 462), that is why the total is not 4007. \\
$^4$Nine \SBcat\ systems are resolved by \Gaia\ (see Table~\ref{tab:resolved}). \\
$^5$One \SBcat\ triple has outer orbit resolved by \Gaia. \\
$^6$One \Gaia\ source also appears in triples (\Gaia\ DR3 4576326312902814208), that is why the total is not 3893. \\
\end{flushleft}
\label{tab:common_sample}
\end{table}

\paragraph{\SBcat\ system duplicates.}
\label{sec:dup}
During the manual checking process, we discovered 18 duplicates (\emph{i.e.} two \SBcat\ system numbers for the same physical binary) that should be turned into new orbit entries instead for the existing system number. 
Such duplicates appear most of the time for SB in clusters. We suggest to keep the smallest \SBcat\ numbers that correspond to the oldest (historical) entries in the catalogue. We suggest not to use anymore the duplicate system number to avoid any confusion in the future. As such the duplicate \SBcat\ system numbers will point toward the original ones. This will allow us to keep track of their duplicate nature, while the highest system number will no longer reflect the total number of physical systems. They are listed in Table~\ref{tab:dup}. These changes will be introduced in the future version of \SBcat\ (see Sect.~\ref{sec:improvements}).

\paragraph{External check with \Simbad.}
Finally, we ultimately compared our cross-matches (considering the previous manual corrections) with an external reference database. By comparing with the cross-match performed by the reference \Simbad\ database \citep{2000A&AS..143....9W, 2019ASPC..521...50}, we further found 15 \SBcat\ systems with erroneous equatorial coordinates in \SBcat\ leading to wrong best matches. This is because these SB belong to higher-order multiples and were referenced by the coordinates of the brightest component of the system, using the \jkt{Washington} Double Star (WDS) coordinates of the multiple, which is a convention followed by many historical databases, not only the \href{http://www.astro.gsu.edu/wds/}{WDS} \citep{2001AJ....122.3466M}, but also the Sixth Catalog of Orbits of Visual Binary Stars \citep[\href{http://www.astro.gsu.edu/wds/orb6.html}{ORB6},][]{ORB6}, and the Multiple Star Catalogue \citep[\href{http://www.ctio.noirlab.edu/~atokovin/stars/stars.php}{MSC},][]{tokovinin2018}.
The latter, unlike WDS, lists different coordinates for resolved components of multiple systems (not the "brightest-component" coordinates) if a given system has several entries in the MSC COMP table. One of those entries is the primary (distinguished by zero separation), the rest are "secondaries" with individual coordinates and other data, where available.
We also noticed in \Simbad\ a few erroneous assignations of \SBcat\ systems that have since been corrected in \Simbad. 
Finally, about 60 SB do not have \Gaia\ DR3 identifications in \Simbad\ whereas we found one for them involved in a double star. The latter is due to the fact that the cross-identification of \Simbad\ with Gaia DR2/3 excluded objects with a neighbour within 3 arcsec of the historical position (in both \Simbad\ and \Gaia). The future version of \SBcat\ and \Simbad\ will be synchronized.\footnote{\url{http://simbad.cds.unistra.fr/guide/catalogues.htx}} All the \SBcat\ coordinates issues reported so far will be replaced by the proper \Gaia~DR3 coordinates.

\subsection{Cross-match results}
\label{sec:xmatch_res}
Now that robust cross\jkt{-}matches have been secured, a statistical analysis may be performed based on the \Gaia~DR3 properties.
Without the 18 duplicates \jkt{discovered}, this version of the \SBcat\ catalogue (2021-03-02) contains 4\,003 physical spectroscopic binaries (see Table~\ref{tab:common_sample}). Among them, 3\,976 have a \Gaia\ DR3 counterpart (99.3\%). Among them 177 \SBcat\ systems do not have a parallax, 16 have a negative one and 91.7\% have a parallax-over-error larger than 10. Among the 3\,976 \SBcat\ systems with a \Gaia DR3 counterpart:
\begin{itemize}
    \item 916 (23\%) have a radial velocity in \Gaia~DR3
    \item 3\,801 have a renormalised unit-weight error (RUWE), among which 49\% have a RUWE > 1.4
    \item  66\% have a `good' astrometric fit from goodness-of-fit statistics
    \item 1\,031 (26\%) are classified as binaries from the Non-Single Star catalogue \citep{2023A&A...674A..34G}:
    \begin{itemize}
        \item Astrometric binaries (AB): 280;
        \item Spectroscopic binaries (SB): 461;
        \item Eclipsing binaries (EB): 62; 
        \item AB+SB: 212;
        \item EB+SB: 16.        
    \end{itemize}
    \item 3\,150 have BP/RP sampled spectra
    \item only 250 have mean RVS spectra
    \item 984 (25\%) show signs of variability and have epoch photometry \citep{2023A&A...674A..16M};
    \item 452 (11\%) are in the variability table of the 2 million EB candidates; 
    \item 3\,731 (94\%) have atmospheric parameters from the Multiple Star Classifier based on BP/RP photometry excess; assuming unresolved co-eval binary systems \citep{2023A&A...674A..26C}.
\end{itemize}

We display the colour-absolute magnitude diagram (CaMD) of the \SBcat\ catalogue in Fig.~\ref{fig:cmd}. 3781 SB in the catalogue (94\%) have \Gaia\ parallaxes, magnitudes and colours. Uncertainties on absolute magnitude were obtained by propagating errors on parallaxes and $G$ magnitudes, the latter being derived following Eq.~(12) of \citet{2020A&A...635A.155M}. 
They are non-negligible for systems outside the main sequence and the red-giant branch. Systems are colour-coded by their RUWE between 1.4 and 15, with 96\% of the systems having RUWE $<$ 15. The \SBcat\ system with the highest RUWE (43.3) is HD~28634 (K2V) in the Hyades cluster, which appears to be an AB \citep{2011AJ....141..172M} and an SB1 \citep{1985AJ.....90..609G}. With a period of 844.6~d, it is also characterized as an AB+SB1 in \Gaia~DR3. 

\begin{table*}
\centering
\scriptsize
\caption{Ten \SBcat\ systems resolved by \Gaia\ with tentative computed inclinations.}
\begin{tabular}{lllrlrlclc}
\hline
\SBcat & \Simbad\ Name & \Gaia$_\mathrm{A}$ & $G_\mathrm{A}$ & \Gaia$_\mathrm{B}$ & $G_\mathrm{B}$ & $\rho_{\SBcat}^{\min}$ [arcsec] & $\rho_{\Gaia}$ [arcsec] & $P$ [d] & $i$ [$^\circ$] \\
\hline
915\footnotemark[1] & $\zeta$ Her A and B & 1312665361415345920 & 2.716 & 1312665361414730624 & 2.763 & 0.38 & 0.26 & 12596.1 & 43 \\
879 & $\xi$ Sco A and B & 4343066192373820800 & 4.851 & 4343066192373234048 & 4.872 & 0.13 & 0.96 & 16326.3 & 8 \\
1022 & 70 Oph A and B & 4468557611984384512 & 3.987 & 4468557611977674496 & 5.539 & 3.95 & 6.41 & 32188.8 & 38 \\
1470\footnotemark[1] & HD 3443 A and B & 2347260998051944448 & 5.954 & 2347272641708189440 & 6.216 & 0.27 & 0.72 & 9165.64 & 22 \\
1472 & BD+42 501 A and B & 339384295642198272 & 9.191 & 339384299937913856 & 9.335 & 0.47 & 0.70 & $116675 \pm 7670$ & 42 \\
1478 & HD 191854 A and B & 2081078269591424896 & 7.756 & 2081078269585872000 & 8.260 & 0.44 & 0.64 & $31125 \pm 44$ & 43 \\
1815\footnotemark[2] & NGC 2682 131 A and B & 604921546365012480 & 11.621 & 604921542069455104 & 11.861 & 0.0002 & 0.42 & $1188.5 \pm 6.7$ & 0 \\
2460\footnotemark[3] & BD+40 883 A and B & 1873073175244118784 & 9.397 & 1873073175249217408 & - & 0.70 & 0.88 & $10777 \pm 241$ & 53 \\
2559\footnotemark[4] & HD 176051 A and B & 2043885295914531072 & 5.174 & 2043885295914530944 & 7.503 & 0.47 & 1.28 & 22423.0 & 22 \\
3657\footnotemark[1] & HD 155876 A and B & 1364668825433448704 & 8.860 & 1364668825433435776 & 9.183 & 0.08 & 1.13 & $4693.5 \pm 18.3$ & 4 \\
\hline
\end{tabular}
\begin{flushleft}
$^1$No \Gaia\ parallax available, \Hipparcos\ used instead \citep{2007A&A...474..653V}. \\
$^2$No \Gaia\ parallax available, \Gaia\ cluster parallax used instead \citep{2018A&A...616A..10G}. Primary (component A) is itself a SB and EB with system number \SBcat\ 1834. \\
$^3$Primary (component A) is itself a SB with system number \SBcat\ 1280. \\
$^4$B component is the host of an S-type exoplanet with $1.5 \pm 0.3$ Jupiter mass \citep{2010AJ....140.1657M}. \\
\end{flushleft}
\label{tab:resolved}    
\end{table*}

\begin{table*}
\centering
\scriptsize
\caption{Fourteen triple-system candidates (\SBcat\ inner binary with an outer \Gaia\ companion). More candidates could exist in \SBcat.}
\begin{tabular}{lllrclll}
\hline
\SBcat & \Simbad\ Name & \Gaia$_1$, \Gaia$_2$ & $G_1$, $G_2$ & $\varpi_1$, $\varpi_2$ [mas] & $\rho_{\SBcat}^{\min}$ [arcsec] & $\rho_{\Gaia}$ [arcsec] & $P$ [d] \\
\hline
619 & HD 91948 & 1047782393124538624, 1047782393123982464 & 6.674, 10.903 & $13.887 \pm 0.027$, $13.835 \pm 0.015$ & 0.0002 & 4.1 & 2.77 \\
1414 & BD+62 2155 A & 2207250454387300224, 2207250458683385472 & 9.847, 11.100 & $1.197 \pm 0.014$, $1.210 \pm 0.065$ & $7 \times 10^{-7}$ & 1.72 & 1.0392 \\
1786 & HD 93161 A & 5350362982543827456, 5350362982543828352 & 8.483, 8.471 & $0.340 \pm 0.028$, $0.386 \pm 0.028$ & $8 \times 10^{-6}$ & 2.01 & $5.63 \pm 0.13$ \\
1934 & MM Eri B & 3185860165825182592, 3185860161531606528 & 10.918, 10.514 & $6.327 \pm 0.0372$, $6.256 \pm 0.0278$ & 0.0008 & 1.07 & $13.5601 \pm 0.0012$ \\
2097 & RX J0451.6+0619 A & 3288074511955863424, 3288074511954439424 & 12.616, 14.944 & $3.863 \pm 0.025$, $3.969 \pm 0.062$ & 0.00003 & 1.16 & 0.6976 \\
2112\footnotemark[1] & HD 75638 A & 604997202213330560, 604997206508854656 & 8.157, 10.116 & $5.623 \pm 0.052$, $5.406 \pm 0.133$ & 0.00009 & 2.15 & $5.8167 \pm 0.0009$ \\
2288 & HD 96345 A & 3968543807205045248, 3968543802915918464 & 8.989, 10.325 & $24.627 \pm 0.104$, $24.406 \pm 0.045$ & 0.0450 & 2.15 & $6089 \pm 1391$ \\
2339 & HD 111306 & 1567523601159749376, 3692343531370523648 & 6.736, 11.807 & $15.536 \pm 0.046$, $16.031 \pm 0.218$ & 0.0031 & 1.33 & $61.504 \pm 0.004$ \\
2376 & GD 319 A & 1570548009752602240, 1570548014049174272 & 12.665, 13.193 & $4.011 \pm 0.040$, $3.979 \pm 0.011$ & $7 \times 10^{-6}$ & 2.62 & 0.6027 \\
3068 & 35 Com A & 3942757338955937408, 3942757334664656512 & 4.772, 7.074 & $12.065 \pm 0.288$, $-$ & 0.01363 & 1.15 & $2908.25 \pm 4.27$ \\
3267\footnotemark[2] & UCAC4 651-076363 & 2076389196098727552, 2076389196100846720 & 14.956, 17.397 & $0.227 \pm 0.083$, $0.314 \pm 0.095$ & 0.00002 & 0.96 & $81.68 \pm 0.12$ \\
3509 & 53 Aqr & 2595463992699783424, 2595463996992115840 & 6.098, 6.213 & $49.50 \pm 1.23$\footnotemark[3] & 0.00467 & 2.22 & $257.31 \pm 0.22$ \\
3813 & LP 395-8 A & 1829571684890816384, 1829571684890816512 & 11.023, 12.875 & $33.898 \pm 0.026$, $33.896 \pm 0.053$ & 0.00036 & 1.67 & 1.1293 \\
3825\footnotemark[4] & HD 159027 & 4595342877592176512, 4595342881890087552 & 7.633, 13.508 & $2.526 \pm 0.069$, $-$ & 0.00226 & 1.29 & $1359.5 \pm 1.0$ \\
\hline
\end{tabular}
\begin{flushleft}
Notes. Indexes 1 and 2 refer to the \SBcat\ system and the suspected outer companion, respectively.\\
$^1$ Reported as a \Gaia\ astrometric binary in NSS.\\
$^2$ Possibly an optical triple because significant different proper motions in declination at $\sim5$~kpc.\\
$^3$ No \Gaia\ parallax available, \Hipparcos\ used instead \citep{2007A&A...474..653V}.\\
$^4$ The brightest component is detected as an astrometric binary in \Gaia\ DR3 NSS; there is also a good wide companion candidate at 28~arcsec, \Gaia\ DR3 4595342877593285632.\\
\end{flushleft}
\label{tab:triple_candidate}
\end{table*}

For reference, the Sun would be located at the intersection of the dotted lines in the CaMD at $G_\mathrm{BP} - G_\mathrm{RP} = 0.82$ and $\mathcal{M}_G = 4.67$ \citep{2018MNRAS.479L.102C}. The direction of the reddening, shown by the black arrow on the CaMD, is obtained from \citet{2018A&A...616A...8A}. The \jkt{loci} of the \SBcat\ binaries in the CaMD have not been corrected for the reddening in Fig.~\ref{fig:cmd}. 
Main-sequence M-type binaries are under-represented in the \SBcat.
It is beyond the scope of this paper to investigate the CaMD diagram of the \SBcat\ catalogue in greater detail, but various avenues for exploration could be pursued.

In the \SBcat~$\cap$~\Gaia\ common sample, the number of unique \Gaia\ DR3 sources is 3\,892, meaning that many spectroscopic multiples in \SBcat\ are unresolved by \Gaia\ DR3. In contrast, a few \SBcat\ systems are resolved \jkt{and} are presented in Table~\ref{tab:resolved}.  

\subsubsection{\SBcat\ systems resolved by \Gaia}
To determine whether some \SBcat\ systems are resolved by \Gaia, we proceed similarly to \citet{1981A&AS...44...47H} who estimated the angular separation for 431 SBs from \citet[SB7 catalogue][]{1978PDAO...15..121B}. First, we compute the absolute projected semi-major axes $a_A\sin{i}$ and, when it is an SB2, $a_B\sin{i}$. Second, we compute the relative semi-major axis $a\sin{i} = (a_A+a_B)\sin{i}$ for SB2, and $a\sin{i} \sim a_A\sin{i}$ for SB1. We further assume that, being observed as SB, $\sin{i}\sim 1$, which provides a lower limit for the semi-major axis $a$. With this estimation, and using \Gaia\ parallaxes, we can guess what would be the separation in arcsec between \Gaia\ sources if they were resolved. To maximize the chance of finding a resolved companion, we set the cone search radius for each system to be three times the estimated semi-major axis. We reliably find 10 systems in \SBcat\ resolved by \Gaia\ (see Table~\ref{tab:resolved}), whose \SBcat\ orbital periods exceed 3~a. Among them, three resolved systems were identified when synchronising \Simbad\ and \SBcat\ catalogues (\SBcat~879, 1815 and 3657). 
The system \SBcat\ 1815 (NGC 2682 131, in the old open cluster M67) is a spectroscopic triple where \jkt{the} \SBcat\ 1815 orbit is the outer one (representing AB pair) and \SBcat\ 1834 orbit is the inner one (component A). Component A (\Gaia\ DR3 04921546365012480) is itself an eclipsing binary while component B (\Gaia\ DR3 04921542069455104) is a blue straggler \citep{2001A&A...375..375V}. The primary of the eclipsing binary is itself a blue straggler \citep{2003AJ....125..810S}. Components A and B are separated by $0.42$ arcsec. It is the only \SBcat\ spectroscopic triple partially resolved by \Gaia.
Finally, we tentatively estimate the inclination for these resolved binaries using $i = \arcsin{(\rho_\SBcat^{\min}/\rho_{\Gaia}})$ where $\rho^{\min}_\SBcat$ is the minimum estimated separation from \SBcat\ orbit and $\rho_{\Gaia}$ is the separation between two \Gaia\ sources.
When parallaxes are not available from \Gaia, we used the \Hipparcos\ ones \citep{2007A&A...474..653V}.

\begin{figure*}
 \includegraphics[width=\linewidth]{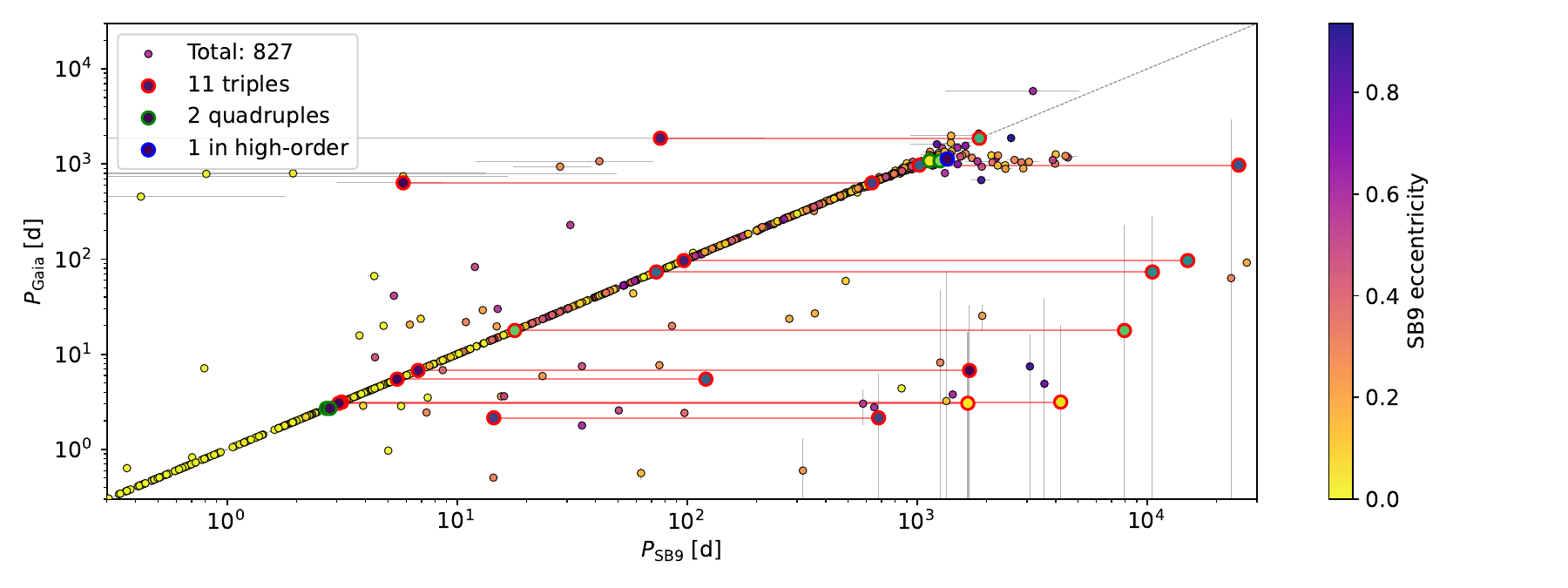}
 \includegraphics[trim=0.4cm 0 0 0, clip, width=\linewidth]{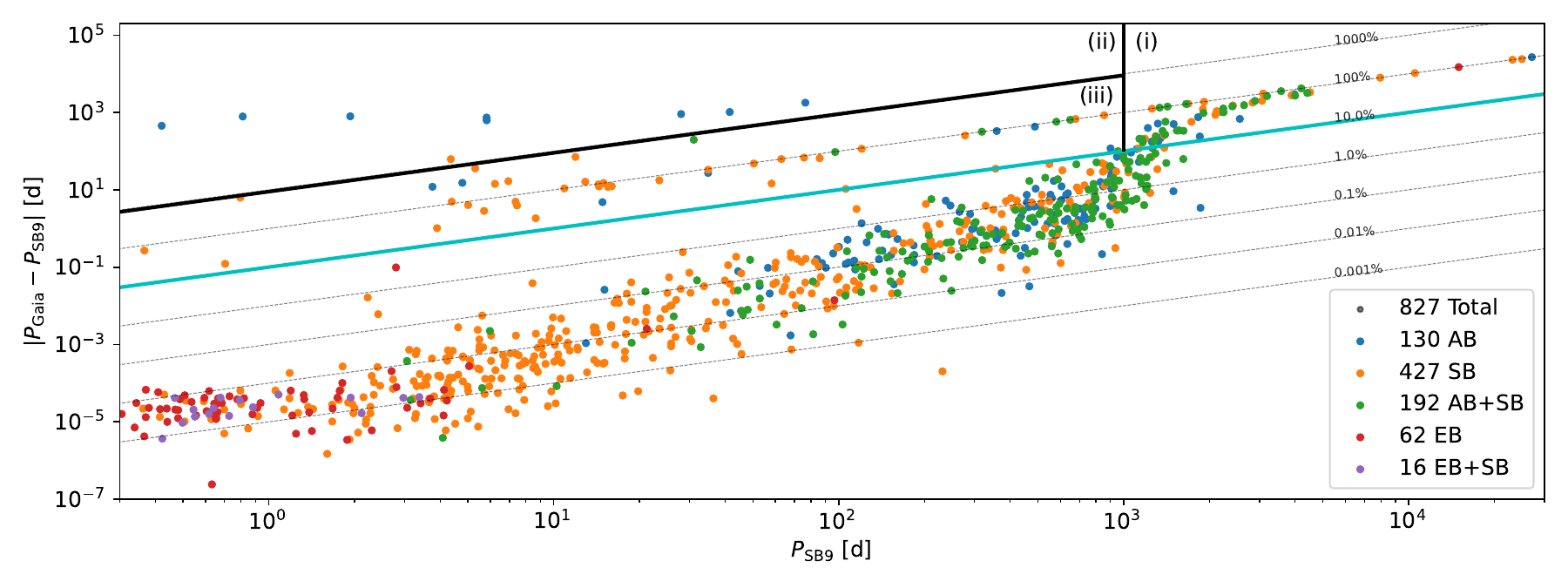}
 \caption{Comparison of orbital periods for the \SBcat\ systems in common with \Gaia~NSS. Top: Log-scale comparison colour-coded with eccentricities. Bottom: absolute period differences on log-log scale. Relative difference from 0.001\% to 1000\% are represented with dotted grey lines. The legend shows the \Gaia\ NSS binary type. The three outliers regions (i), (ii) and (iii) above the cyan line (relative period error larger than 10\%) are discussed in the text.}
 \label{fig:comp_gaia_period}
\end{figure*}
\begin{figure*}
\centering
    \includegraphics[width=\linewidth]{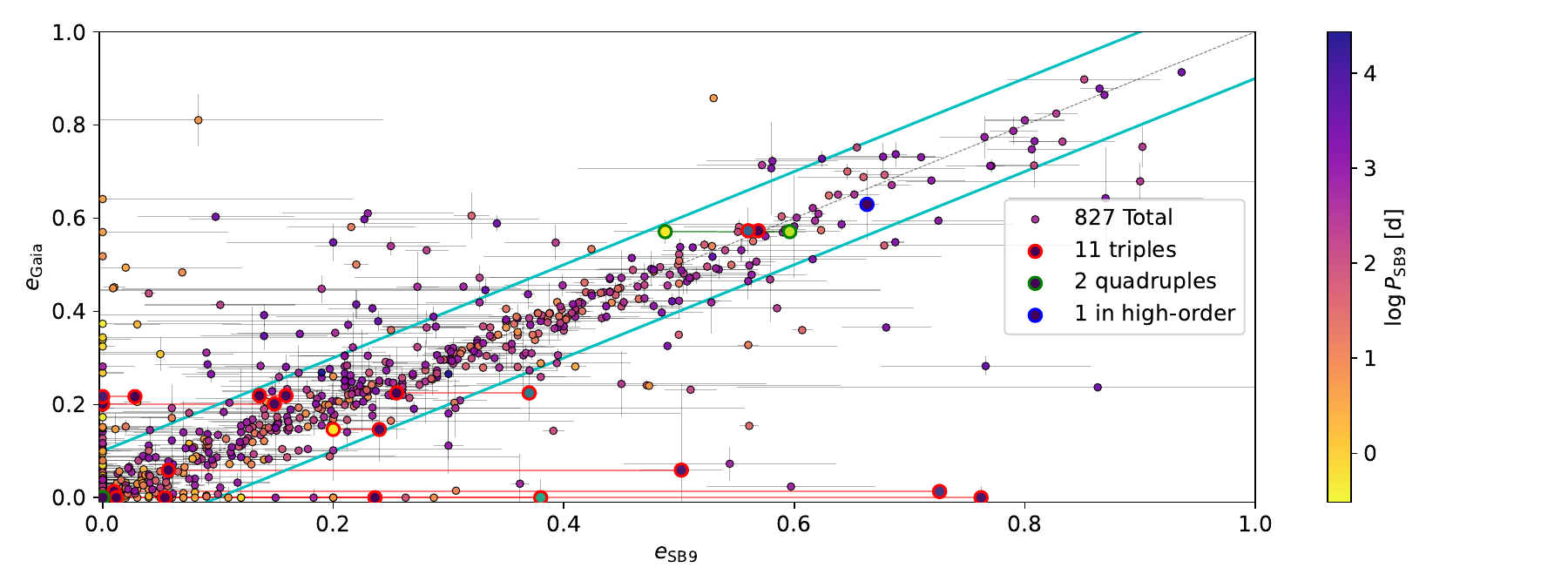}
    \caption{Comparison of eccentricities for the \SBcat\ systems in common with \Gaia~NSS, colour-coded with the decimal logarithm of the period.}
    \label{fig:comp_gaia_eccentricity}
\end{figure*}

\begin{figure*}
\centering
\includegraphics[trim=0 0.5cm 0 0.5cm, clip, width=0.8\linewidth]{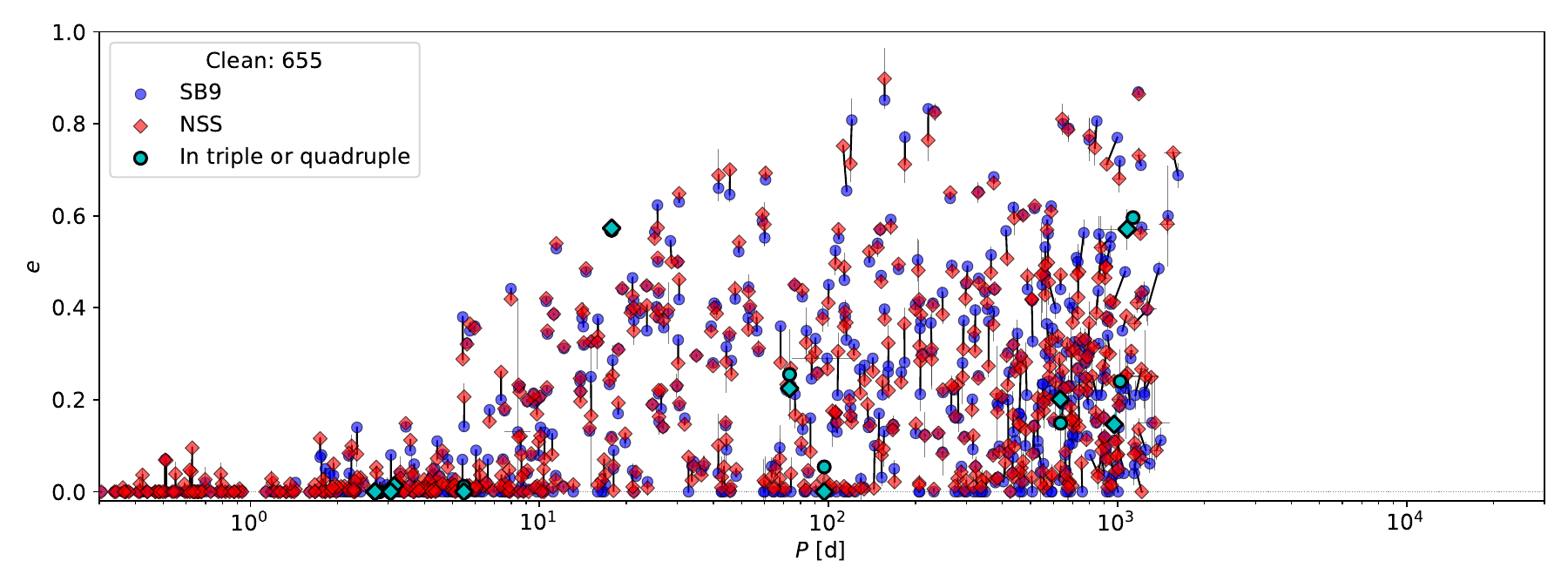}
\includegraphics[trim=0 0.5cm 0 0.5cm, clip, width=0.8\linewidth]{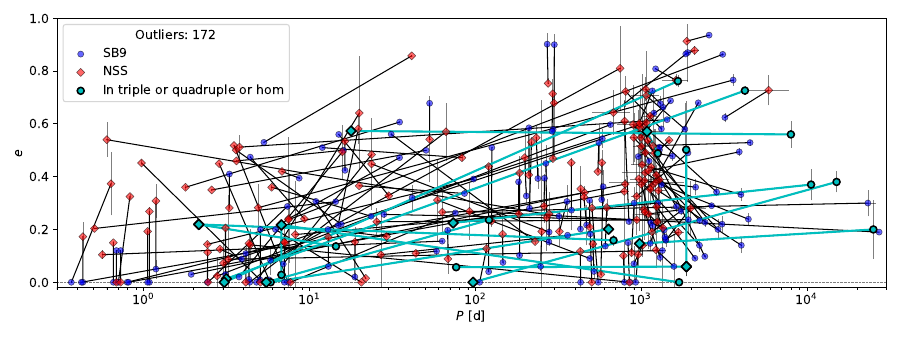}
\caption{Eccentricity vs decimal logarithm of the period for the common \SBcat-NSS sample of 827 binaries, 655 in the clean sample (middle), and 172 in the outliers (bottom). Orbital parameters from \SBcat\ are in blue and from \Gaia~NSS in red. We connected matched binaries with black lines. Binaries in triple, quadruple or higher-order multiples (hom) are shown in cyan.}
\label{fig:elogp_comp}
\end{figure*}

\subsubsection{Triple candidates with \Gaia}
In the course of identifying resolved \SBcat\ systems by \Gaia, we uncovered a certain number of triple-system candidates. We did not search extensively for them as it is not the primary goal of this paper, but report the \SBcat\ systems where the minimum estimated separation $\rho^{\min}_\SBcat$ are orders of magnitude smaller than the separation between two close \Gaia\ sources $\rho_{\Gaia}$. They are presented in Table~\ref{tab:triple_candidate}. The \SBcat\ periods of the inner pairs are reported confirming that such systems are not resolved by \Gaia. \jkt{A} difference between $\rho^{\min}_\SBcat$ and $\rho_{\Gaia}$ on one hand, and \jkt{short} periods of \SBcat\ systems on the other hand\jkt{,} make these \SBcat\ systems with close \Gaia\ sources good triple system candidates. None but one (\SBcat\ 2112) is detected as an astrometric binary itself in NSS with a period of $743.6\pm10.8$~d. In the \SBcat\ notes, it is mentioned that this system is a triple-lined system, but only the SB1 orbit of $5.8167\pm0.0009$~d for the short-period pair is given in \SBcat. This make the \SBcat\ 2112 a probable quadruple system with the \Gaia\ companion at 2.15 arcsec. We excluded one triple candidate from Table~\ref{tab:triple_candidate} because it is known as a quadruple \citep{2001AJ....122..402L} made of \jkt{an} SB1 (HD~119931~A) and an SB2 contact system (\SBcat~2029, HD~119931~B). In Table~\ref{tab:triple_candidate}, \SBcat\ 3068, 3509 and 3825 are less reliable candidates because only one parallax is available, making the companion a possible `optical' triple.

\subsubsection{Higher-order SB (not) resolved by \Gaia}
The presence of a unique \Gaia\ source for two or more \SBcat\ systems means that the multiple system is not resolved by \Gaia. 
\emph{E.g.} \SBcat\ 1524 (SB2) and 1525 (SB1) form a spectroscopic quadruple system with two visible components \citep{2006AJ....131.1702T}, previously detected as a triple \citep{2001A&A...374..227T}, this quadruple being component B of system $\chi$~Tau, the primary being a B9V star. High-order systems with \SBcat\ orbits are shown in Tables~\ref{tab:triple} and \ref{tab:quadruple}. Among the 76 spectroscopic triples (Table~\ref{tab:triple}),  we found only one to be partially resolved by \Gaia, \emph{i.e.} the outer companion is resolved, not the inner pair. One spectroscopic triple (\SBcat\ 157 \& 158, Algol) is too bright to be observed by \Gaia. Among the 34 spectroscopic quadruples (Table~\ref{tab:quadruple}), half (17 quadruples) are resolved by \Gaia, \emph{i.e.} the two pairs are resolved, not the individual components. Two of them have inner binaries identified as binaries in \Gaia DR3 NSS, \emph{i.e.} TYC 1212-473-1 (\SBcat 1607) as an NSS SB and HD~179332 (\SBcat\ 3839) as an NSS AB (see Table~\ref{tab:quadruple}). Among \SBcat\ systems belonging to higher-order multiples (Table~\ref{tab:hom}), we found about 14 \SBcat\ systems belonging to six higher-order multiples. In  the multiple system WDS 7346+3153, Castor~A and Castor~B are too bright for \Gaia (see Table~\ref{tab:too_bright}).

\section{Comparison with \Gaia\ DR3 NSS}
\label{sec:gaia}

We compared the \SBcat\ catalogue with the \Gaia\ Non-Single Star (NSS) catalogue \citep{2023A&A...674A..34G}, which contains the largest homogeneous sample of SB ($\sim277\,000$) but also astrometric binaries (AB, $\sim508\,000$) and eclipsing binaries\footnote{We do not mention the \jkt{approximately} $2\,184\,000$ EB from another table, \texttt{vari\_eclipsing\_binary}, which provides detections and partial orbital characterisations.} (EB, $87\,000$). There are overlaps between AB, SB and EB: $\sim58\,200$ binaries are AB+SB, 424 are SB+EB, 262 AB+EB and 17 are AB+SB+EB. We focus the cross-match on the table \texttt{nss\_two\_body\_orbit} which provides the complete set of orbital parameters, while tables \texttt{nss\_acceleration\_astro} and \texttt{nss\_non\_linear\_spectro} provide tentative and partial orbital solutions. 

In \SBcat, we select the best orbit to compare with, \emph{i.e.} the ones with the highest grade (from 0 to 5\footnote{An exception exist for $\alpha$~Cen (system 815), orbit number  3, with a grade of 5.1.}, the higher corresponding to the best one) or the most recent reference. Half of the \SBcat\ orbits have no grade. 
In NSS, a \Gaia\ source could have several orbital solutions as well, distinguished by their \texttt{nss\_solution\_type}. We selected as the best NSS orbit the one with the largest significance on the orbital period, \emph{i.e.} the largest period-over-error value.

\begin{figure*}
    \centering
    \includegraphics[width=0.33\linewidth]{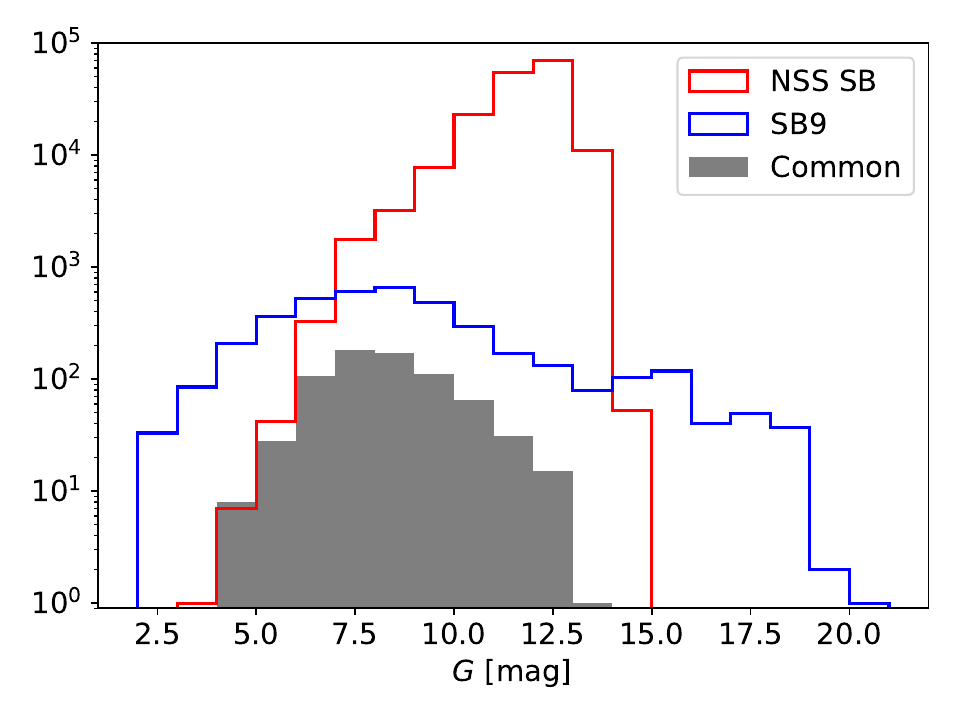}
    \includegraphics[width=0.33\linewidth]{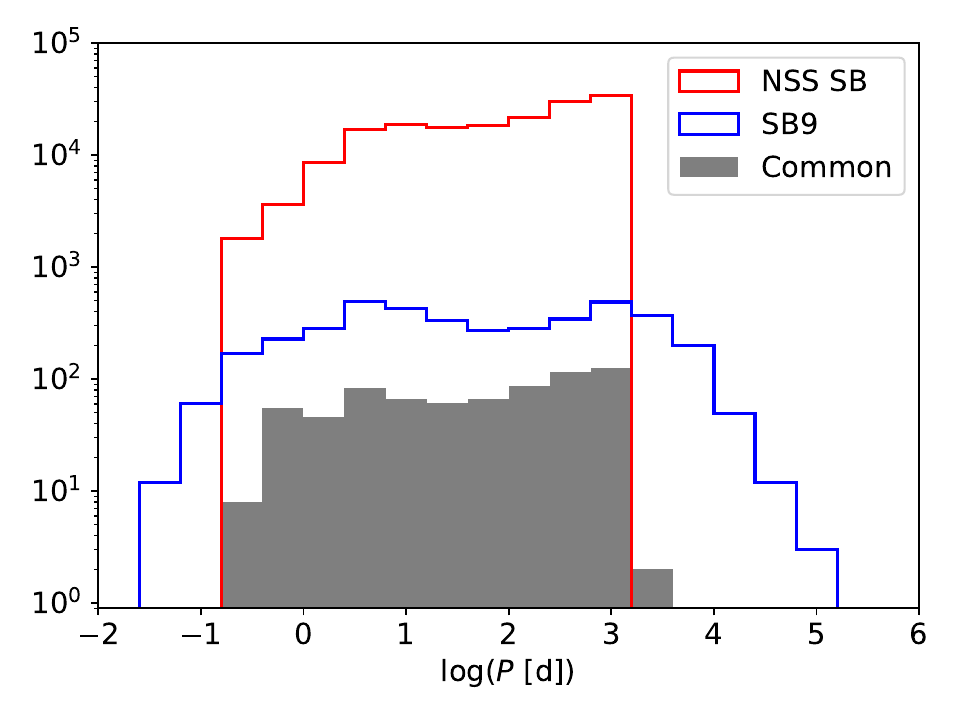}
    \includegraphics[width=0.33\linewidth]{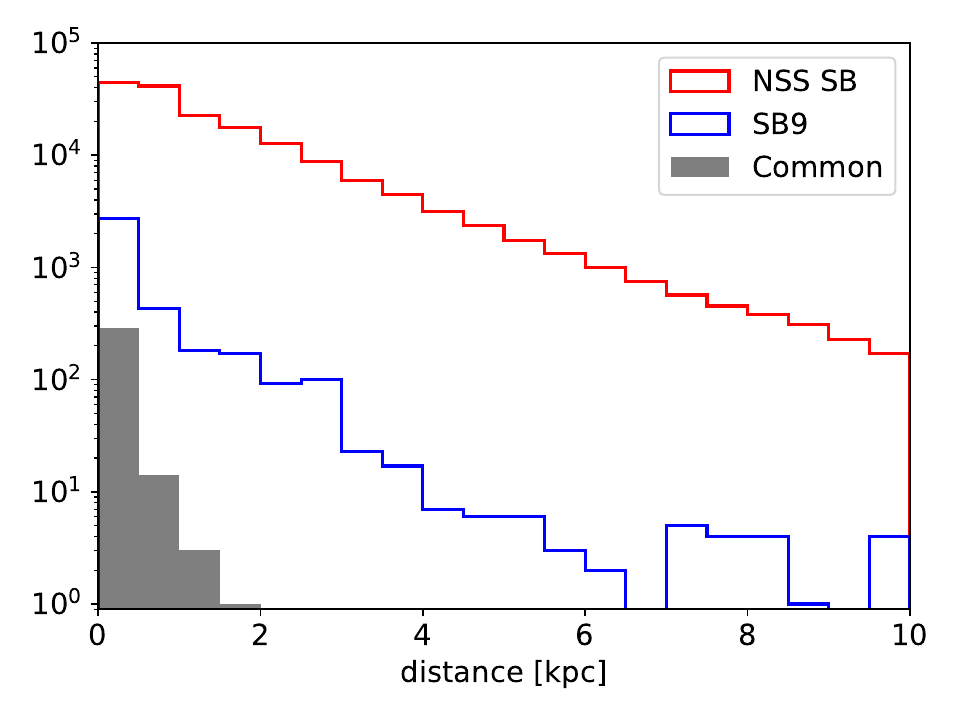}
    \caption{Magnitudes (left), period (middle) and distance (right) histograms of \Gaia~NSS SB with full orbital solutions (172k), \SBcat\ (4k) and the clean common SB sample (0.66k).}
    \label{fig:histo_mag_distance}
\end{figure*}

The intersection between \SBcat\ and \Gaia~NSS produces a common sample of 827 binaries (see Table~\ref{tab:common_sample}) which only represents $\sim1/5$ of binaries in \SBcat. 
Among them 14 are not resolved by \Gaia: 11 triples, \jkt{two} quadruples and \jkt{one}  higher-order \jkt{system}. For triple and quadruple systems, their \SBcat\ and \Gaia\ identifiers can be found in Tables~\ref{tab:triple} and \ref{tab:quadruple}.  
In this common sample, 62 have combined NSS orbital solutions: AB+SB for most of them and a few EB+SB.
We compare \SBcat\ periods with \Gaia\ NSS ones in two different ways as displayed in Fig.~\ref{fig:comp_gaia_period}. For this common sample, the tightest binary is GZ~And (\SBcat~1961) with an orbital period of 0.3~d which is also an EB. The widest pair is 26~Dra (\SBcat~2557) with a period of 27087~d, while it is estimated to be $92.04\pm0.29$~d in NSS. The binary with the largest \Gaia\ NSS period is BD+15~2570 with $5845.7\pm1868.9$~d (\SBcat~2301 with $3186\pm29$~d).
We also compare the \SBcat\ and \Gaia~NSS orbital eccentricities as shown in Fig.~\ref{fig:comp_gaia_eccentricity}, colour-coded by period. In general, the uncertainties are larger in \SBcat. The scatter is weaker at high eccentricities. We noticed two populations on the horizontal and vertical axes corresponding to circular binaries in one catalogue that have non-null eccentricities in the other catalogue. More \SBcat\ circular systems have non-null \Gaia~NSS eccentricities than vice versa.

\subsection{Identifying SB outliers}
\label{sect:outliers}
Half of this common NSS -- \SBcat\ sample has a relative precision on the orbital period better than 0.1\%. Among them, there are 16 systems both measured as eclipsing and spectroscopic binaries (EB+SB) with excellent solutions where the relative period difference is lower than 0.01\%. 115 systems (14\%) have relative period differences larger than 10\% and are considered as outliers (systems above the cyan line in the bottom panel of Fig.~\ref{fig:comp_gaia_period}). We can distinguish three categories among these outliers: (i) the ones due to the threshold effect (aliasing effect), (ii) outliers above the 1000\% relative difference, and (iii) outliers around the 100\% relative difference, \emph{i.e.} between 10 and 1000\%. The category (i) is due to a threshold effect around 1000~d, \emph{i.e.} \Gaia\ NSS binaries have periods saturating around 1000~d. This corresponds to the time span of \Gaia~DR3 based on 34 months of data collection (from 25 July 2014 to 28 May 2017). We identified 61 systems is this category. In category (ii) with relative difference larger than 10 (1000\%), nine systems are detected as astrometric binaries in \Gaia\ but one identified as SB1. They are all candidates triple systems with the inner binary measured as SB in \SBcat\ and outer system identified by astrometry with \Gaia (see Table~\ref{tab:unresolved_triples}). Two of them are spectroscopic triples in \SBcat, namely \SBcat\ 1624 \& 1625 and \SBcat\ 3727, 3728. The ratio of outer over inner periods ranges from 15 to 1000. The smallest ratio is the SB1 with \Simbad\ identifier NGC 2682 176 in the open cluster M67. The category of outliers (iii) with relative period difference between 10 and 1000\% (and \SBcat\ periods lower than 1000~d) represents 45 systems mainly consisting of \Gaia SB and AB+SB. The period ratio (\Gaia\ over \SBcat) ranges from 0.002 to 9. An illustrative example is given for \SBcat\ 231 with a period ratio of 0.1 on right panel of Fig.~\ref{fig:orbit2} as discussed in Sect.~\ref{sec:comparing_orbits}.

\begin{table*}
\centering
\caption{Nine unresolved triple candidates with \SBcat\ inner spectroscopic orbit and \Gaia\ NSS outer astrometric orbit.}
\begin{tabular}{lrcclrcc}
\hline
\Simbad\ & \SBcat & $P_\SBcat$ [d] & $e_\SBcat$ & \Gaia\ DR3 & $G$ & $P_{\Gaia}$ [d] & $e_{\Gaia}$ \\
\hline
63 Gem & 454 & 1.9327 & 0.03 & 865209037088705024 & 5.099 & $798.4 \pm 11.4$ & $0.372 \pm 0.020$ \\
HD 83270 & 1624\footnotemark[1] & 5.8214 & 0 & 1070770191963666816 & 7.466 & $633.6 \pm 2.8$ & $0.201 \pm 0.021$ \\
NGC 2682 176 & 1821 & $4.3556 \pm 0.0002$ & 0 & 604916422468464768\footnotemark[2] & 12.476 & $66.38 \pm 0.28$ & $0.570 \pm 0.109$ \\
$\phi$ Phe & 1916 & $41.489 \pm 0.019$ & $0.32 \pm 0.05$ & 4957620109130881664 & 5.096 & $1070.4 \pm 29.5$ & $0.605 \pm 0.015$ \\
V899 Her & 2024 & 0.4212 & 0 & 1325067886936436224 & 7.773 & $454.3 \pm 1.4$ & $0.339 \pm 0.024$ \\
HD 75638A & 2112 & $5.8167 \pm 0.0009$ & $0.0830 \pm 0.056$ & 604997202213330560 & 8.157 & $743.5 \pm 10.8$ & $0.810 \pm 0.161$ \\
HD 150710 & 2677 & $28.01 \pm 0.012$ & $0.257 \pm 0.014$ & 1312345747129055744 & 6.938 & $938.5 \pm 10.5$ & $0.237 \pm 0.005$ \\
V335 Peg & 2977 & 0.8107 & 0 & 2717981134566374272 & 7.149 & $787.7 \pm 48.4$ & $0.344 \pm 0.346$ \\
BD+60 585 & 3727\footnotemark[1] & $76.433 \pm 0.014$ & $0.057 \pm 0.016$ & 464891250151058816 & 8.779 & $1865.4 \pm 143.2$ & $0.059 \pm 0.0365$ \\
\hline
\end{tabular}
\begin{flushleft}
Notes. They all appear as outliers of the common sample and are consequently out of the clean common sample. \\
$^1$ Member of a \SBcat\ spectroscopic triple (1624, 1625) or (3727, 3728). \\
$^2$ Identified as SB1 in NSS.
\end{flushleft}
\label{tab:unresolved_triples}
\end{table*}

\begin{table*}
\centering
\scriptsize
\caption{Seven SB1 in the clean sample that are SB2 in \Gaia~NSS.}
\begin{tabular}{llllllrrll}
\hline
\Simbad\ & $K_1$ [km/s] & $P$ [d] & $e$ & Ref. & \Gaia~DR3 & $K_1$ [km/s] & $K_2$ [km/s] & $P$ [d] & $e$ \\
\hline
149 & 58.1 & 6.6383 & 0.07 & (1) & 461559592481317760 & $67.01 \pm 0.21$ & $79.57 \pm 0.3$ & $6.637771 \pm 0.0000065$ & $0.0096 \pm 0.0022$ \\
197 & 32.4 & 16.725 & 0 & (2) & 66507469798631808 & $39.86 \pm 0.38$ & $70.23 \pm 0.68$ & $16.7402 \pm 0.0012$ & $0.0698 \pm 0.0088$ \\
270 & 25.8 & 39.2807 & 0.36 & (3) & 3184295870015339264 & $54.09 \pm 0.26$ & $52.28 \pm 0.10$ & $39.2839 \pm 0.0016$ & $0.3485 \pm 0.0021$ \\
900 & 38.7 & 39.888 & 0.28 & (4) & 5828317422956035072 & $38.26 \pm 0.15$ & $46.95 \pm 0.20$ & $39.8429 \pm 0.0011$ & $0.2736 \pm 0.0033$ \\
906 & 20.3 & 31.846 & 0.16 & (5) & 6017724140678769024 & $34.33 \pm 0.14$ & $30.32 \pm 0.15$ & $31.8004 \pm 0.0011$ & $0.1464 \pm 0.0025$ \\
1118 & 86.5 & 1.8172 & 0.05 & (6) & 2153474650639369600\footnotemark[1] & $157.73 \pm 1.95$ & $112.94 \pm 7.16$ & $1.817470 \pm 0.000035$ & 0 \\
1382 & 84.9 & 2.341 & 0.02 & (7) & 1882432630526896000 & $124.63 \pm 0.48$ & $85.71 \pm 0.63$ & $2.340895 \pm 0.000013$ & $0.0318 \pm 0.0040$ \\
\hline
\end{tabular}
\begin{flushleft}
Notes. All these values appear as outliers of the common sample and are therefore outside the clean common sample. \\
$^1$ Circular SB2 in NSS.\\
(1) \citet{1960ZA.....49..206W} --
(2) \citet{1975PDAO...14..319P} --
(3) \citet{1945ApJ...101..370C} --
(4) \citet{1928AnCap..10....8S} --
(5) \citet{1970MNRAS.147..355B} --
(6) \citet{1968PhDT........23Y} --
(7) \citet{1976PASP...88..195B}
\end{flushleft}
\label{tab:sb1tosb2}
\end{table*}

When looking at the eccentricities, we consider as outliers systems that have an absolute eccentricity difference larger than 10\%, \emph{i.e.} outside the area defined by the cyan lines in Fig.~\ref{fig:comp_gaia_eccentricity}. They represent 129 systems (16\%). The intersection and combination of these two criteria (i) on relative period differences of 10\%, and (ii) on absolute eccentricity differences larger than 10\% provide samples of 72 and 172 systems, respectively. We will use the combination of these two criteria to determine the final set of 172 outliers representing about 21\% of the common sample. Figure~\ref{fig:elogp_comp} shows how this affects the eccentricity-decimal logarithm period diagram of this common sample (827 binaries), what results as the clean sample (top panel, 655 golden binaries), and the outliers themselves (bottom panel, 172 outliers). We also noticed that all systems with periods larger that 1500~d are excluded. When checking for higher-order mutliples, we identified one triple unresolved by \Gaia\ (476270508301993344) where the two \SBcat\ systems (216 \& 2052) have very similar periods of 2.6982 and 2.7966~d. \Gaia\ (476270508301993344) is identified as an EB with a period of 2.6984~d. This is why we set \SBcat\ 2052 as an outlier. This clean common sample can serve as a validation sample for future surveys. It will also extend toward longer orbital periods with future \Gaia~releases.  

Seven SB1 from the clean sample of 655 binaries turn out to be SB2 in NSS. Table~\ref{tab:sb1tosb2} reports the semi-amplitudes, periods and eccentricities in each catalogue. For these systems in \SBcat, no uncertainties are reported for semi-amplitude, period and eccentricity by the authors provided in the reference column, because studies are quite old. We noticed that the semi-amplitudes are significantly different between NSS and \SBcat. Among these seven SB1, four are variable: \SBcat~270 (RZ Eri) and \SBcat~900 ($\iota$~TrA) of spectral type F; and \SBcat~1118 (BH ~Dra) and \SBcat~1382 (GX~Peg) of spectral type A.  
Five false positives also popped up among the outliers, because they are all belonging to \SBcat\ spectroscopic triples unresolved by \Gaia, matching the outer \SBcat\ pair with the NSS inner pair. An interesting example is SS~Lac, known as a variable since 1915, and member of the open cluster NGC~7209. It is indeed an EB with a period of 14.4~d but on an eccentric orbit ($e=0.136$). After 1950, the eclipses fully disappeared, puzzling astronomers about the causes. Intensive investigations by \citet{2000AJ....119.1914T} revealed that the inclination changed by 6° over 40 years, due to the gravitational effect of a distant companion (Kozai-Lidov effect) on an outer orbit of $678.8\pm4.1$~d. This spectroscopic triple system has both inner and outer orbits in \SBcat~(\SBcat 2456 \& 2458).
It is not only unresolved by \Gaia,  but the inner SB2 orbital solution provides a period of 2.1~d with an eccentricity of 0.22 which is at odds with \citet{2000AJ....119.1914T} determination.

A similar study was conducted on a sample of about 100 triple stars monitored through the CHIRON spectroscopic program \citep{2023AJ....165..220T}. Among these, only about 30 systems have counterparts in the \Gaia\ DR3 NSS catalogue. The comparison reveals notable discrepancies: nine of these systems show different radial velocity amplitudes between the CHIRON spectrograph ($R\sim80\,000$) and Gaia NSS solutions, indicating an underestimation of the velocity amplitudes in \Gaia\ likely due to blending or insufficient resolution of spectral lines. Furthermore, some systems that CHIRON identifies as SB2 remain unresolved as SB2 in \Gaia\ data, affecting the completeness and accuracy of orbital characterization. One system exhibits \jkt{a} period alias in \Gaia\ NSS, where the reported period is much shorter than the true period determined from ground-based observations. These findings confirm that\jkt{,} while \Gaia\ NSS is invaluable for large-scale surveys, dedicated ground-based spectroscopic monitoring remains essential for detailed and reliable orbital solutions in cases of complex hierarchical systems such as triples, where \Gaia's automated pipeline struggles with blended or multiple-component spectra, and period aliasing.

\begin{figure*}
    \centering
    \includegraphics[width=0.43\linewidth]{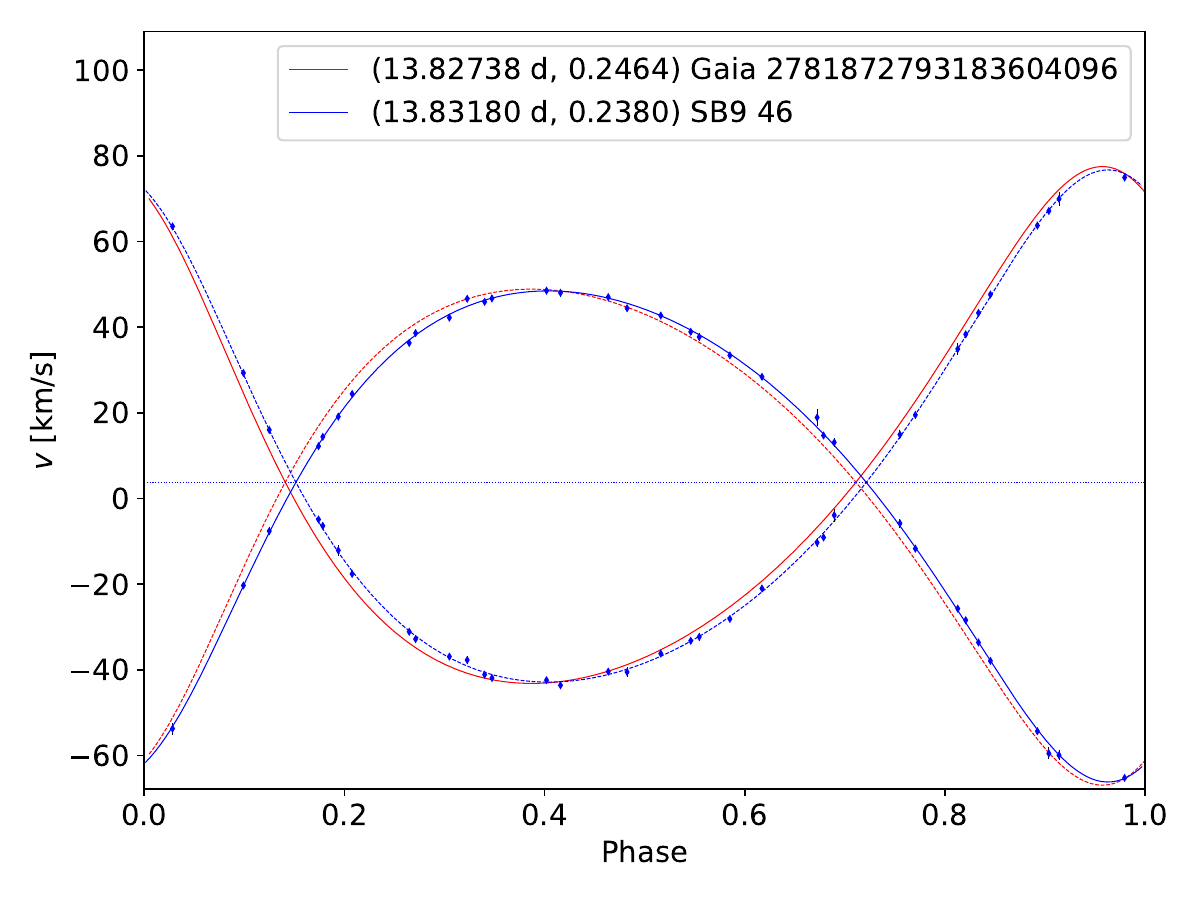}
    \includegraphics[width=0.43\linewidth]{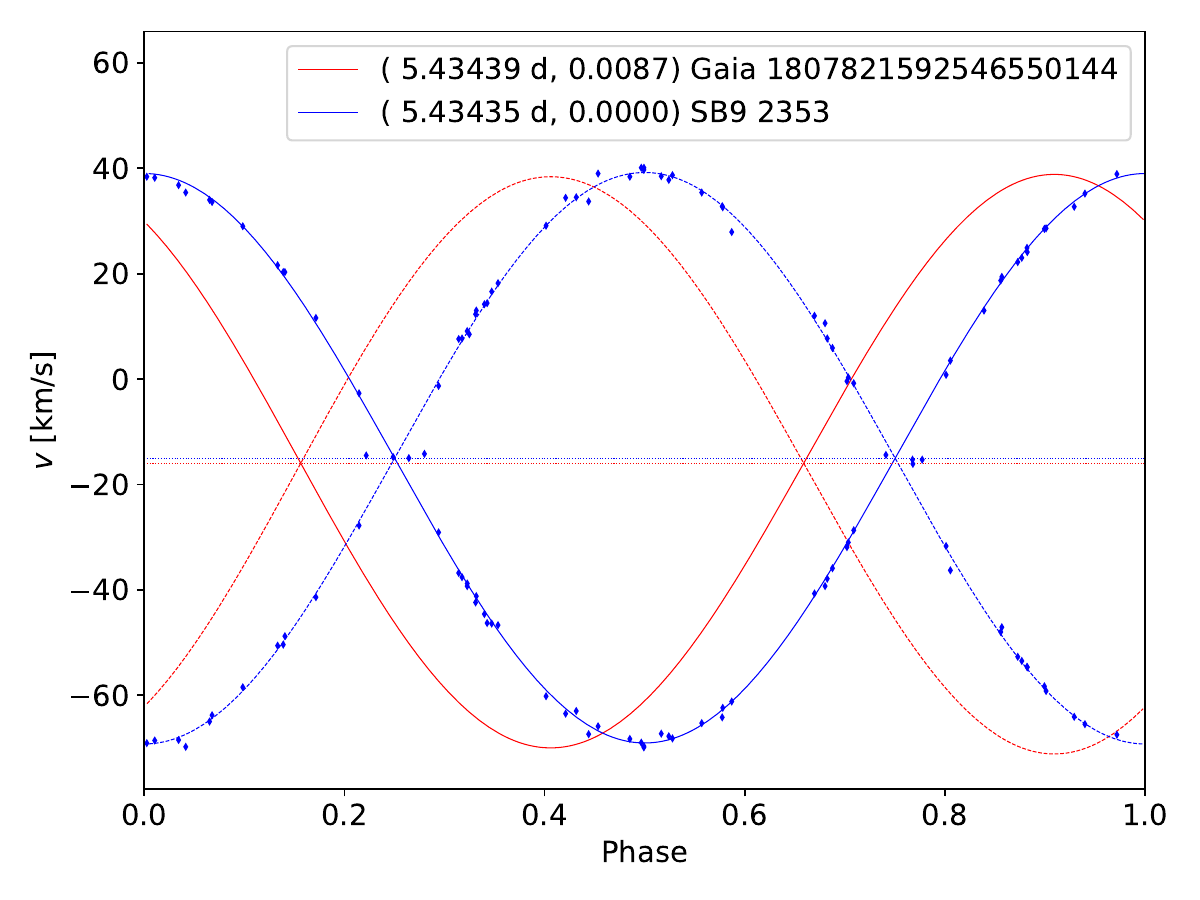}
    \caption{Comparison of RV curves for two close SB2 in the clean sample. Period and eccentricity are given in parenthesis in the legend. Only RV measurements from \SBcat\ are available. Horizontal dotted lines are the center-of-mass velocities. Left: 64~Psc (HIP~3810) with an \SBcat\ orbital solution from \citet{1991A&A...248..485D}. Right: V1423 Aql  (HIP~99210) with an \SBcat\ orbital solution from \citet{1990JApA...11...43G}.}
    \label{fig:orbit1}
\end{figure*}

\begin{figure*}
    \centering
    \includegraphics[width=0.43\linewidth]{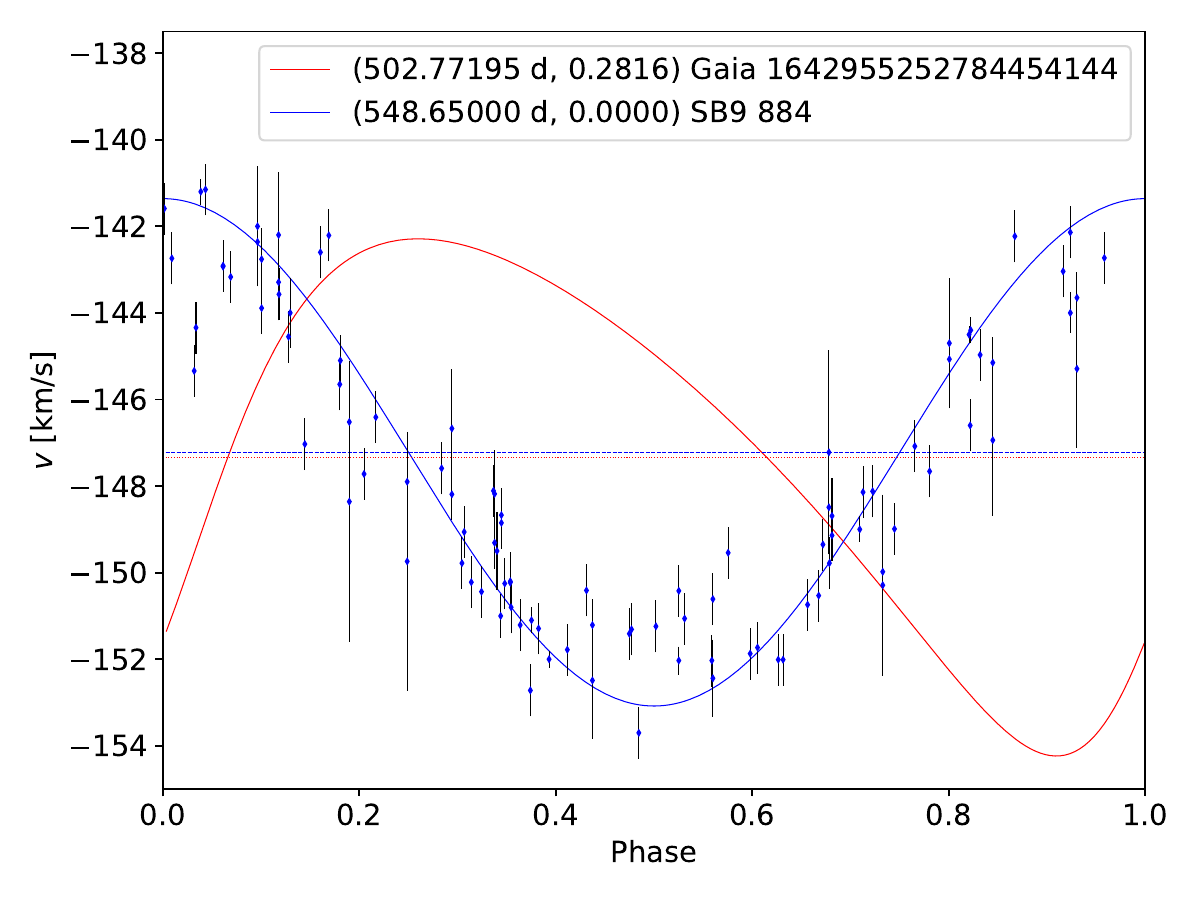}
    \includegraphics[width=0.43\linewidth]{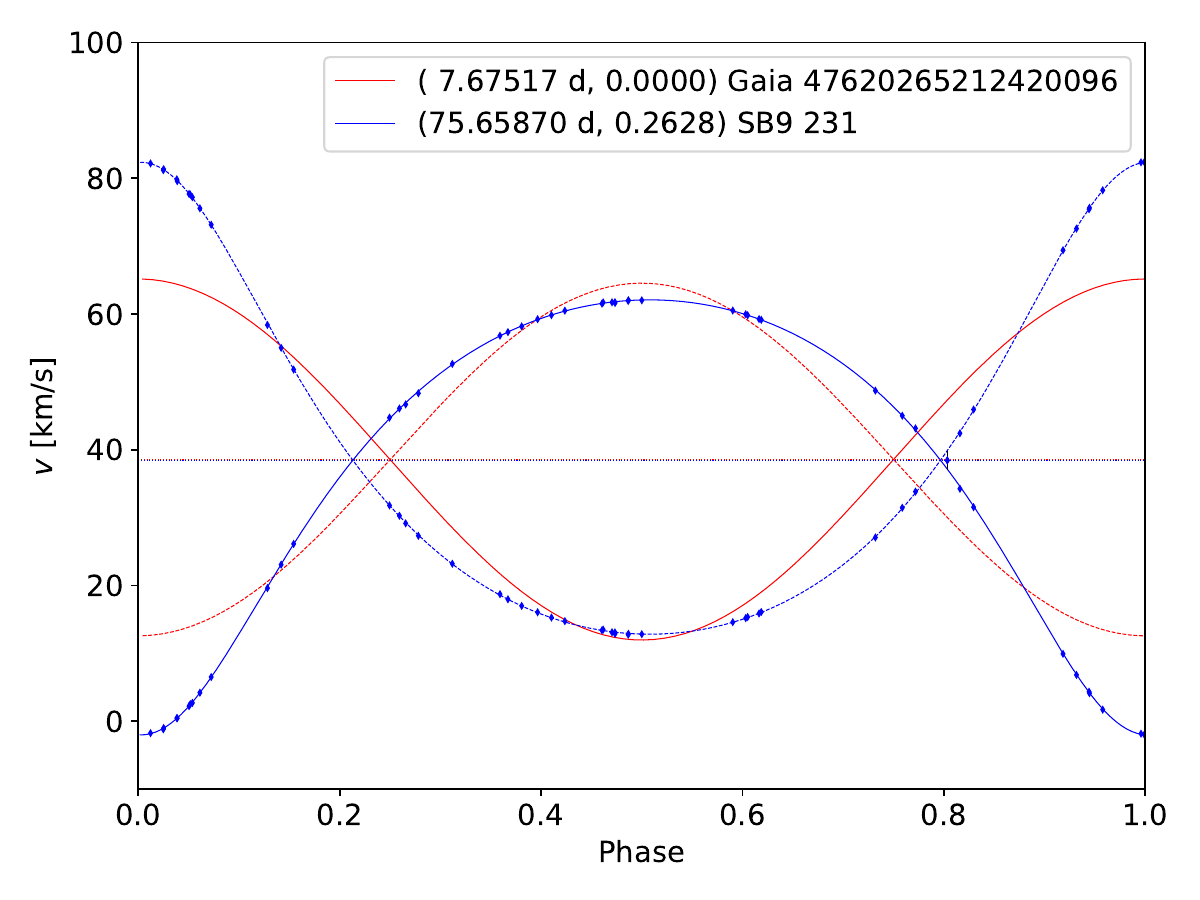}
    \caption{Comparison of RV curves for two SB identified as outliers. Left: AG~Dra (HIP 78512), a symbiotic star, with an orbital solution from \citet[][4 orbits exist in \SBcat]{2000AJ....120.3255F}. Right: V1232 Tau, member of the Hyades open cluster, with an orbital solution from \citet{2003Obs...123....1T}.}
     \label{fig:orbit2}
\end{figure*}

\subsection{Comparing orbital solutions}
\label{sec:comparing_orbits}
In Fig.~\ref{fig:comp_gaia_period}, we showed that about 50\% of the common binaries have a relative precision better than 0.1\% on the orbital period. We illustrate two cases from the clean sample at moderate and null eccentricities (Fig.~\ref{fig:orbit1}). 64~Psc (left panel) is an SB2 with a period of two weeks and a moderate eccentricity of 0.24. The orbital parameters of \SBcat\ and NSS are in excellent agreement, with more precise parameters from NSS which relies on 24 epochs, compared to 35 observations in \citet{1991A&A...248..485D}, but spanning 4 months only. A second example from the clean sample, V1423~Aql (right panel of Fig.~\ref{fig:orbit1}), is an SB2 on a quasi-circular orbit of 5.4~d. We notice here a phase shift which could be explained by the fact that the \SBcat\ solution \citep{1990JApA...11...43G}  assumed a circular orbit (null eccentricity and periastron argument) while NSS provides a non-null eccentricity and periastron argument ($e = 0.0087\pm0.0011$ and $\omega=33.3\pm5.9^\circ$).

We also show two cases from the outliers in Fig.~\ref{fig:orbit2}. While the centre-of-mass velocities are in excellent agreement, the orbital elements are different. The first example is a symbiotic star AG~Dra (left panel, an SB1). The relative difference in orbital period is lower than 10\% but while \citet{2000AJ....120.3255F} assumes a circular orbit, NSS derived an excentric orbit with $e = 0.282\pm0.028$. A phase shift is also present ; the time span of the observations is about 6600~d, covering about 12 cycles. The second outlier example is V1232~Tau (HIP~20056) from the Hyades open cluster, which is an SB2 with a period of 76~d and an eccentricity of 0.263, identified to have a 7.7~d period with a null eccentricity in NSS (with 20 good observations). The \SBcat\ solution covers a time span of about 2400 d, involving about 30 orbital cycles. The semi-amplitudes of the NSS solution are significantly smaller. This \jkt{is} a typical example \jkt{where} the \SBcat\ solution seems more reliable than the \Gaia one.

\begin{figure*}
    \centering
    \includegraphics[trim=4cm 0 3cm 0, clip, width=0.9\linewidth]{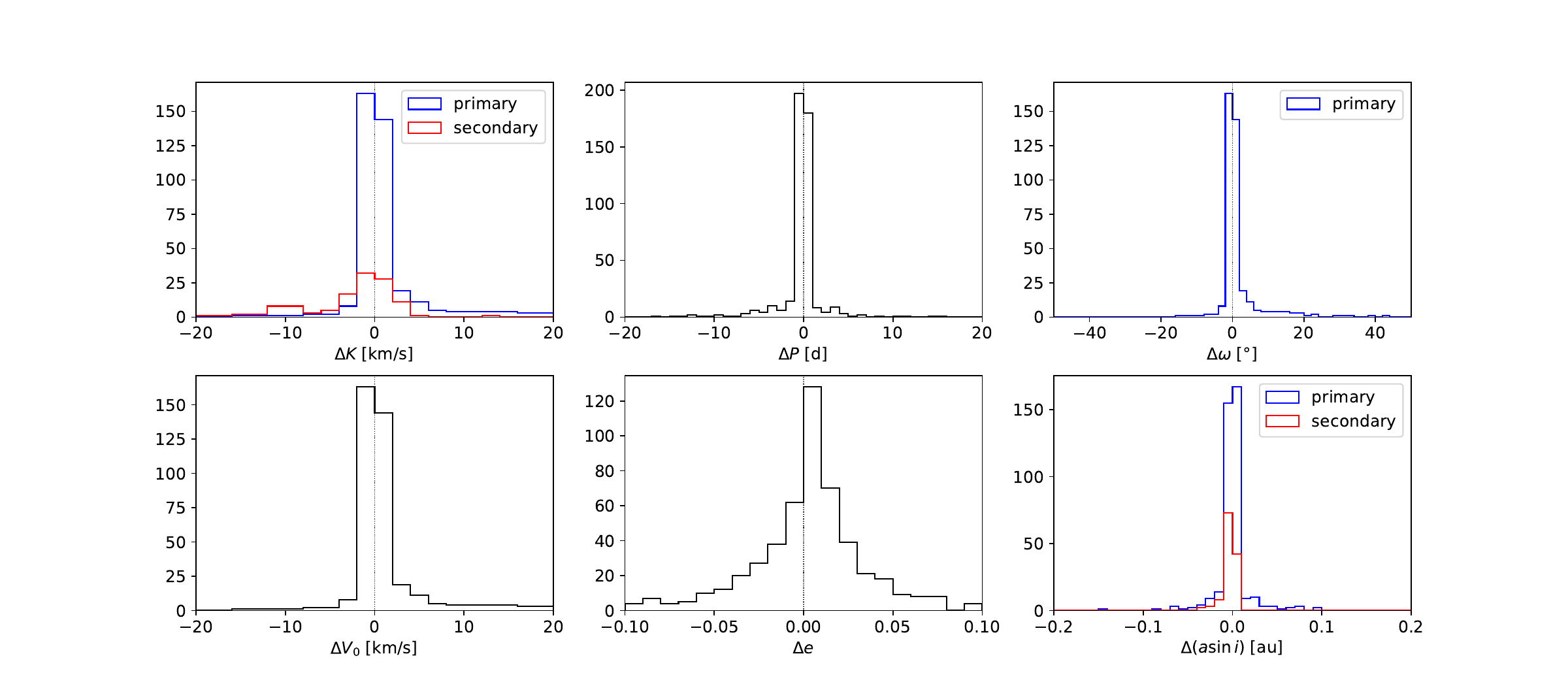}
    \caption{Histograms of the differences (\Gaia\ $-$ \SBcat) in orbital parameters for a subsample of 494 SB of the clean common sample.}
    \label{fig:histo_delta_op}
\end{figure*}

Finally we compare the spectroscopic orbital parameters for a subsample (494) of the clean common sample. We selected all common binaries with \texttt{nss\_solution\_type} of SB1 (250), SB1 with astrometric solution (104), SB2 (110), circular SB2 (25) and spectroscopic binaries with eclipses (5). For them, we plotted the differences (\Gaia\ $-$ \SBcat) in semi-amplitudes $K$, centre of mass velocities $V_0$, periods $P$, eccentricities $e$, periastron argument of the primaries $\omega$, and in projected semi-major axis $a\sin{i}$ on Fig.~\ref{fig:histo_delta_op}. Overall, thanks to the selection in relative period and absolute eccentricity difference of 0.1, the agreement \jkt{is} rather good with $\langle \Delta K_1 \rangle = 2.0\pm9.3$~\kms\ and $\langle \Delta K_2 \rangle = -4.7\pm13.8$~\kms; $\langle\Delta V_0\rangle = 0.1\pm2.6$~\kms; $\langle\Delta P\rangle = -1.9\pm14.3$~d; $\langle\Delta e\rangle = (2.4\pm31.0)\times 10^{-3}$; $\langle\Delta \omega\rangle = 7\pm69$~°; and $\langle\Delta (a_1 \sin{i})\rangle = (0\pm18)\times 10^{-3}$~au and  $\langle\Delta (a_2 \sin{i})\rangle = (-3\pm7)\times10^{-3}$~au. 

\subsection{Why is the common sample so small?}
The clean common sample of 655 golden binaries is composed, in NSS, of 81~AB, 351~SB, 151~SB+AB, 57~EB, and 15~SB+EB.
We show in Fig.~\ref{fig:histo_mag_distance} the comparison between the magnitudes (left), the periods (middle), and the distances (right) of \Gaia~NSS, \SBcat\ and the clean common sample of SB. The NSS SB are the ones with full spectroscopic orbital solutions (\emph{i.e.} excluding SB with non\jkt{-}linear trends from the \Gaia \texttt{nss\_non\_linear\_spectro} catalogue). There are about 172k and cover a magnitude range of [3.5, 14.4], smaller than the SB from \SBcat. Indeed, \SBcat\ has 4k SB in the V magnitude range [$-1.46$, 26.9], but in the plotted histogram, this range is reduced to $G$ magnitudes [2.1, 20.9] excluding the too bright and too faint stars that are not in \Gaia, and summarized in Tables~\ref{tab:too_bright} and \ref{tab:too_faint}. 

The histogram of periods in the middle panel of Fig.~\ref{fig:histo_mag_distance} clearly show that many \SBcat\ systems have shorter or longer periods compared to SB from NSS. The common sample covers the same range of orbital periods as the parent NSS SB distribution. There are two SB from the clean common sample with periods larger than 1400~d: BD+75~523, \SBcat\ 2263 ($1620\pm7$~d for \SBcat\ and $1558\pm110$~d for NSS) with an NSS AB+SB solution; and  the red giant 28~Ori, \SBcat\ 2844 ($1496\pm15$~d for \SBcat\ and $1487\pm100$~d for NSS) with an NSS AB solution. This is why they appear with larger periods compare to the NSS SB sample in red, in the middle panel of Fig.~\ref{fig:histo_mag_distance}.

\begin{figure}
    \centering
    \includegraphics[trim=1.5cm 0 1.5cm 1.2cm, clip, width=\linewidth]{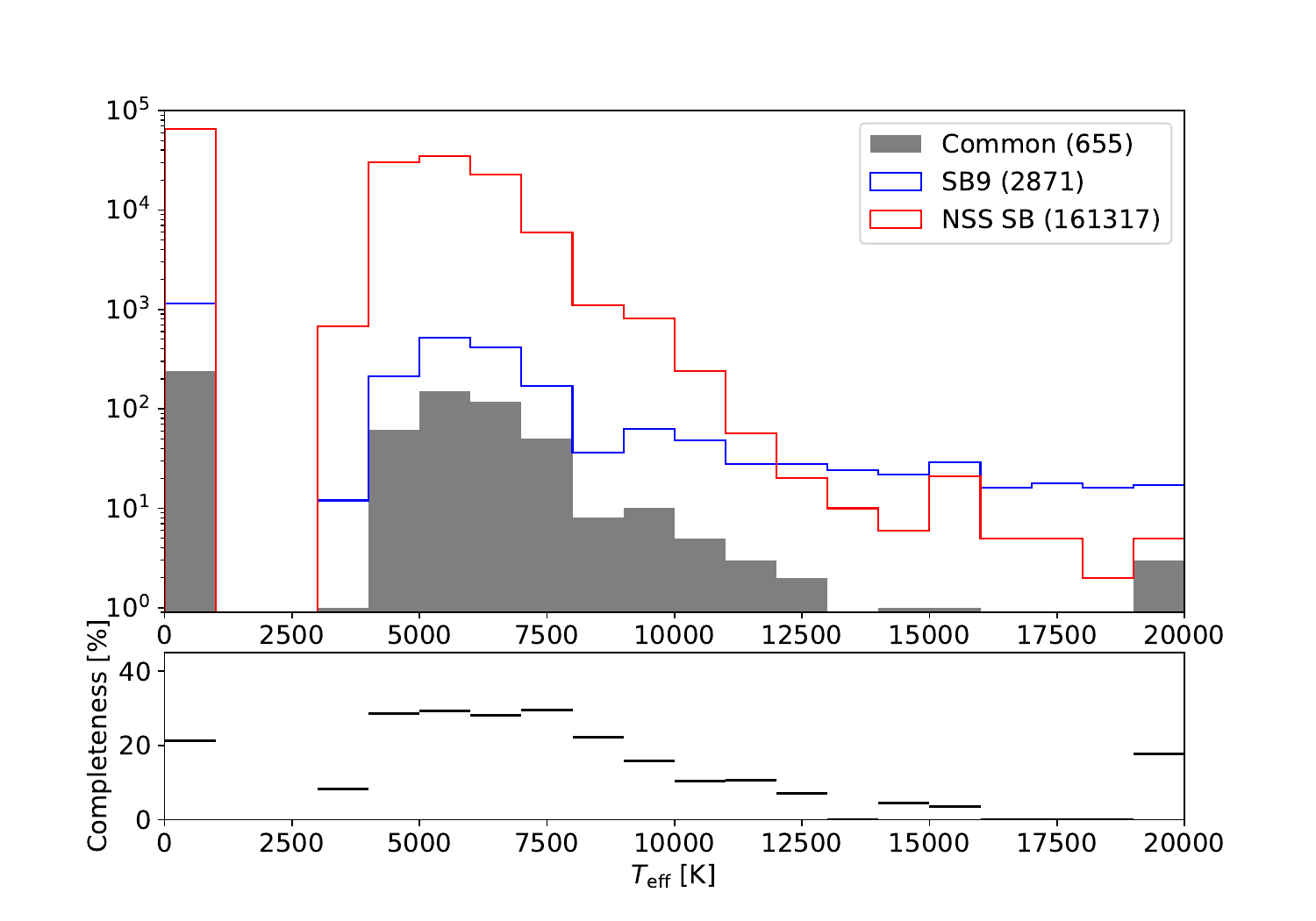}
    \caption{Top: comparison of $T_\mathrm{eff}$ (from Gaia DR3 GSPphot) distribution for NSS spectroscopic binaries (red), \SBcat\ systems (blue) and the clean common sample of binaries (grey). The samples are selected on the same $G$ magnitudes [4.2, 13.0] and period [0.3, 1558]~d ranges of the clean common sample. At 0 K are shown the amount without any $T_\mathrm{eff}$. Bottom: completeness in \SBcat. }
    \label{fig:histo_teff}
\end{figure}

\begin{figure}
    \centering
    \includegraphics[width=\linewidth]{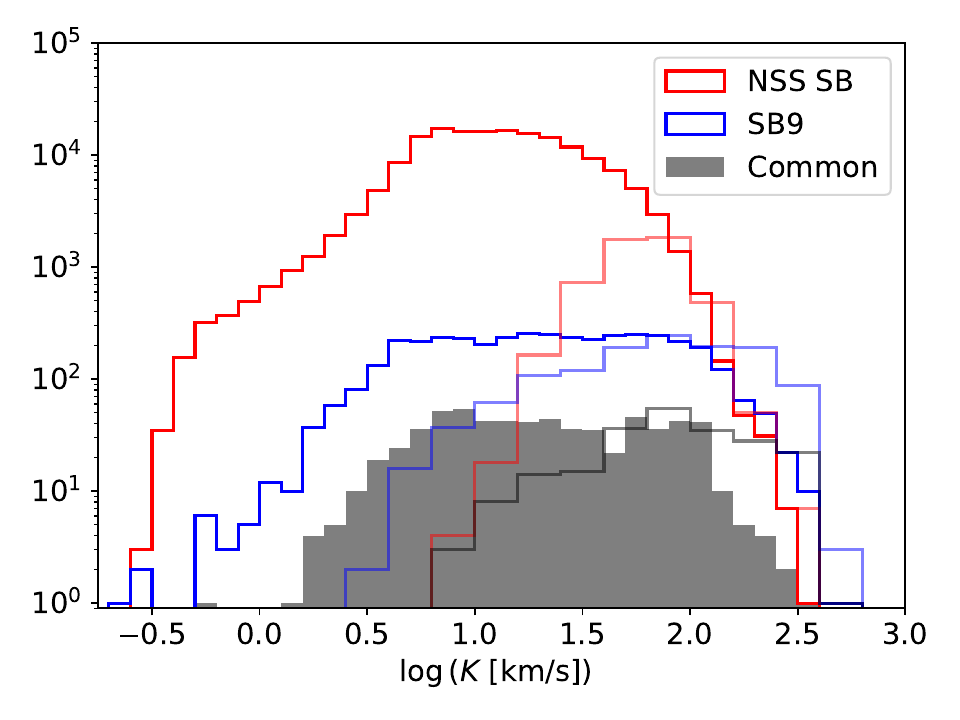}
    \caption{Semi-amplitude of the primaries for NSS spectroscopic binaries (red), \SBcat\ systems (blue) and the clean common sample of binaries (grey). Semi-amplitudes of the secondaries are shown with light red, blue and grey, for the corresponding samples.}
    \label{fig:histo_K}
\end{figure}

The clean common sample represents only 23\% of the \SBcat\ binaries on the same period-magnitude-limited sample with G magnitudes in [4.2, 13.0] and periods in [0.3, 1558] d range.
We investigated why the common fraction between \SBcat\ and \Gaia\ NSS is so small. There is not a unique answer and it turns out that many factors are responsible for this. 
(i) \SBcat\ systems brighter than $G\sim3$ represents $\sim 130$ systems, \emph{i.e.} 3\% of \SBcat\ systems. 
(ii) 11\% of \SBcat\ binaries have $G>13$ and hence are impossible to be observed as SB with the \Gaia RVS. 
(iii) 22\% of \SBcat\ binaries have periods exceeding 1000~d and consequently are possibly not in NSS or appearing as outliers with relative period difference > 10\%.
(iv) Early-type binaries (with $T_\mathrm{eff} \ge 10\,000$~K) are over\jkt{-}represented in \SBcat\ and under-represented in NSS, as clearly depicted in Fig.~\ref{fig:histo_teff}. The bottom panel shows that the completeness, defined as the ratio of the common sample over \SBcat\ systems per bin of effective temperature, is about 30\% for FGK primaries; it strongly decreases for M and early-type primaries. Most of them also appear as short period binaries, and \SBcat\ have more SB with $K_1\gtrsim 150$~\kms\ as seen \jkt{in} Fig.~\ref{fig:histo_K}.
(v) We identified one important source of incompleteness: binaries for which an astrometric motion is present. According to \citet{2025A&A...693A.124G}: \emph{"Since the RVS is a slitless spectrometer, the calibration in wavelength of the spectra is requesting an epoch position of the object that is accurate enough. Therefore, for the objects that are deviating from the single star behaviour, we could expect that the wavelength calibration is not correct, and thus the measured transit RVs are somewhat biased and this creates artefacts in the time series. No correction of this effect is offered for DR3."} Subsequent strong selection cuts are performed in the analysis of SB1, with a conservative approach, rejecting binary solutions if $F_2 > 3$, if the phase coverage has \jkt{a} gap larger than 0.3, if $K\ge 250$~\kms, etc. \citep[see][for more details]{2025A&A...693A.124G}. 
(vi) Finally, when we cross-match \SBcat\ with \Gaia\ spectroscopic orbital trends which include about 56.8k systems, we only found 56 common systems. Among the 338k \Gaia\ systems with astrometric acceleration solution, \emph{i.e.} systems with non-linear proper motion, we found 206 systems in common. Comparing with these partial orbital solutions only account for 7\% of \SBcat. \jkt{Contrary} to \cite{2024A&A...682A...7B}, we never reached a completeness of 40\%, even in the same period-magnitude-limited sample\footnote{We have the same G magnitude range but they used a maximum period of 1100~d.} for which we only obtained 24\% instead of 23\% completeness. This is mainly due to the fact that they only identified 3413 unique \SBcat\ (while we have 3976  unique \SBcat\ systems corresponding to 3892 unique \Gaia\ identifiers, see Table~\ref{tab:common_sample}, because 10 \SBcat\ are resolved by \Gaia, see Table~\ref{tab:resolved}). 

The clean common SB sample is represented on the CaMD (Fig.~\ref{fig:cmd}) with magnitudes brighter than 13 and not only cover\jkt{s} the main sequence but also the red giant branch. In addition, all the SB systems in the clean common sample are closer than 2~kpc, while 73\% of NSS SB and 92\% of \SBcat\ systems meet this criterion. We also note that 99.6\% of \SBcat\ systems and 99.2\% of Gaia NSS SB are located within 10~kpc.

\begin{figure*}
\centering
    \includegraphics[width=0.75\linewidth, trim=0 2.5cm 0 2.5cm, clip]{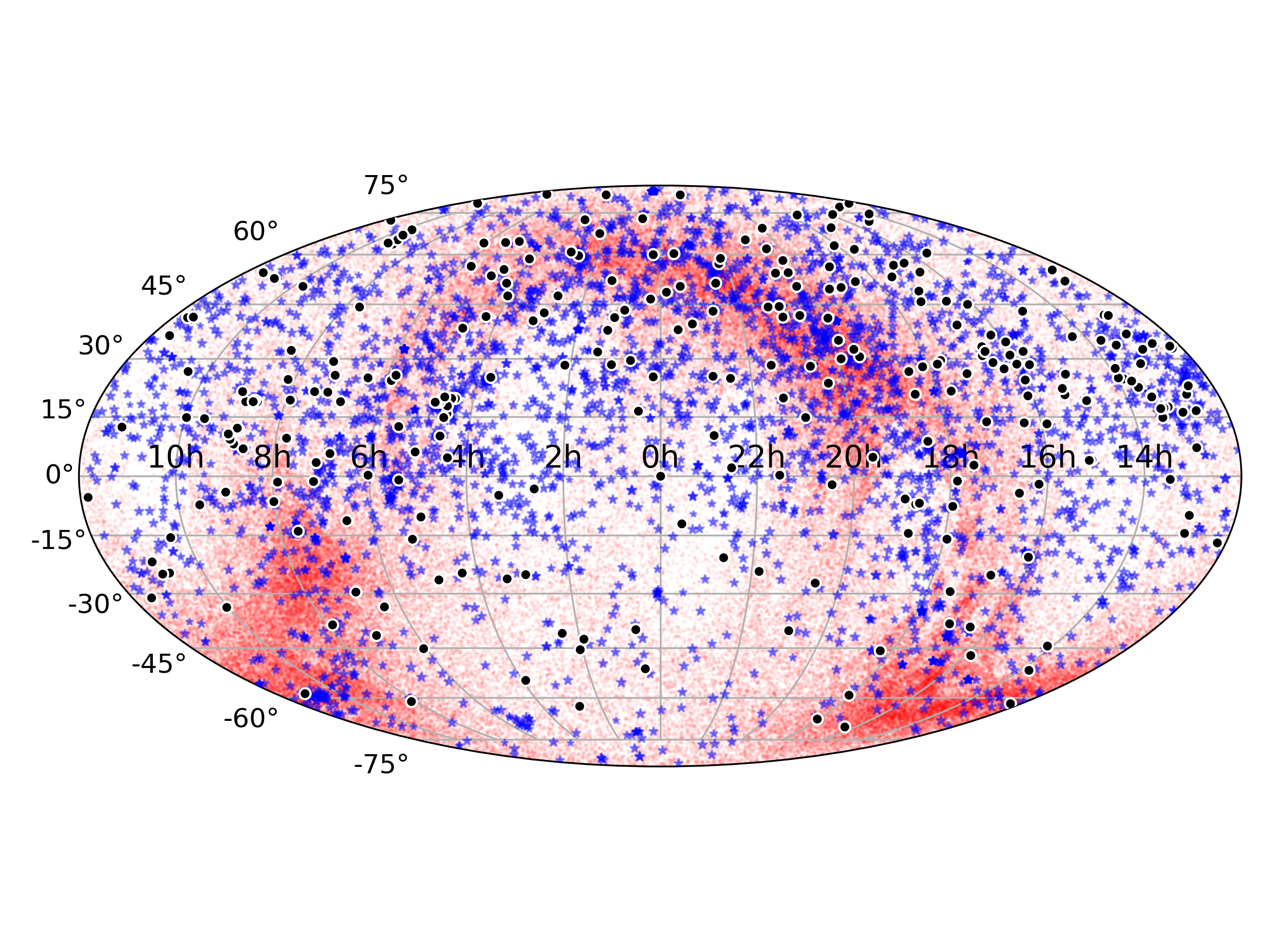}
\caption{Mollweide projection of \Gaia~NSS (light red dots), \SBcat (blue stars) and clean common sample (black circles) of SB in equatorial coordinates.}
\label{fig:skymap}
\end{figure*}

Finally we show the distribution on the sky of \Gaia~NSS, \SBcat\ and the clean common sample of SB in a Mollweide projection in equatorial coordinates (Fig.~\ref{fig:skymap}). While the \Gaia SB show \jkt{an} over-density in the Galactic plane, the \SBcat\ systems show an under-density in the southern hemisphere, which is also the case for the clean common sample, and is mainly due to historical observations made in the northern hemisphere.

\section{Evolution of \SBcat\ towards a new version: \SBX}
\label{sec:improvements}
The modernization and extension of the \SBcat\ catalogue has recently taken a decisive step forward. Building on the limitations identified in the legacy \SBcat\ infrastructure (namely, flat file storage, inconsistent data formatting, limited search capabilities, and inflexible user interfaces), our work has led to the creation of the \SBX\ (\emph{The eXtended Catalogue of Spectroscopic Binary Orbits}\footnote{\jkt{Already} available at \url{http://astro.ulb.ac.be/sbx}}), a scalable, standards-compliant platform designed to support the next decades of research on binary stars.

At the core of this transformation is the migration of the entire catalogue to a modern PostgreSQL relational database. The new schema, designed with domain best practices, enforces primary/foreign key constraints, normalization of system and orbit tables, and type safety for all physical and bibliographic parameters. Historical inconsistencies, ranging from missing values and typographical errors, were systematically resolved through custom preprocessing scripts and cross-validation against source literature and external catalogues. This new version enables more reliable and complex queries, eliminates duplicates, and establishes a solid foundation for future extensions. Details of this migration and the new implementation can be found on the \SBX\footnote{\url{http://astro.ulb.ac.be/sbx/intro}}.

A key enhancement in \SBX\ is the systematic enrichment of astrometric and photometric parameters using external resources, mainly \Gaia\ DR3. High-precision positions, proper motions, parallaxes, and $G$ magnitudes are now included for nearly all systems for which cross-identification is possible; fallbacks to \Hipparcos\ and \Simbad\ ensure completeness for the brightest (Table~\ref{tab:too_bright}) and faintest binaries (Table~\ref{tab:too_faint}). 
Data provenance is explicitly tracked, and new data fields document the source and reliability of each entry.

Complementing the new database, \SBX\ features a modern, Python-based web application (using Django framework\footnote{\url{https://djangoproject.com}}) that vastly expands search, visualization, and data export capabilities. The new interface allows: (i) robust and resilient object lookup by name, coordinate (cone search), or alias, including fuzzy matching and integration with the CDS Sesame resolver; (ii) immediate access to all orbital solutions, notes, and bibliographic references for a given system; (iii) interactive and downloadable plots of radial velocity curves, with uncertainties and phase folding, generated dynamically on the server; (iv) direct download of measurements in CSV format; and (v) a portal (under development) for user-submitted orbit solutions with workflow for data validation and review.
This architecture ensures a clear separation data, logic, and presentation, making future updates and extensions straightforward.

Recognizing the necessity for seamless integration with virtual observatory (VO) tools, \SBX\ provides a Table Access Protocol (TAP) service compliant with International Virtual Observatory Alliance (\href{https://www.ivoa.net/}{IVOA}) recommendations \citep{2008ivoa.spec.1030O}. Powered by a dedicated TAP/ADQL middleware (using the \texttt{vollt} library, \citealt{2023ivoa.spec.1215M}), this service enables standardized batch querying and cross-matching for external platforms, pipelines, and visualization tools (e.g., \href{https://www.star.bris.ac.uk/~mbt/topcat/}{Topcat}). This opens the catalogue to large-scale, federated, and automated scientific workflows, and reduces integration friction for external astronomical data \jkt{centres}.

While \SBX\ represents a major leap in technical capability and scientific usability, ongoing and future efforts will focus on several fronts. (i) Complete the missing RV data for 910 \SBcat\ systems (23\% of the full catalogue). On a larger effort basis, it would also be welcome to add the list of $\sim300$ references selected from 1987 to 2002 into \SBX. 
 (ii) Where possible, provide the missing uncertainties on the orbital parameters; for example, 35\% of the periods and 47\% of the eccentricities lack quoted uncertainties.
(iii) Implement a new grading scheme for the quality of the orbits. The previous grading scheme was empirical and subjective to the contributor to the \SBcat\ catalogue. The idea is to use objective criteria to assess the quality of an orbit: small uncertainties on the radial velocities, and good sampling of the orbital phase. The last one may be tricky to define for eccentric orbits. We could rely on the eccentric anomaly by imposing an anomaly gap smaller than a given value. Another option could be to define four grades: A -- all orbital parameters constrained; B -- some orbital parameters constrained; C -- orbital parameters unconstrained; D -- wrong parameters/no SB. Various options still need to be more carefully investigated before concrete implementation. (iii) Complete spectral type and luminosity class and add atmospheric parameters (effective temperature, gravity and metallicity for one (SB1) or both (SB2) components when available elsewhere (\emph{e.g.} from the \Gaia\ atmospheric parameters, or other large spectroscopic surveys). The future addition of interactive front-end components, advanced filtering, and visualization will improve user experience for data exploration.

The \href{http://www.astro.ulb.ac.be/sbx/}{\SBX website} ensures that the catalogue of spectroscopic binaries has evolved from a legacy static archive to an open, dynamic, and interoperable research infrastructure, fully aligned with big data astronomy and VO best practices. This transformation puts the new \SBX\ catalogue at the heart of future developments in stellar astrophysics, binary star population studies, and multi-catalogue synergy.

\section{Conclusion}
The \SBcat\ catalogue is a comprehensive compilation of approximately 4\,000 spectroscopic binary systems, consisting of about 2\,800 single-lined (SB1) and 1\,200 double-lined (SB2) binaries gathered from literature. This work presents the current status of the \SBcat\ catalogue as of 2021-03-02, \emph{i.e.} more than 20 years after the reference article by \citet{2004A&A...424..727P}, including its detailed statistical properties such as distributions of orbital periods, eccentricities, spectral types and luminosity classes of both primary and secondary stars. A major part of this study is the careful cross-match of \SBcat\ with the \Gaia\ DR3 catalogue, revealing 3\,976 common sources and highlighting the complexities of matching binaries across catalogues with issues such as duplicates and unresolved multiples. The comparison with \Gaia\ NSS shows partial overlap with insights into orbital parameters consistency and discrepancies. We identified a clean common sample of 655 systems which includes \jkt{eight} binaries in spectroscopic triples and \jkt{two} binaries in quadruple systems. We identified higher-order multiples (76 triples, 34 quadruples, and six higher order systems in which SB are present) within \SBcat\ and discussed systems resolved or unresolved by \Gaia. Importantly, this work results in the transition from \SBcat\ to the modernized \SBX\ (extended) catalogue, featuring a relational SQL database, improved web interface, and Virtual Observatory access standards, aiming to enhance accessibility, data quality, and analysis capabilities for the binary and multiple star community. The evolution of \SBcat\ into \SBX\ addresses data completeness, includes coordinates and astrometric parameters from \Gaia\ DR3,
as well as a new table for parents and children binaries in hierarchies, paving the way for future large-scale binary star research.

\section*{Acknowledgements}
\footnotesize
We thank the referee for constructive remarks that help to improve the readibility of the manuscript.
We thank E. Gosset, C. Babusiaux and O. Malkov for useful discussion and user feedback on the \SBcat\ catalogue.
T.M.\ is granted by the BELSPO Belgian federal research program FED-tWIN under the research profile Prf-2020-033\_BISTRO. 
J.S.\ acknowledges support from STFC under grant number ST/Y002563/1. 
This research has made use of the \Simbad\ database, operated at CDS, Strasbourg, France \citep{2000A&AS..143....9W}.
All the previous contributors to \SBcat\ are warmely thanked: R. Griffin, J.M. Carquillat, M. Imbert, L. Szabados, D. Stickland, E. Glushkova, and R. Leiton. 
Deep gratitude to D. Pourbaix, who passed away too soon, for bringing the \SBcat\ catalogue into the digital age and enriching it beyond all other contributors.
This work has made use of data from the European Space Agency (ESA) mission\Gaia\ (\url{https://www.cosmos.esa.int/gaia}), processed by the\Gaia\ Data Processing and Analysis Consortium (DPAC, \url{https://www.cosmos.esa.int/web/gaia/dpac/consortium}). Funding for the DPAC has been provided by national institutions, in particular the institutions participating in the \Gaia\ Multilateral Agreement.

\vspace{-0.5cm}
\section*{Data Availability}
Table~\ref{tab:common_sample} will be available online.
All the data of this work can be retrieved on the \SBX\ website and its associated TAP service at \url{https://astro.ulb.ac.be/sbx}.
\vspace{-0.5cm}
 



\bibliographystyle{mnras}
\bibliography{sb9gaia_mnras} 



\appendix

\section{\SBcat\ distributions for dwarf and giant components}

\begin{figure*}
\centering
    \vspace{1em}
    \includegraphics[width=0.49\linewidth]{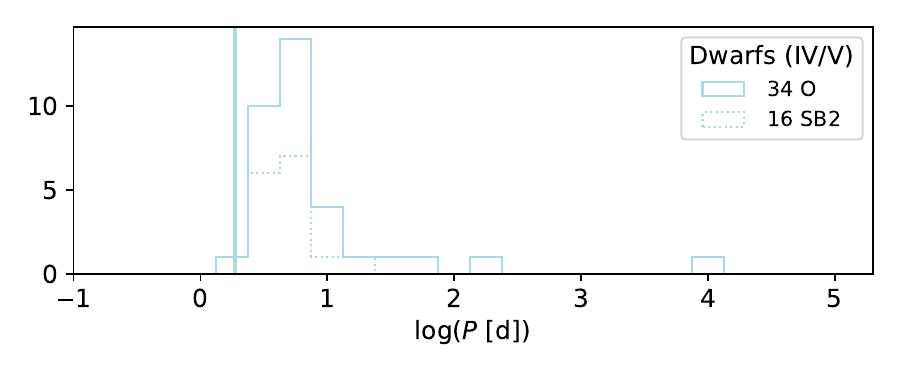}
    \includegraphics[width=0.49\linewidth]{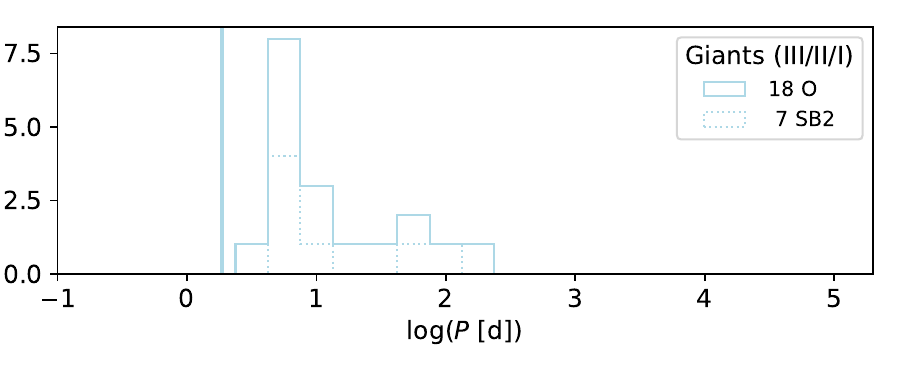}
    \vspace{-2em}
    \includegraphics[width=0.49\linewidth]{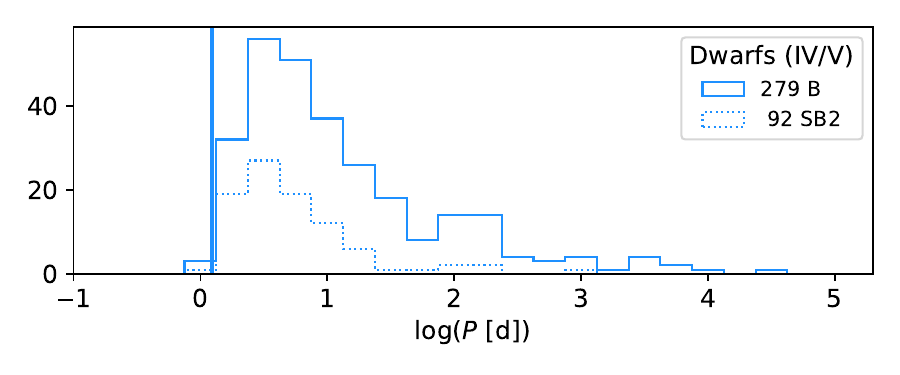}
    \includegraphics[width=0.49\linewidth]{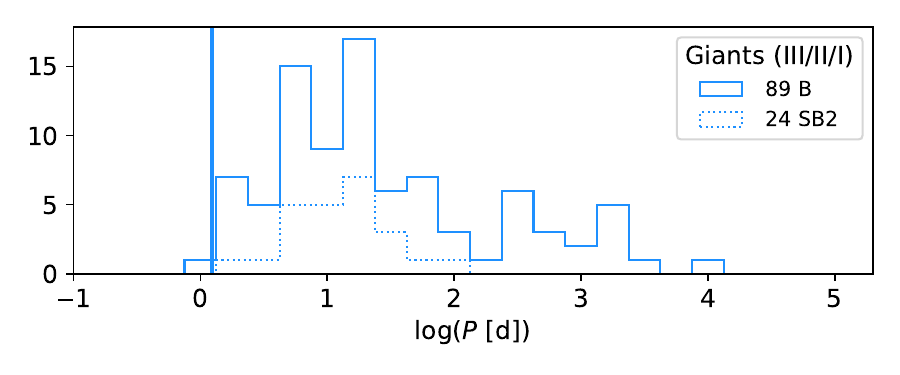}
    \vspace{-2em}
    \includegraphics[width=0.49\linewidth]{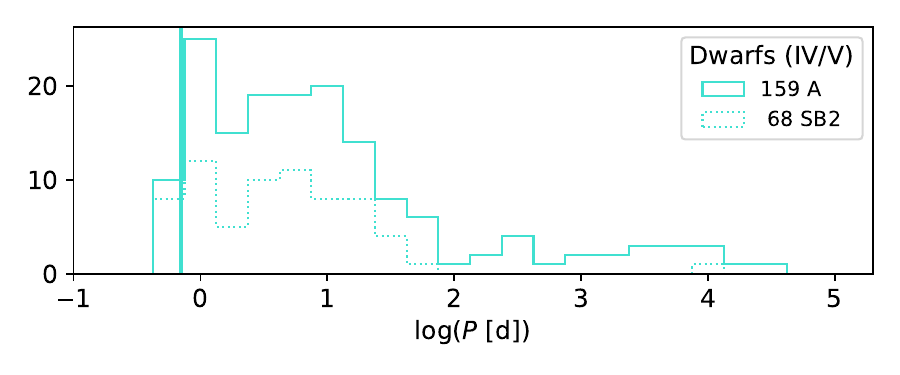}
    \includegraphics[width=0.49\linewidth]{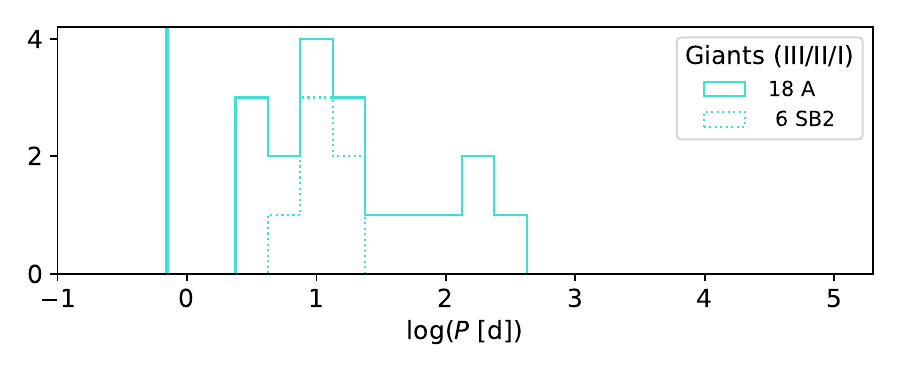}
    \vspace{-2em}
    \includegraphics[width=0.49\linewidth]{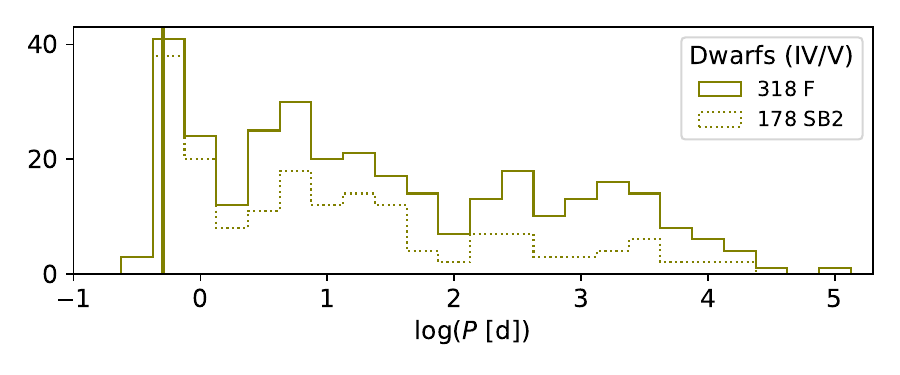}
    \includegraphics[width=0.49\linewidth]{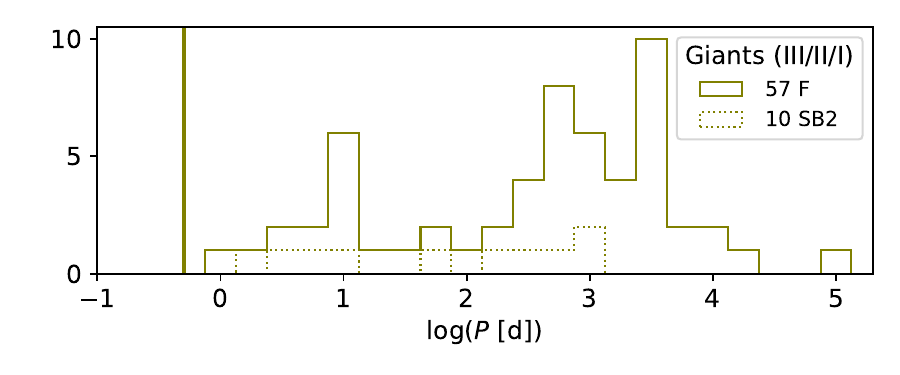}
    \vspace{-2em}
    \includegraphics[width=0.49\linewidth]{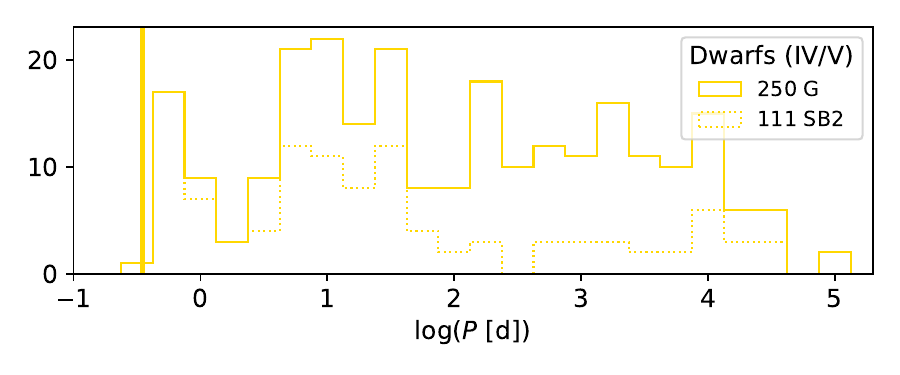}
    \includegraphics[width=0.49\linewidth]{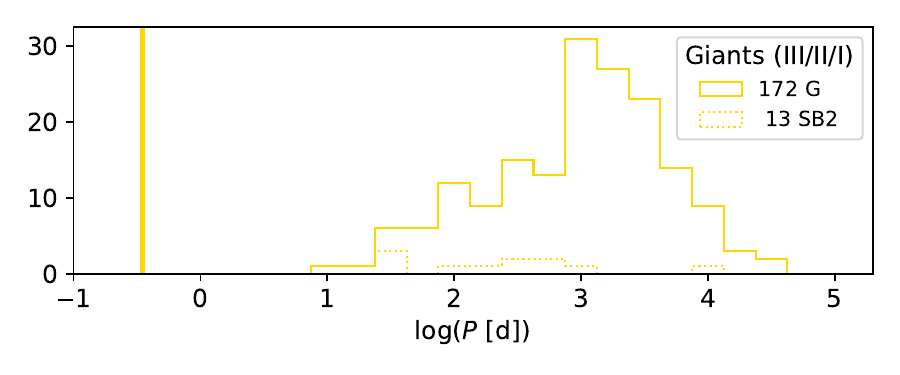}
    \vspace{-2em}
    \includegraphics[width=0.49\linewidth]{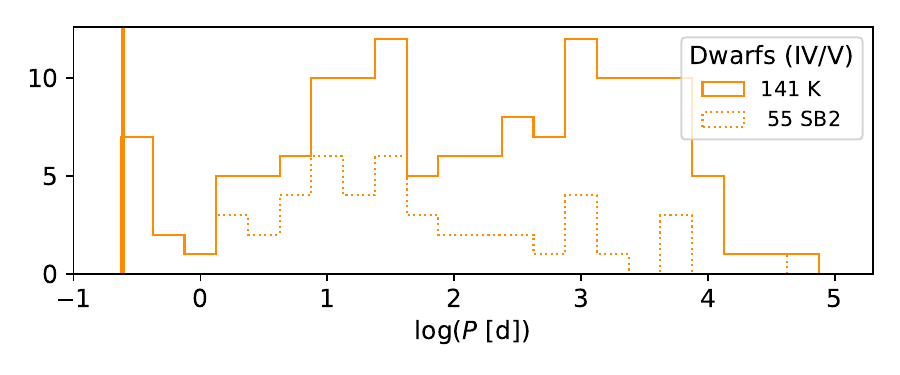}
    \includegraphics[width=0.49\linewidth]{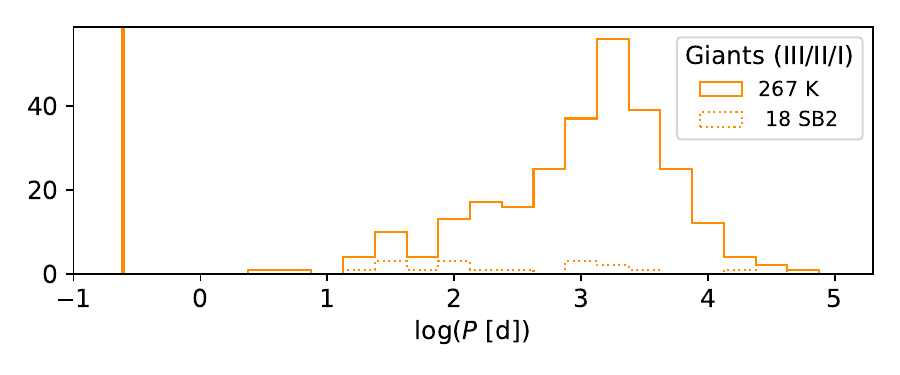}
    \vspace{-2em}
    \includegraphics[width=0.49\linewidth]{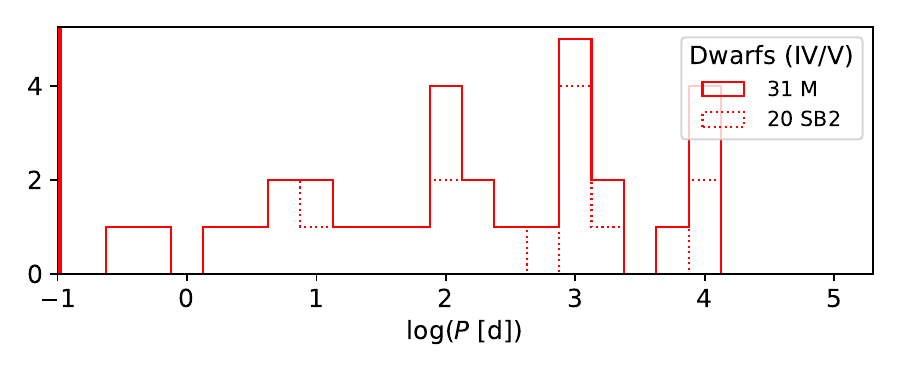}
    \includegraphics[width=0.49\linewidth]{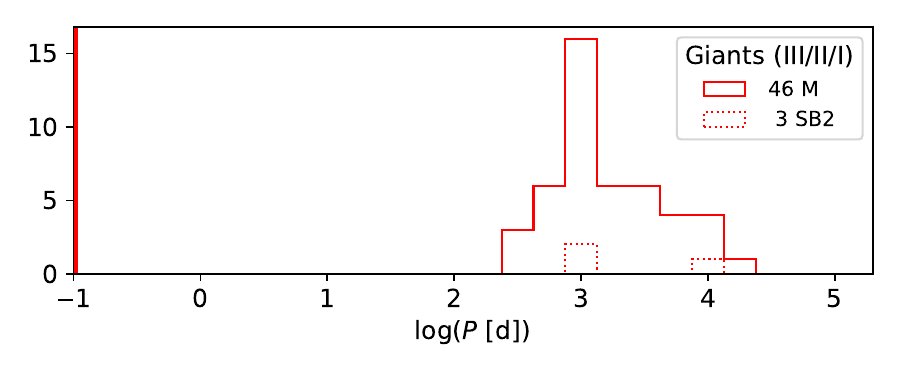}
    \vspace{1em}
    \caption{Distribution, in absolute numbers, of the \SBcat\ orbital periods per spectral type for dwarfs, corresponding to luminosity classes of IV and V (left) and for giants, corresponding to luminosity classes of III, II and I (right) for the SB primary component. Colour codes spectral type. Vertical lines correspond to the minimum period estimated for a contact equal-mass binary of a given  spectral type.}
    \label{fig:histo_periods_details}
\end{figure*}

\begin{figure*}
    \centering
    \vspace{1em}
    \includegraphics[width=0.49\linewidth]{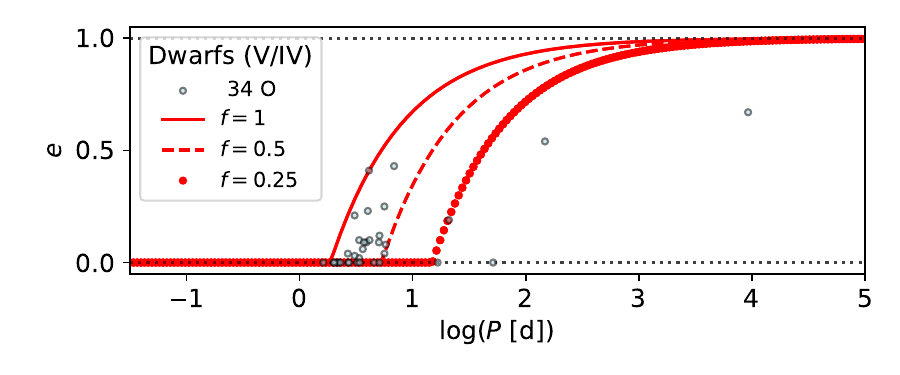}
    \includegraphics[width=0.49\linewidth]{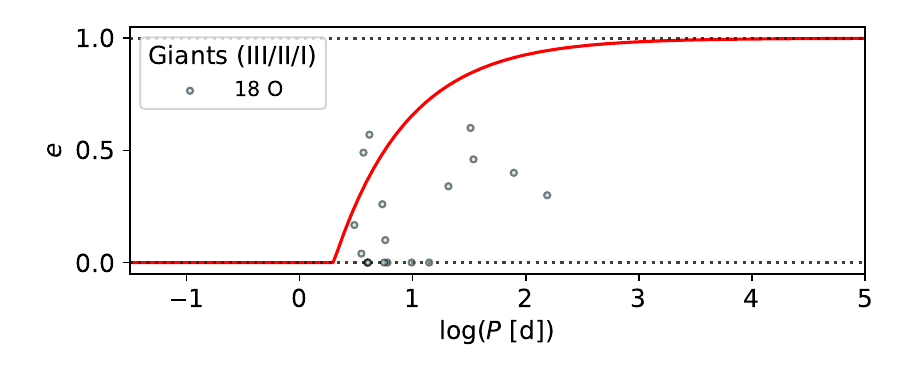}
    \vspace{-2em}
    \includegraphics[width=0.49\linewidth]{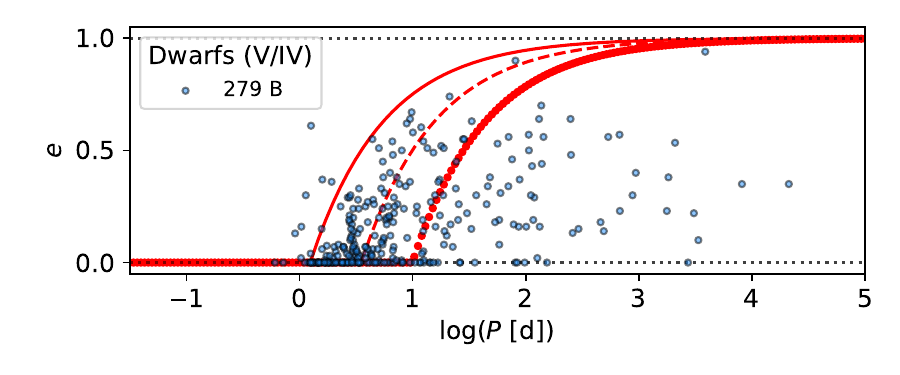}
    \includegraphics[width=0.49\linewidth]{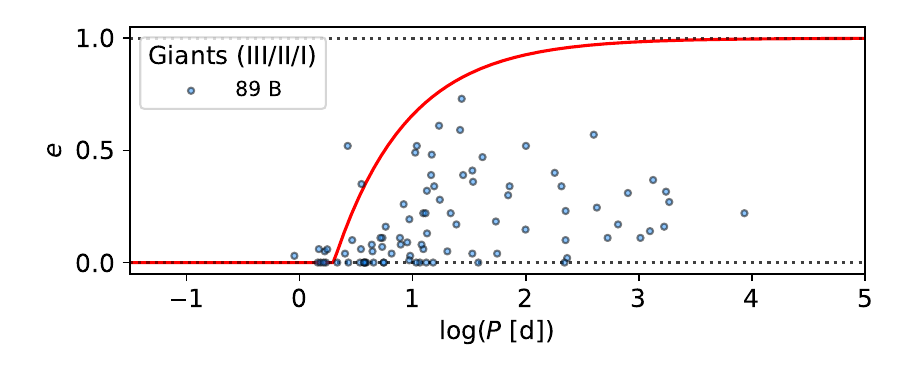}
    \vspace{-2em}
    \includegraphics[width=0.49\linewidth]{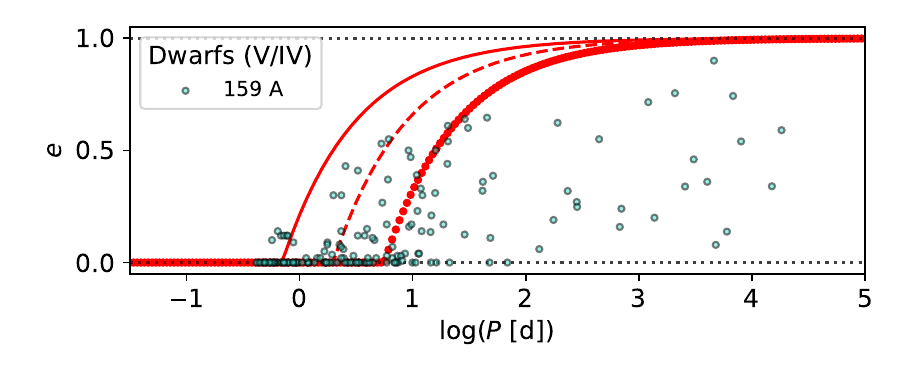}
    \includegraphics[width=0.49\linewidth]{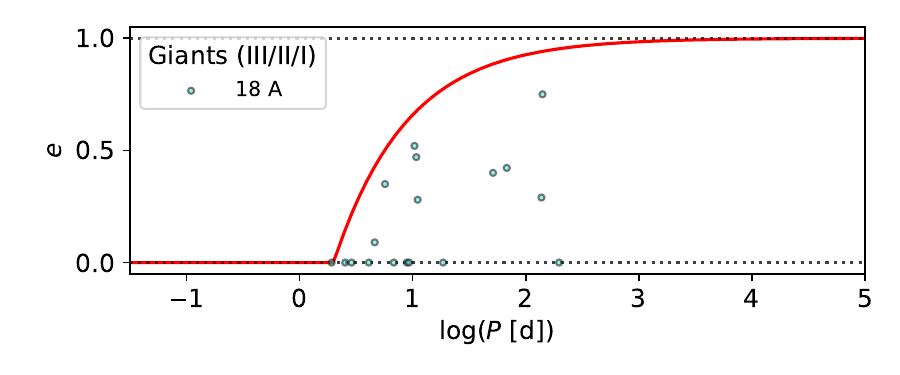}
    \vspace{-2em}
    \includegraphics[width=0.49\linewidth]{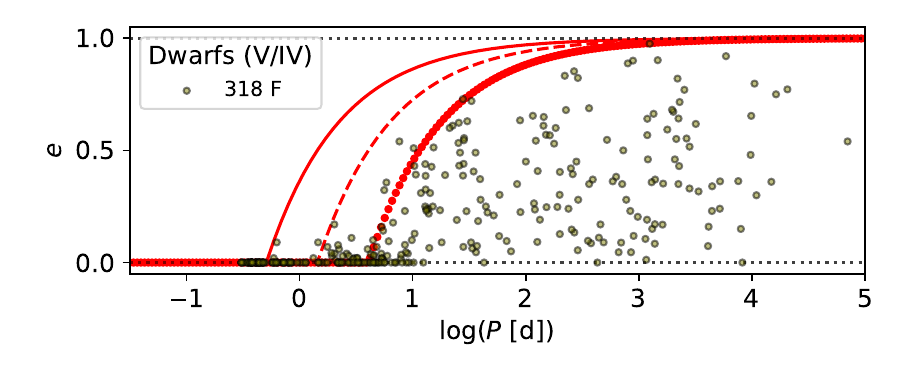}
    \includegraphics[width=0.49\linewidth]{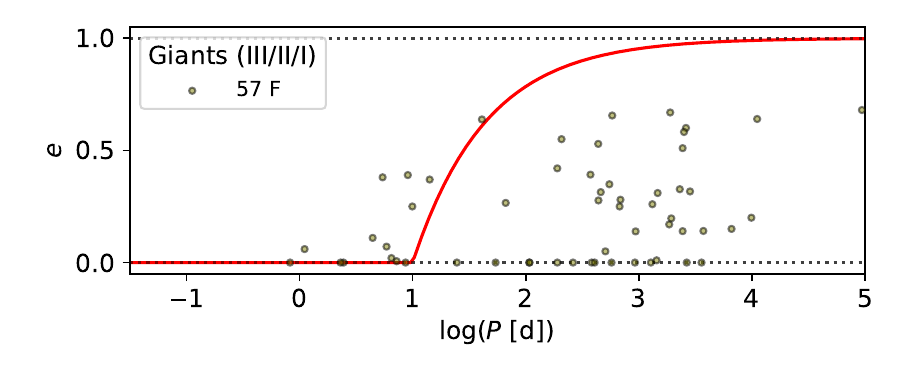}
    \vspace{-2em}
    \includegraphics[width=0.49\linewidth]{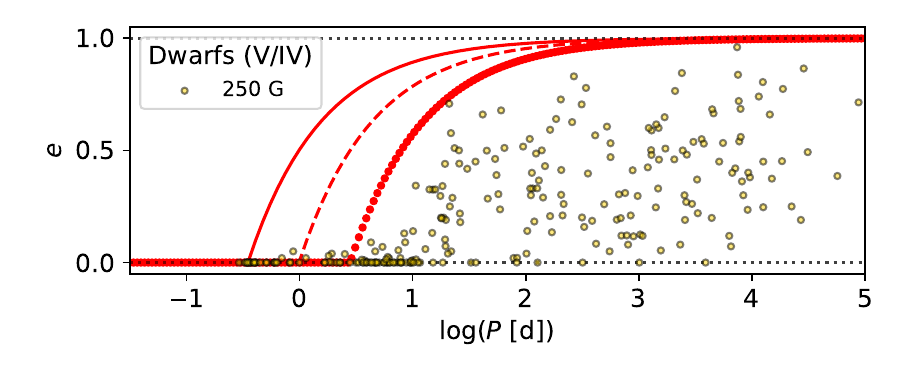}
    \includegraphics[width=0.49\linewidth]{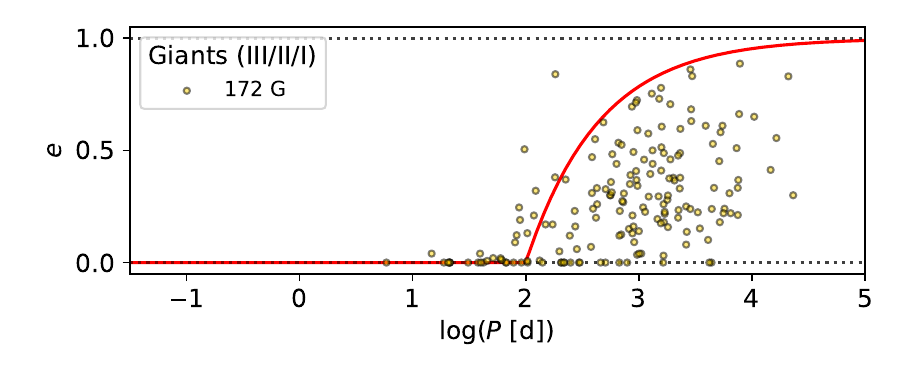}
    \vspace{-2em}
    \includegraphics[width=0.49\linewidth]{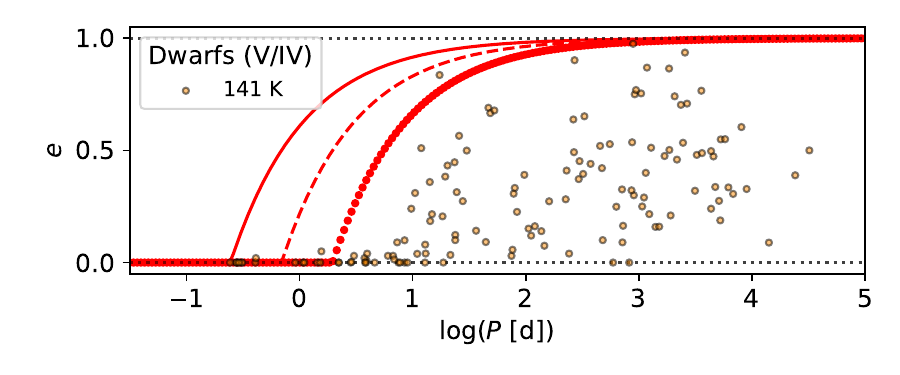}
    \includegraphics[width=0.49\linewidth]{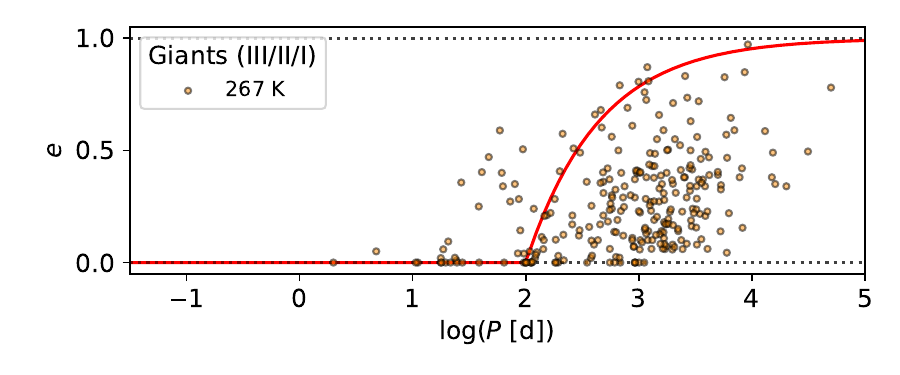}
    \vspace{-2em}
    \includegraphics[width=0.49\linewidth]{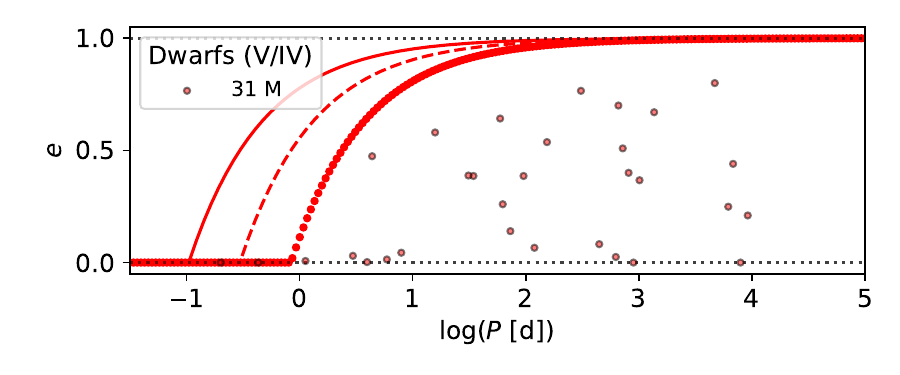}
    \includegraphics[width=0.49\linewidth]{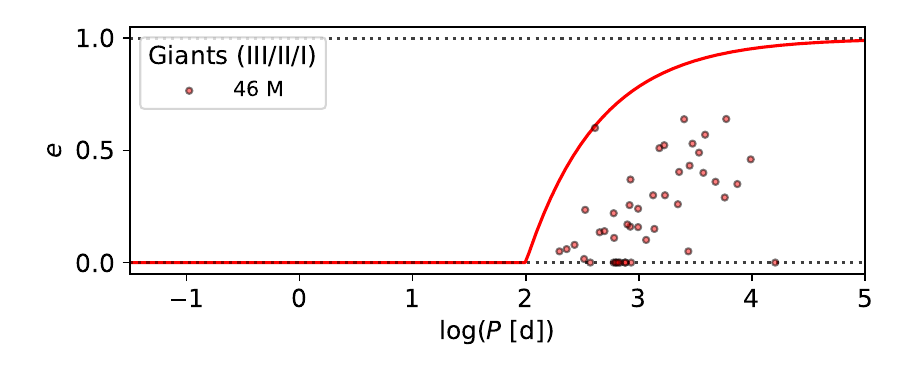}
    \vspace{1em}
    \vspace{1em}
    \caption{Eccentricity-period diagrams of \SBcat\ binaries per spectral type for dwarfs, corresponding to luminosity classes V and IV (left) and for giants, corresponding to luminosity classes III, II and I (right) for the SB primary component. Colour codes spectral type. \emph{Periastron} envelopes for three different values of the filling factor $f$ and their corresponding circularisation periods are displayed (Eq.~\ref{eq:Pc_f} and Table~\ref{tab:Pc_st}) for main sequence primaries (left panels). For giant primaries,  $f=1$ \emph{periastron} envelopes are displayed for $P_\mathrm{c}=2$~d (OBA giants), $P_\mathrm{c}=10$~d (F giants), and $P_\mathrm{c}=100$~d for GKM giants. }
    \label{fig:elogP_details}
\end{figure*}

\section{Cross-match Strategy}
\label{sec:strategy}
\begin{figure*}
    \includegraphics[width=\linewidth]{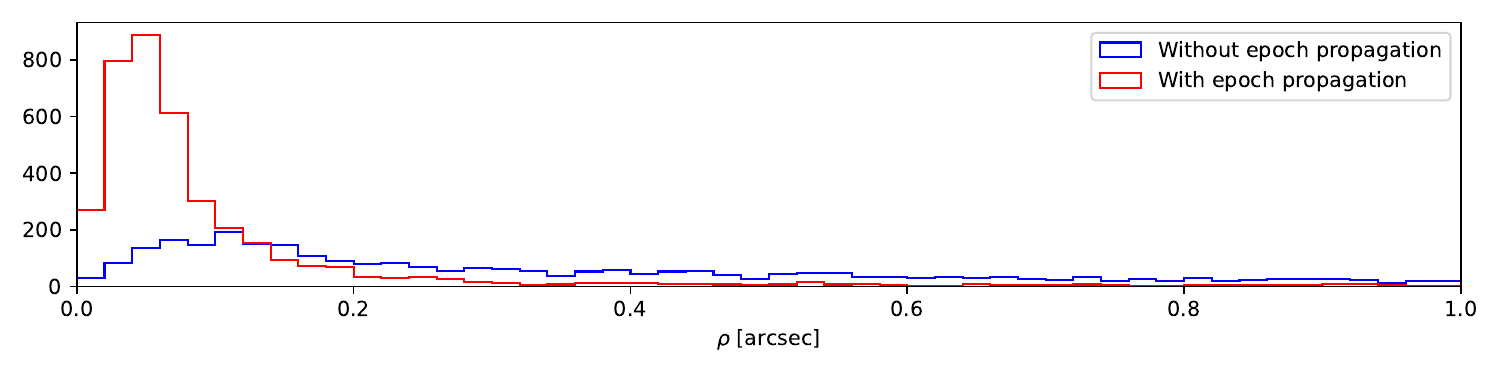}
    \caption{Cross-matched angular separation between \SBcat\ and \Gaia\ with (red) and without (blue) epoch propagation. The median angular separation is 0.24~arcsec without back propagation of the \Gaia\ epoch, and it is reduced to 0.06~arcsec with epoch propagation.}
    \label{fig:histo_sep}
\end{figure*}
\begin{figure}
    \centering
    \includegraphics[width=\linewidth]{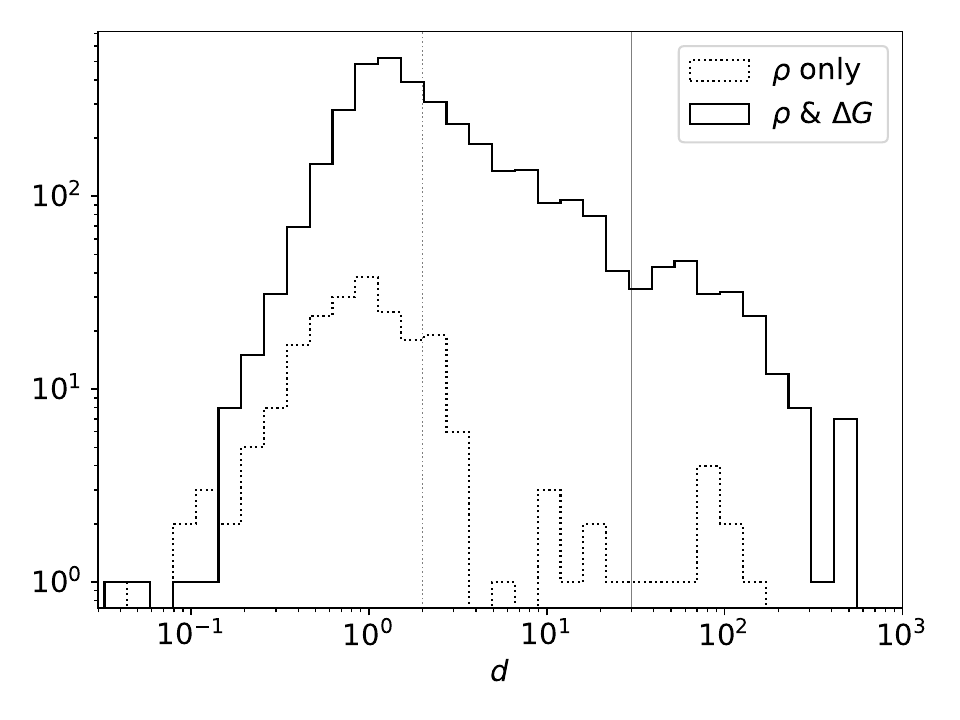}
    \caption{Distribution of Euclidean distances for one-to-one and one-to-many `best' matches, corresponding to 3712 systems. We manually check all systems with $d>2$ when only separation match $\rho$ is available; and with $d>30$ when both separation match and magnitudes difference $\Delta G$ are available.}
    \label{fig:de}
\end{figure}
Before performing the cross-match, we first uploaded the \SBcat\ catalogue from external TAP (CDS VizieR) to the \Gaia archive (in Advanced ADQL tab). Then, we used the  query:
\begin{verbatim}
SELECT *  
FROM "B/sb9/main"  
WHERE Seq <= 4021  
ORDER BY Seq
\end{verbatim}
where the condition on \texttt{Seq} is done to select the 2021-03-02 version of the catalogue. We reiterate this query for \texttt{"B/SB9/orbits"} and \texttt{"B/SB9/alias"}. We thus obtain 4021 systems and 5042 orbits as expected. From the resulting queries, we create user tables to allow ADQL queries from both \Gaia and \SBcat\ catalogues.
The  \SBcat\ coordinates from VizieR are claimed to be in the FK5 system, and are computed by VizieR\footnote{\url{https://cdsarc.cds.unistra.fr/viz-bin/cat/B/sb9}}. The difference between the FK5 (equinox J2000.0) and the ICRS systems is of the order of 50 mas \citep{1998A&A...331L..33F}, which is much larger than the  \Gaia\ DR3 median position uncertainties of 0.05 mas at $G<17$ (magnitudes in \SBcat\ reach $G=19$), but remains reasonably low compared to the size of the cone search radius used in the cross-match.

Prior to the cross-match, we select a first subset of \Gaia\ DR3 sources around the \SBcat\ coordinates with a cone search radius of 170 arcsec. Indeed, the displacement over 16 years of the Barnard's star, the star with the highest proper motion \citep{1916AJ.....29..181B}, is 166.133~arcsec, computed by retro-propagating its position from epoch J2016 to J2000.
With such a large cone, we obtain a subset of more than 1.8 million \Gaia\ matches, no limits on magnitude being imposed. 
The advantage of this method allows to compute the back-propagation only for this subset.
We make the epoch back-propagation with the ADQL function \texttt{EPOCH\_PROP\_POS} which relies on the 6 astrometric parameters (right ascension, declination, parallax, the two proper motions and the radial velocity) from the \Gaia\ DR3 epoch (J2016.0) to the \SBcat\ epoch (J2000.0). We note that the ADQL function, while returning right ascension, declination, parallax and the two proper motions in the same units as the function inputs, does  return the propagated radial velocity\footnote{It is actually the proper motion in the radial direction which is computed by the  \texttt{EPOCH\_PROP\_POS} ADQL function.} in mas~a$^{-1}$ and not in \kms; the returned quantity needs to be multiplied by $A/\varpi$, where $A$ is the astronomical unit ($A=4.740470446$~\kms~a) and $\varpi$ the parallax in mas, to obtain the radial velocity in \kms, as explained in the \Gaia~DR3 documentation\footnote{\href{https://gea.esac.esa.int/archive/documentation/GDR2/Data_processing/chap_cu3ast/sec_cu3ast_intro/ssec_cu3ast_intro_tansforms.html}{3.1.7 Transformations of astrometric data and error propagation}}.

Finally we performed the cross-match at epoch J2000.0 with a cone search radius of 10 arcsec. We obtain at least one cross-match pair for 3981 out of the 4021 \SBcat\ systems and 8710 unique \Gaia\ sources. Thanks to the epoch back-propagation, we recovered 26 \SBcat\ systems. In addition, the histograms of Fig.~\ref{fig:histo_sep} shows that the median of the cross-match separation $\rho$ is decreased by a factor of four (from 0.24 to 0.06 arcsec) when epoch propagation is considered. 
With this strategy, there were some \SBcat\ systems without \jkt{a} \Gaia\ match within 10 arcsec. 
Among them, we identified 21 \SBcat\ systems without \jkt{a} match in \Gaia~DR3 because \jkt{they are} brighter than $G=3$ (see Table~\ref{tab:too_bright}). \jkt{These} include many \jkt{of the} brightest binaries \jkt{in their} constellations, \jkt{such} as $\alpha$~Cen, Sirius, Procyon; triple systems like \jkt{Algol}; and even higher-order systems \jkt{such as} Capella ($\alpha$~Aur) and Castor A and B ($\alpha$ Gem).
At the opposite end, there are six \SBcat\ systems that are too faint to be observed by \Gaia (see Table~\ref{tab:too_faint}). They include a low-mass X-ray binary, UY Vol \citep{1986ApJ...306..599C}, a binary with a pulsar,  PSR~J0823+0159 \citep{1980ApJ...236L..25M} and two faint SB identified in SDSS \citep{2005A&A...444..643P}. At later stages, we also add the Hulse-Taylor binary pulsar \citep{1975ApJ...195L..51H} and PSR~J2305+4707 \citep{1985ApJ...294L..21S}. 

A basic search\footnote{Based on the Sesame resolver from \Simbad\ \citep{2000A&AS..143....9W}.} \jkt{in the} \Gaia\ archive for the three remaining systems without \Gaia\ matches revealed that the angular separation was larger than the 10 arcsec threshold chosen (\SBcat~562, 2098 and 2454 with approximately, 16, 11 and 42 arcsec). The treatment of these cases is performed separately and manually.

\section{Best Match Sanity checks}
\label{sec:sanity_checks}
For each \SBcat\ system, we select as `best' match the one with the lowest Euclidean distance. We display the distribution of these Euclidean distances in Fig.~\ref{fig:de}.  
We performed a manual check of \emph{oto} and \emph{otm} matches with the largest Euclidean distances. For the 245 \SBcat\ systems without magnitude (6\%), the Euclidean distance only relies on the angular separation.
For those, we manually check all the best matches with Euclidean distance $d>2$ corresponding to 40 system numbers. Among them we identify 3 \SBcat\ systems with erroneous `best' matches. For all the other systems with $\rho$ and $\Delta G$ available, we investigated all the \emph{oto} and \emph{otm} systems with $d>30$, which concern 232 system numbers ($\sim6$\%). Many of them are cataclysmic binaries, classical novae, symbiotic stars, hot subdwarfs, white dwarfs, etc. showing a large photometric discrepancy between \Gaia\ and \SBcat. Among them, we identify 26 \SBcat\ systems with erroneous `best' \emph{oto} and \emph{otm} matches. For instance \SBcat~1503 has wrongly been attributed the coordinates of HD~895, a Tycho visual binary \citep{2002A&A...384..180F}, while it actually is HD~895 C, a close SB2, making the overall system at least a quadruple. 

We also performed a manual check of all best matches that involve two or more \SBcat\ systems per \Gaia\ source (corresponding to \emph{mto} and \emph{mtm} matches). This involves inspecting the best matches attached to 124 unique \Gaia\ sources, corresponding to 253 unique \SBcat\ system numbers. It reveals the presence of many SB belonging to high-order systems (Sec.~\ref{sec:hom}) as well as some duplicates (Sec.~\ref{sec:dup}).

\section{\SBcat\ system duplicates}
\begin{table}
\centering
\caption{18 \SBcat\ system duplicates sorted per original system number.}
\begin{tabular}{lll}
\hline
duplicate & \SBcat\ system & Name \\
number & number & \\
\hline
2544 & 374\footnotemark[1] & $\mu$ Ori AB \\
2169 & 376\footnotemark[2] & 64 Ori \\
3779 & 1568 & HD 196673A \\
2917 & 1981 & NGC 3960 50 \\
2904 & 2359 & NGC 2489 25 \\
2927 & 2364 & NGC 5822 312 \\
2897 & 2489 & BD $-$14 2111 \\
2938 & 2484 & NGC 6940 111 \\
2939 & 2485 & NGC 6940 130 \\
2890 & 2487 & NGC 2360 51 \\
2893 & 2488 & BD $-$15 1742 \\
2923 & 2491 & NGC 5822 11 \\
2928 & 2494 & NGC 5823 34 \\
2875 & 2498 & NGC 2099 149 \\
2877 & 2503 & HD 248448 \\
2878 & 2506 & NGC 2099 966 \\
2879 & 2507 & NGC 2099 1505 \\
2902 & 2530 & NGC 2477 430 \\
\hline
\end{tabular}
\begin{flushleft}  
$^1$ Member of a spectroscopic quadruple with \SBcat~372 (SB1) and 373 (SB2); see Table~\ref{tab:sb4}.\\  
$^2$ Member of a spectroscopic triple with \SBcat~375.
\end{flushleft}
\label{tab:dup}
\end{table}

\section{Triple systems with orbits in \SBcat}
\begin{table*}
\centering
\scriptsize
\caption{List of 76 triple systems with \SBcat\ orbits.}
\begin{tabular}{lllll}
\hline
\SBcat & \Gaia & \Simbad\ & in NSS & Separation [arcsec] \\
\hline
27, 2545 & 2528433919872925312 & *  13 Cet\\
49, 1726 & 426512590631144704 & HD   5408\\
122, 123 & 457464770661735296 & V* DM Per\\
157, 158 & too bright & * bet Per\\
169, 2464 & 118986060277836160 & V* UX Ari\\
211, 212 & 3305012316783533056 & * lam Tau\\
226, 227 & 270632490692926720 & * b Per\\
262, 1629 & 3313241714640886400 & HD 285947\\
319, 320 & 3214077516844988800 & * eta Ori A\\
332, 333 & 3217626018825418752 & V* VV Ori\\
375, 376 & 3374718330328109440 & *  64 Ori\\
430, 431 & 3046036062400216576 & V* FZ CMa\\
490$^*$, 3941 & 654808519221768192  & V* UU Cnc & EB\\
609, 2617 & 5668792270155162240 & V* HS Hya\\
718, 1727 & 3695061317957036928 & * eta Vir\\
733$^*$, 3874 & 3962579368940811904 & HD 110195 & SB2\\
789, 790 & 6293889335197400064 & V* DL Vir\\
817, 818 & 1241620894326361344 & HD 129132\\
828, 2993 & 1391924687894823296 & V* TZ Boo\\
856, 2386 & 1374695306328813568 & *   7 CrB A\\
962, 963 & 4490117076492339712 & HD 157978\\
1023, 1024 & 4576326312902814208 & HD 165590A, HD 165590 \\
1028, 2463 & 4590455518346609792 & HD 166181\\
1055, 1056 & 4273231054929181824 & * d Ser A\\
1134, 1135 & 6770084041623596032 & * psi Sgr\\
1172, 2142 & 2031776202613700480 & V* SU Cyg\\
1223, 1224 & 6875990375999580544 & * bet02 Cap, * bet01 Cap\\
1253, 3632 & 1735416587180214528 & HD 196795 A, HD 196795\\
1280, 2460 & 1873073175244118784 & BD+40   883 A, BD+40   883\\
1288, 1289 & 1852355455594429440 & HD 201433A\\
1293, 1294 & 2191133748536265216 & HD 203025\\
1296, 1297 & 1745244296985069184 & HD 203345\\
1319, 1320 & 1741318868675197184 & V* EE Peg\\
1390, 1391 & 1982245024789617280 & HD 214608\\
1431, 1432 & 2658510028089009792 & HD 219018\\
1493, 1494 & 4472789731012285824 & BD+04  35622\\
1508, 3080 & 2231007778225666944 & HD 214511\\
1595, 1596 & 368060765779919360 & BD+38   118\\
1624, 1625$^*$ & 1070770191963666816 & HD  83270 & AB\\
1659, 1660$^*$ & 992435451683384320 & V* V455 Aur & SB2\\
1723, 1724 & 1683524861026802176 & HD 105287\\
1729, 1730 & 1746335802793081344 & HD 202908\\
1731, 1732 & 1346932740806216960 & V* V819 Her\\
1752, 1753 & 1652148132065326720, & HD 158209\\
1769, 1770 & 5350358069101053184 & V* V572 Car\\
1815, 1834 & 604921542069455104, 604921546365012480 & NGC  2682   131, NGC  2682   131 A & & 0.4\\
1835, 1836 & 6731200706257840384 & V* TY CrA\\
1919, 1977 & 664288577196528128 & HD  73174\\
2408, 2409 & 2768404669096691072 & V* LN Peg\\
2411, 2412 & 3582095053777917952 & V* HU Vir\\
2456, 2458 & 1975001785790477312 & V* SS Lac & SB2\\
2477, 2906 & 5727490488683081216 & BD-12  2381\\
2561, 2592 & 2791062373929675904 & HD   7119 & SB2\\
2562, 2591 & 859117605230828032 & AG+59  808\\
2588, 2594 & 2058736124602103552 & HD 191588\\
2602, 2603 & 5452239641134855168 & HD  97131\\
2604$^*$, 2605 & 1518695423639372160 & HD 109648 & SB2\\
2606, 2607 & 2169527108088349952 & V* V1061 Cyg\\
2682, 3927 & 4031448894654813696 & HD 103613\\
2683, 2684 & 4339465085630200960 & HD 152751\\
2754, 2755 & 478368788803049600 & HD  32893\\
2806, 2807 & 3916285149112698240 & HD 100518 A\\
3449$^*$, 3450 & 661306319407420160 & BD+19  2077 & SB2\\
3500, 3501 & 3465682748358563968 & HD 104471 A, HD 104471\\
3518, 3519 & 4618008180223988352 & HD  10800\\
3520, 3521 & 4755117967402573056 & HD  18198 A, HD  18198\\
3522, 3523 & 4614931093854496640 & HD  35877\\
3552$^*$, 3805 & 3378123861436589952 & HD  48565 & SB1\\
3615, 3616 & 458319228639270016 & HD  14039\\
3699, 3700 & 6898265656937739648 & HD 204236\\
3701, 3702 & 2619653390242676096 & HD 211276\\
3727, 3728 & 464891250151058816 & BD+60   585 & AB\\
3870, 3871 & 1517909994379553152 & HD 109803\\
3872, 3873 & 3959246783557460096 & HD 109954\\
3966, 3967$^*$ & 242935724068826368 & HD  20577 & SB1\\
4016, 4017 & 2653803053170408832 & HD 214169\\
\hline
\end{tabular}
\begin{flushleft}
Notes. \Gaia~DR3 separations are computed at epoch J2016.0.  
An asterisk (*) indicates membership in the clean common sample.
\end{flushleft}
\label{tab:triple}
\end{table*}

\section{Quadruple systems with orbits in \SBcat}
\begin{table*}
\centering
\scriptsize
\caption{List of the 34 quadruple systems with \SBcat\ orbits including two probable optical quadruples.}
\begin{tabular}{lllll}
\hline
\SBcat & \Gaia & \Simbad\ & in NSS & Separation [arcsec] \\
\hline
87, 1504 & 292304861302115072, 292304861302114816 & HD  10308, HD  10308 B  & & 10.3\\
112, 113 & 300312157810876544, 300312157810876160 & * iot Tri A, * iot Tri B & & 3.9\\
181, 1523 & 3263936692671872384, 3263936658312134272 & HD  22468, HD  22468 B & & 6.7\\
216$^*$, 2052 & 476270508301993344 & HD  25638 & EB\\
372, 373, 374 & 3329650894896257024 & * mu. Ori A, * mu. Ori B, * mu. Ori\\
461, 462 & too bright & * alf Gem B, * alf Gem A\\
514, 515 & 3072894691921171456 & HD  71663\\
532, 533 & 583151353574583552, 583151353574583424 & * eps Hya, * eps Hya C & & 2.8 \\
575, 1506 & 813111839701758080, 813111835405980928 & * A Hya, AG+40 1074 & & 25.0\\
593, 594 & 623076270045202048 & V* XY Leo\\
1948, 635 & 5350346970905044480, 5350346970887961984 & HD  93206, HD  93206 B & & 1.0\\
667, 668 & 756853643637996160, 756853643638639104 & * ksi UMa B, * ksi UMa A & & 1.8\\
734, 735 & 3528976341368900096, 3528976341368899712 & V* VV Crv, HD 110318 & & 5.2\\
764, 765 & 1563590579347125632, 1563590510627624064 & * zet01 UMa, * zet02 UMa & & 14.4 \\
858, 859 & 1217683254878932352& HD 140629A, HD 140629 B\\
885, 2403 & 4456633958128648832 & HD 144515\\
1524, 1525 & 150461092453290240 & * chi Tau B\\
1559, 1607 & 97746175687968768, 97746137032809984 & BD+21 255, TYC 1212-473-1 & -, SB &  53.5 (optical?)\\
1747, 1748 & 3115721215281475840, 3115721219581645952 & BD+02  1483 A, BD+02  1483 B & & 0.9\\
1773, 1774 & 5350357862947802880 & CPD-59  2636\\
1838, 1844 & 401249764784080000, 401249764784080384 & HD   6645, HD   6645 B & & 8.7\\
2394, 2461 & 4661730461966674304 & SK -67   18\\
2415, 2416 & 2869876947957928576 & HD 221264\\
2449, 2450 & 3534414590303807232, 3534414594600352896 & HD  98800 A, HD  98800 B & & 0.5\\
2620$^*$, 2621 & 3365396498949922432 & HD  49635 & AB\\
2700, 2701 & 389688601879413248, 389688606176165248 & HD   4134 A, HD   4134 B\\
2984, 3210 & 1587395899441029760 & V* ET Boo\\
2985, 3214 & 1700166038233698048 & HD 133767\\
2989, 2994 & 4175947159460432384 & HD 162905\\
3213, 4003, 4004 & 733549185449902208 & HD  95660\\
3539, 3540 & 3708802460231857664, 3708802464524486656 & HD 110555 A, HD 110555 B & & 0.6\\
3649, 3650 & 294090502544836992, 294090536905167744 & * phi Psc B, * phi Psc & & 7.6 \\
3839, 3840 & 2155746340444188544, 2155746413460589696 & HD 179332, BD+60 1892 & AB, - & 47.8 (optical?)\\
3875, 3876, 3877 & 1441569596393755648 & HD 117078 \\
\hline
\end{tabular}
\begin{flushleft}
Notes. \Gaia~DR3 separations are computed at epoch J2016.0.  
An asterisk (*) indicates inclusion in the clean common sample.
\end{flushleft}
\label{tab:quadruple}
\end{table*}

\section{High-order systems with orbits in \SBcat}
\begin{table*}
\centering
\scriptsize
\caption{List of the six higher-order multiple systems with \SBcat\ orbits.}
\begin{tabular}{llllll}
\hline
\SBcat & \Gaia & \Simbad\ & in NSS & Separation [arcsec] & MSC \\
\hline
257, 1527 & 3293714285052444800, 3293714491209540352 & *  88 Tau, *  88 Tau B & --, AB & 69.5 & sextuple\\
305, 1484 & 180702438220217344, 180702438220219904 & *  14 Aur, *  14 Aur B&&14.2& quintuple\\
340, 341 & 3017364132050194688 3017364132049943680 & * tet01 Ori A, * tet01 Ori B && 8.9 & Orion Trapezium\\
461, 462, 463 & too bright, too bright, 892348454394856064 & * alf Gem B, * alf Gem A, V* YY Gem & & & septuple\\
1510, 1511 & 1374656960860690560, 1374656960861310336 & HD 139691 A, NAME AG+36 1364 B & & 15.0 & sextuple\\
1023, 1024, 1986 & 4576326312902814208, 4576326312902814208, 4576326312901650560 & HD 165590 A, HD 165590, V* V885 Her & & 28.5 & quintuple\\
\hline
\end{tabular}
\label{tab:hom}
\end{table*}

\section{The common sample \SBcat $\cap$ NSS}
\begin{table*}
\centering
\scriptsize
\caption{The common sample of 827 binaries between \SBcat\ and \Gaia\ DR3 NSS.}
\begin{tabular}{lllllllll}
\hline
\texttt{sb9} & \texttt{gaia} & \texttt{simbad} & \texttt{outlier} & \texttt{in\_triple} & \texttt{in\_quadruple} & \texttt{in\_hom} & \texttt{nss\_solution\_type} & \texttt{family} \\
\hline
11 &	4924503334299333888	&HD 1273& False	& False	& False	& False & SB1\\
13 &	4689821167286415744	&V* AQ Tuc& False	& False	& False	& False	& EclipsingBinary\\
15 &	428246623544412416	&V* TV Cas& False	& False	& False	& False	& EclipsingBinary\\
16 &	2857501394830370944	&V* LR And& False	& False	& False	& False	& SB1\\
18 &	2862259565759107200 &HD 2019& True	& False	& False	& False	& SB2\\
...\\
216	& 476270508301993344 & HD 25638	& False	& False	& True	& False	& EclipsingBinary &	(216, 2052)\\
...\\
490	& 654808519221768192 & V* UU Cnc & False	& True	& False	& False	 &EclipsingBinary &	(490, 3941)\\
...\\
1527	& 3293714491209540352	&* 88 Tau B	&True	&False	&False	&True	&Orbital	&(257, 1527)\\
\hline
\end{tabular}
\begin{flushleft}
Notes. The column definitions are as follows:  
\texttt{sb9} is the \SBcat\ identifier, \texttt{gaia} is the \Gaia~DR3 \texttt{source\_id}, \texttt{simbad} is the main \Simbad\ identifier.  
\texttt{outlier} is true if the relative-period or absolute-eccentricity difference exceeds 10
\texttt{in\_triple}, \texttt{in\_quadruple}, and \texttt{in\_hom} flag multiplicity categories.  
\texttt{nss\_solution\_type} follows the DR3 Non-Single Star schema.  
The clean common sample corresponds to \texttt{outlier = False} and contains 655 systems.
\end{flushleft}
\label{tab:common_sample}
\end{table*}


\bsp	
\label{lastpage}
\end{document}